\documentclass[12pt,a4paper]{article}            
 \usepackage[skins,theorems]{tcolorbox}
\tcbset{highlight math style={enhanced,
  colframe=red,colback=white,arc=0pt,boxrule=1pt}}
  \usepackage[bookmarksopen, bookmarksnumbered, bookmarksopenlevel=2]{hyperref}
  \usepackage{tikz}
  \usepackage{tikz-3dplot}
 \usetikzlibrary{calc}
 \usetikzlibrary{decorations} %
 \usepackage[UKenglish]{babel}
 \usepackage[toc,page]{appendix}
 \usepackage{amsmath}
 \usepackage{amssymb}
 \usepackage{graphicx}
 \usepackage{hhline}
 \usepackage[bf]{caption}
\usepackage{cite}
\usepackage[vcentermath]{youngtab}
\usepackage{geometry}
\usepackage{slashed}
\usepackage{color}
\usepackage{stackrel}
\usepackage{tikz-cd} 
\usepackage{cancel} 

\usepackage{empheq}
\usepackage{arydshln}

\newcommand{\bqa}{\begin{eqnarray}}
\newcommand{\eqa}{\end{eqnarray}}

\newcommand{\nn}{\nonumber}

\def\hor{{\rm hor}}
\def\ver{{\rm vert}}
\def\ttw{{2,2}}
\def\del{\partial}
\def\hbas{G}
\def\Gfour{G}
\def\Gauge{{\CG}}
\def\qr{r}
\def\mq{\mathfrak{q}}

\hypersetup{
    pdftitle={},
    pdfauthor={},
    pdfsubject={}
}
\numberwithin{equation}{section}
\numberwithin{table}{section}

\makeatletter

\newtheorem{Lemma}{Lemma}

\newtheorem{proposition}{Proposition}

\newcommand{\be}{\begin{equation}}
\newcommand{\ee}{\end{equation}}
\newcommand{\beq}{\begin{equation}}
\newcommand{\eeq}{\end{equation}}
\newcommand{\ba}{\begin{aligned}}
\newcommand{\ea}{\end{aligned}}

\newcommand{\bea}{\begin{eqnarray}}
\newcommand{\eea}{\end{eqnarray}}

\newcommand{\cO}{\mathcal{O}}

\newcommand{\cC}{\mathcal{C}}

\newcommand{\cL}{\mathcal{L}}
\newcommand{\cS}{\mathcal{S}}

\newcommand{\cF}{\mathcal{F}}
\newcommand{\cI}{\mathcal{I}}

\newcommand{\cR}{\mathcal{R}}

\newcommand{\cM}{\mathcal M}

\def\unit{{1\kern-.65ex {\rm l}}}
\def\1{{1\kern-.65ex {\rm l}}}

\def\slash#1{\ooalign{\hfil/\hfil\crcr$#1$}}

\def\IZ{\mathbb{Z}}
\def\IP{\mathbb{P}}

\def\CG{{\cal G}}

\def\IR{{\mathbb{R}}}

\newcount\hour \newcount\minute
\hour=\time \divide \hour by 60
\minute=\time
\def\now{%
\ifnum \hour<13
  \ifnum \hour=0 \advance \hour by 12 \number\hour:\else \number\hour:\fi%
     \ifnum \minute<10 0\fi%
     \number\minute%
\ A.M.%
\else \advance \hour by -12 \number\hour:%
  \ifnum \minute<10 0\fi%
  \number\minute%
  \ P.M.%
\fi%
}

\def\fnote#1#2{\begingroup\def\thefootnote{#1}\footnote{#2}
     \addtocounter{footnote}{-1}\endgroup}

\makeatother

\begin{document}

\begin{flushright}
{\tt\normalsize CERN-TH-2018-280}\\
\end{flushright}

\vskip 40 pt
\begin{center}
{\large \bf Modular Fluxes, Elliptic Genera, and  \vspace{2mm} \\Weak Gravity Conjectures in Four Dimensions
} 

\vskip 11 mm

Seung-Joo Lee${}^{1}$, Wolfgang Lerche${}^{1}$,
and Timo Weigand${}^{1,2}$

\vskip 11 mm

\small ${}^{1}${\it CERN, Theory Department, \\ 1 Esplande des Particules, Geneva 23, CH-1211, Switzerland} \\[3 mm]
\small ${}^{2}${\it PRISMA Cluster of Excellence and Mainz Institute for Theoretical Physics, \\
Johannes Gutenberg-Universit\"at, 55099 Mainz, Germany}

\fnote{}{seung.joo.lee, wolfgang.lerche,  
timo.weigand @cern.ch}

\end{center}

\vskip 7mm

\begin{abstract}

We analyse the Weak Gravity Conjecture for chiral four-dimensional F-theory compactifications with $N=1$ supersymmetry.
Extending our previous work on nearly tensionless heterotic strings in six dimensions, we show that under certain assumptions
a tower of asymptotically massless states arises in the limit of vanishing coupling of a $U(1)$ gauge symmetry coupled to gravity.
This tower contains super-extremal states whose charge-to-mass ratios 
are larger than those of certain extremal dilatonic Reissner-Nordstr\"om black holes, precisely as required by the
Weak Gravity Conjecture. Unlike in six dimensions, the tower of super-extremal states does not always populate a 
charge sub-lattice.

The main tool for our analysis is the elliptic genus of the emergent heterotic string in the chiral $N=1$ supersymmetric effective 
theories. This also governs situations where the heterotic string is non-perturbative.
We show how it can be computed in terms of BPS invariants on elliptic four-folds, by making use of various dualities and mirror symmetry.
Compared to six dimensions, the geometry of the relevant elliptically fibered four-folds
is substantially richer than that of the three-folds, and we classify the possibilities for obtaining critical, nearly tensionless heterotic strings.
We find that the (quasi-)modular properties of the elliptic genus crucially depend on the choice of flux background.
Our general results are illustrated in a detailed example. 
\end{abstract}

\vfill

\thispagestyle{empty}
\setcounter{page}{0}
\newpage

\tableofcontents

\thispagestyle{empty}
\setcounter{page}{1}
\newpage

\section{Introduction}

It is one of the celebrated properties of string theory that it relates deep properties of quantum gravity and field theory in dimensions lower than ten to the geometry
of the space on which the theory is compactified.
Recently, various general conjectures about the structure of any consistent theory of quantum gravity have attracted considerable interest
of string theorists and non-string theorists alike. String compactifications offer an outstanding opportunity to test these conjectures 
and to uncover the mathematical structures of the compactification space to which they point. 

Among the earliest such conjectures is the Weak Gravity Conjecture (WGC) \cite{ArkaniHamed:2006dz}, which asserts that in any gauge theory coupled to quantum gravity there should exist some set of particles whose charge-to-mass ratio exceeds that of 
an extremal black hole. 
In various stronger forms, the set of such particles must be an infinite tower of states, possibly even populating a (sub-)lattice of the charge lattice \cite{Cheung:2014vva,Heidenreich:2015nta,Heidenreich:2016aqi,Andriolo:2018lvp}.
The WGC is related to other conjectures, such as the Swampland Distance Conjecture \cite{Ooguri:2006in}, further refined and investigated in \cite{Klaewer:2016kiy,Palti:2017elp,Heidenreich:2017sim,Andriolo:2018lvp,Heidenreich:2018kpg,Grimm:2018ohb,Blumenhagen:2018nts,Lee:2018urn}, in that the infinite tower of particles in question becomes light at infinite distance in moduli space where the gauge coupling asymptotes to zero \cite{Grimm:2018ohb,Lee:2018urn,Lee:2018spm,Grimm:2018cpv,Corvilain:2018lgw}.
In turn, this can be viewed as a microscopic obstruction against
 the coexistence of a global symmetry \cite{Banks:1988yz,Banks:2010zn,Harlow:2018tng} with quantum gravity.
Related versions of the WGC have potentially important consequences for large-field inflation as studied e.g. in \cite{Rudelius:2014wla,delaFuente:2014aca,Rudelius:2015xta,Montero:2015ofa,Brown:2015iha,Bachlechner:2015qja,Hebecker:2015rya,Brown:2015lia,Heidenreich:2015wga,Kooner:2015rza,Ibanez:2015fcv,Hebecker:2015zss,Baume:2016psm,Heidenreich:2016jrl,Blumenhagen:2017cxt,Valenzuela:2017bvg,Ibanez:2017vfl,Aldazabal:2018nsj,Blumenhagen:2018nts,Blumenhagen:2018hsh,Reece:2018zvv,Hebecker:2017uix,Montero:2017yja,Hebecker:2018ofv,Hebecker:2018yxs} and many other aspects of high energy physics including \cite{Reece:2018zvv,Ooguri:2016pdq,Ibanez:2017kvh,Hamada:2017yji,Ibanez:2017oqr,Lust:2017wrl,Gonzalo:2018dxi,Klaewer:2018yxi,Craig:2018yvw}.

Various general arguments have been put forward \cite{Nakayama:2015hga,Harlow:2015lma,Cottrell:2016bty,Benjamin:2016fhe,Montero:2016tif,Fisher:2017dbc,Cheung:2018cwt,Andriolo:2018lvp,Hamada:2018dde,Montero:2018fns} to prove the WGC beyond the original heuristic idea that extremal black holes should be able to decay. Direct evidence in string compactifications with more than eight supercharges has been given in \cite{ArkaniHamed:2006dz,Heidenreich:2016aqi}.

In \cite{Lee:2018urn,Lee:2018spm} we have started to quantitatively verify the WGC from the perspective of string compactifications on non-trivial
 backgrounds with eight supercharges.
In six-dimensional compactifications of F-theory, the WGC can be proven to hold near the asymptotic weak coupling point, with the crucial help of modularity of the underlying theory \cite{Lee:2018urn}.
The tower of states satisfying the conjecture indeed contains a sublattice of the charge lattice, as conjectured in \cite{Heidenreich:2016aqi}. As additional input from string theory, the rank of the sublattice is determined geometrically in terms of the height-pairing of a rational section of an elliptic fibration. The states arise as  excitations of a solitonic string in F-theory, which becomes tensionless near the weak coupling point.
In that limit, the string is dual to the critical, weakly coupled heterotic string compactified to six dimensions.
The tower of excitations relevant for proving the WGC lies in a subsector of the spectrum, which is encoded in an appropriately defined elliptic genus of the string.
This elliptic genus is in general a ratio of quasi-modular weak Jacobi forms, and extends the classic perturbative results of \cite{Schellekens:1986xh} that are
based on conformal field theory. 
Its (quasi-)modular properties are the key to proving the WGC: A theta-function decomposition yields directly the charge-to-mass ratio for a subset of the string excitations, which is in perfect agreement with the charge-to-mass ratio of extremal dilatonic \cite{Heidenreich:2015nta,Lee:2018spm} charged black holes. The latter point 
also reflects the importance of taking into account the effect of scalar fields \cite{Palti:2017elp}. 

One aim  in this paper is to extend our previous work to four-dimensional compactifications with $N=1$ supersymmetry. 
While at first similar, the four-dimensional theories
turn out to be substantially different compared to the six-dimensional ones, and this is why it is worthwhile to analyze them.
 A  discussion of swampland conjectures in four-dimensional, $N=1$ supersymmetric compactifications has recently been given, with different focus, in \cite{Gonzalo:2018guu}.

Specifically, we will consider compactifications of F-theory on elliptically fibered Calabi-Yau four-folds, $Y_4$, and identify 
solitonic strings that are obtained from  D3-branes wrapping certain curves $C_0$ in their base, $B_3$.
As we will show, such strings are guaranteed to become asymptotically tensionless in the limit where the gauge coupling of a $U(1)$ gauge group vanishes (while the Planck mass is kept fixed). 
Under certain conditions, the curves $C_0$ are rational with trivial normal bundle; as a result, the solitonic strings correspond to critical heterotic strings, now
compactified to four dimensions.
One of our goals is to show that a subsector of their excitations, namely the one which is counted by the elliptic genus, indeed
satisfies the Weak Gravity Conjecture.

In view of the special role played by modularity in six dimensions, our first task is therefore 
to analyze the modular properties of the elliptic genus of this kind of four-dimensional, critical heterotic strings.
Concretely,  as compared to six dimensions, the four-dimensional compactifications we consider have the following special features:
\begin{enumerate}
\item 
A non-trivial background flux is necessary to obtain a chiral four-dimensional, $N=1$ supersymmetric
 effective theory with non-vanishing elliptic genus.
The possible fluxes are constrained by the requirement that the theory can be uplifted from
two to four dimensions, and also by
certain quantization and supersymmetry conditions. 
\item 
The elliptic genus is proportional to ${\rm Tr} \, Q$, and so it can be non-vanishing only in theories
with an ``anomalous'' $U(1)$ gauge group.  For perturbative heterotic strings based on conformal field theory, it must be a
 Jacobi modular form, $\varphi_{w,m}$, of weight $w=-1$ and of some integral index $m\geq2$ (which depends on the model).
For a more general, non-perturbative effective string arising from F-theory, the modular properties of the elliptic genus 
crucially depend on the flux, potentially rendering it quasi-modular or worse.
\item
The fluxes induce a $D$-term potential, which may obstruct the weak coupling limit.
A special subset of those fluxes which do not lead to such an obstruction are the ones which lead to a modular or quasi-modular elliptic genus. 
Viewed from a heterotic duality frame, this means that the $D$-term potential is generated at one-loop only and hence automatically vanishes in the weak coupling limit. 
Moreover, if the $D$-term is independent of NS5-brane moduli \cite{Honecker:2006dt,Blumenhagen:2006ux}, the heterotic string becomes perturbative and hence its elliptic genus is fully modular as opposed to only quasi-modular.
A tree-level involvement of the K\"ahler moduli in the $D$-term, by contrast, leads to severe distortions of modularity.

\item
An important property of odd weight Jacobi forms is that they are anti-symmetric in the fugacity parameter (i.e., the $U(1)$ field strength).
As a consequence, their theta-expansion leads to certain gaps in the charge spectrum. 
Thus,  for the special subset of (quasi-)modular fluxes, the super-extremal states do not form a charge sublattice, at least  as far as their index is concerned, while they still form an infinite tower. On the other hand, for generic non-modular fluxes,
for which the elliptic genus is not a weak Jacobi form, we expect 
these gaps to disappear, and we have verified this in examples.
Despite this fact, the super-extremal states generically continue not to form a charge sublattice (though there can be exceptions).

\end{enumerate}

The actual computation of the elliptic genus proceeds by exploiting the duality 
between F-theory on a circle and M-theory. 
This is similar to the logic for six-dimensional compactifications of F-theory \cite{Klemm:1996hh,Haghighat:2013gba,Haghighat:2014vxa}.
Indeed also in four dimensions the (non-BPS) excitations of the string
relate to BPS particles that appear when the theory is reduced on $S^1$.  This theory in three dimensions is dual to M-theory compactified on $Y_4$,
and here the BPS particles arise from M2-branes wrapping 2-cycles in the four-fold $Y_4$.
Counting (non-BPS) string excitations in four dimensions is therefore equivalent to counting the BPS invariants of 
these 2-cycles in three dimensions. Note that these invariants do crucially depend on the background four-flux, which is the main new
feature in going from six to four dimensions.
This computation is performed with the help of mirror symmetry, by considering Type IIA string theory on $Y_4$ in the presence of fluxes,
 and determining the flux-induced free energy of the two-dimensional effective theory. 
 The BPS invariants can then be inferred from the instanton expansion of this free energy.

While the first part of this article is concerned with investigating the modular properties of the elliptic genus of a heterotic  string that arises
from F-theory, we will subsequently make the connection between this string and the Weak Gravity Conjecture in the four-dimensional $N=1$ supersymmetric
field theory.
Despite similarities to the six-dimensional analysis in \cite{Lee:2018urn,Lee:2018spm}, we will  find a number of important differences:

\begin{enumerate}
\item
There are two qualitatively different ways to take the weak coupling limit of a gauge theory in four-dimensional F-theory in the presence of gravity.
In the first type, a heterotic string becomes asymptotically tensionless and weakly coupled, similar to the six-dimensional situation. 
We are able to prove that it contains an infinite tower of states which satisfy the super-extremality condition with respect to certain charged dilatonic black holes.

\item
The second type of limit is guaranteed to contain a {\it classically} asymptotically tensionless string, which in general need not be a heterotic string because it arises from a D3-brane on a curve  which can have a positive normal bundle. 
This phenomenon has no analogue in six dimensions.
Analysing this type of string is harder, because it is a priori
not clear  if it is weakly coupled. Nevertheless, at the level of classical geometry, the tensionless string leads to a breakdown of the effective field theory as we approach the limit at infinite distance where the gauge coupling vanishes. Since this is what is required by the swampland conjectures, it is therefore natural to speculate that this behaviour persists upon taking into account quantum corrections to the volume of the wrapped curve. 

\item
The $U(1)$ gauge field acquires a St\"uckelberg mass via the Green-Schwarz anomaly cancelling term, $S_{GS}\sim\int B \wedge F$,
which is related by supersymmetry to the $D$-term \cite{Lerche:1988zy}. Both terms necessarily appear at one-loop order,
if the elliptic genus is (quasi-)modular and non-zero.  In this case 
the St\"uckelberg mass is parametrically smaller than the Kaluza-Klein scale in the asymptotic weak coupling limit, and for the purpose of discussing the Weak Gravity Conjecture,
we can consider the $U(1)$ gauge field as effectively being massless. On the other hand,
for generic non-modular fluxes, no such suppression occurs, and the St\"uckelberg mass and D-terms may appear already at tree level.
The presence of an unsuppressed St\"uckelberg mass for the $U(1)$ gauge field makes 
the interpretation of the Weak Gravity Conjecture bound particularly interesting. 

\end{enumerate}

This article is structured as follows: In Section \ref{subsec_warmup} we give an overview of fluxes and mirror symmetry pertinent to this discussion, followed by a 
recapitulation of the modular properties of the four-dimensional elliptic genus in Section \ref{subsec_ellgen}.
In Section \ref{sec_ellgen4foldflux}, we describe in more detail the solitonic strings that arise in four-dimensional F-theory,  which are dual to critical heterotic strings.
In particular, we will motivate the relation between their elliptic genera and the flux-dependent Gromov-Witten invariants.
In Section \ref{sec:fluxes}, we discuss the constraints on $U(1)$ fluxes in order for the elliptic genus to be modular or quasi-modular. 
These conditions are first analysed from the F-theory perspective and then translated to the heterotic duality frame. 
We present a detailed  example encompassing these aspects of fluxes versus modularity in Section \ref{sec_Example1}. 
The  relation to the Weak Gravity Conjecture is analysed in Section \ref{sec_WGC}: After showing that  -- modulo the caveats above -- the weak coupling limit gives rise to an asymptotically tensionless, critical heterotic string,
we use the (quasi-)modularity properties of its elliptic genus to prove the Weak Gravity Conjecture whenever a limit of this type is available.
We then analyze the parametric behaviour of the St\"uckbelberg mass for the $U(1)$ gauge field for different types of gauge fluxes.
 In the interest of readability, many technical details and further illustrative examples are relegated to appendices.
We conclude with a discussion of open questions in Section \ref{sec_conc}.

\section{From mirror symmetry to the elliptic genus} \label{Mirrorandellgen}

We consider F-theory compactified on an elliptically fibered Calabi-Yau 4-fold $Y_4$, whose base space we denote by $B_3$. As is well-known, this 
leads to a theory in four dimensions with $N=1$ supersymmetry; some aspects of this theory that are most relevant to our discussion will be reviewed in Section \ref{subsec_Geomset}.
Upon an additional compactification on a circle, the theory becomes dual to M-theory compactified on the same $Y_4$. A further circle compactification relates this to
Type IIA string theory compactified on $Y_4$, which yields $N=(2,2)$ supersymmetry in two dimensions: \\

\begin{minipage}{4cm}
F-theory on \vspace{1mm}\\
$\mathbb R^{1,1} \times S^1_A \times S^1_B \times Y_4$  
\end{minipage}
\begin{minipage}{2cm}
$\longleftrightarrow$
\end{minipage}
\begin{minipage}{4cm}
M-theory on \vspace{1mm}\\
$\mathbb R^{1,1} \times S^1_B \times Y_4$  
\end{minipage}
\begin{minipage}{2cm}
$\longleftrightarrow$
\end{minipage}
\begin{minipage}{4cm}
Type IIA theory on \vspace{1mm} \\
$\mathbb R^{1,1}  \times Y_4$ 
\end{minipage}
\\

Our primary goal is to compute the elliptic genus of a certain solitonic string in the four-dimensional F-theory compactification. 
The string arises from a D3-brane wrapping a distinguished curve in the base space $B_3$, and can be interpreted as a critical heterotic string in four dimensions.
More details on the solitonic string and its interpretation as a heterotic string will be given in Section \ref{sec:D3-heterotic}.
The elliptic genus represents a certain index of (non-BPS) excitations of this four-dimensional string. 
Via the chain of dualities depicted above, the index-like multiplicities of these (non-BPS) string excitations can be  mapped to certain BPS invariants  of  the two-dimensional theory shown on the right. These can
 in turn be computed via mirror symmetry of the Type IIA string compactification \cite{Greene:1993vm}.

In the next section, we will informally introduce the objects that are relevant for this application of two-dimensional mirror symmetry. 
Subsequently we will introduce the elliptic genus in a similar vein.

\subsection{Mirror symmetry with fluxes and the free energy} \label{subsec_warmup}

As is well-known \cite{
Witten:1996md,
Mayr:1996sh,
Klemm:1996ts,
Gukov:1999ya},
in order to fully define the four-dimensional  F-theory (or two-dimensional Type IIA) compactification on $Y_4$, we must specify a background four-form flux, $\Gfour\in H^{4}(Y_4)$, 
in addition to the geometry of the elliptic fibration. In the sequel it is always assumed that such a flux has been activated.  
 Many properties of the effective theory will depend on the choice of flux, in particular whether the theory is chiral and/or supersymmertry is broken. 
The structure of $H^{4}(Y_4)$ is quite intricate, and in particular 
the piece that is relevant to us splits into a horizontal and a vertical part \cite{Greene:1993vm},\footnote {There may be a further part that 
is neither horizontal nor vertical \cite{Braun:2014xka}, but this is not important for our purposes.}
\be
H^\ttw(Y_4)\ =\  H^\ttw_\hor(Y_4)\oplus H^\ttw_\ver(Y_4)\ ,
 \ee
 which exchange under mirror symmetry. 
 A given horizontal flux  $\Gfour^{(i)}_{\rm hor}\in H^\ttw_\hor(Y_4)$, $i=1,...,h^\ttw_\hor$, generates a superpotential 
 \cite{Gukov:1999ya} 
 \be\label{fluxpot}
\cF^{(i)}(\alpha)\ =\ \int G^{(i)}_{\hor}\wedge \Omega^{4,0}(\alpha)\,
 \ee
 both from the perspective of the four-dimensional F-theory and the two-dimensional Type IIA theory.
As indicated the superpotential  is a function of the complex structure moduli, $\alpha$, of $Y_4$,
 and the generation of such flux-induced potentials has been extensively discussed in the literature (see e.g. 
\cite {Mayr:1996sh,
Klemm:1996ts,
Lerche:1997zb,
Gukov:1999ya,
Blumenhagen:2006ci,
Jockers:2009ti,
Alim:2009bx,
Grimm:2009ef,
Grimm:2009sy,
Grimm:2010ks,
Braun:2011zm,
Grimm:2011fx,
Krause:2011xj,
Intriligator:2012ue,
Braun:2014pva,Braun:2014xka}).
 
 In this paper we will be interested in different fluxes, namely in vertical fluxes
 \be
 \Gfour^{(j)}_{\rm vert}\in H^\ttw_{\rm vert}(Y_4),\qquad  j=1,...,h^\ttw_{\rm vert} \,.
 \ee  
 More precisely,
 we will focus on vertical fluxes that are associated with extra $U(1)$ factors in the four-dimensional F-theory, and accordingly we split
 \be\label{H22split}
H^\ttw_\ver(Y_4)\ =\  H^\ttw_{\ver,0}(Y_4)\oplus \widetilde H^\ttw_{\ver,U(1)}(Y_4)\ .
 \ee
 While both types of vertical fluxes can be switched on for compactification of Type IIA string theory on $Y_4$,
only the second component of $H^\ttw_\ver(Y_4)$ can be lifted to four-dimensional F-theory, where it leads to
 a chiral spectrum and an anomalous $U(1)$ gauge symmetry. 
 This will be reviewed in more detail in Section \ref{subsec_Geomset}.
 Moreover we can always choose  an integral basis of
 $H^\ttw_\ver(Y_4)$ such that the intersection metric takes the following symmetric, block diagonal form
 \be\label{etaform}
 \eta\ =\    \left(\begin{array}{c;{2pt/2pt}r}
    \begin{matrix} & * \\ * &  \end{matrix}    & \begin{matrix}  \\  \end{matrix} \\ \hdashline[2pt/2pt]
    \begin{matrix}  &  \end{matrix} & *
    \end{array}
    \right),
 \ee
which matches the splitting  (\ref{H22split}).  That is, $\widetilde H^\ttw_{\ver,U(1)}(Y_4)$  is equipped with
 a non-vanishing self-intersection form, while 
$ H^\ttw_{\ver,0}(Y_4)$ splits into further pieces that have an off-diagonal intersection form. The significance of this basis will be
that it corresponds to an eigenbasis under modular transformations. This was first observed in \cite{Haghighat:2015qdq,Cota:2017aal}  for fluxes in $H^\ttw_{\ver,0}(Y_4)$, while the use of fluxes in $\tilde H^\ttw_{\ver,U(1)}(Y_4)$
has not been considered in the context of mirror symmetry before.

As indicated at the beginning of this section, one of the goals in this paper is to investigate the elliptic genus of a heterotic string in a 
chiral four-dimensional, $N=1$ supersymmetric F-theory compactification
 in the presence of $U(1)$ fluxes
 \be
 \Gfour_j\equiv {\Gfour}^{(j)}_{\ver,U(1)}\in \widetilde H^\ttw_{\ver,U(1)}(Y_4) \,.
 \ee 
The multiplicities of string excitations that contribute to the elliptic genus are encoded in a family of free energies, $\cF_{\Gfour_j}(t)$, which are labelled by
these fluxes.
In terms of the flat K\"ahler moduli and their exponentials
\be
t_a, \qquad  a=1,..,h^{1,1}(Y_4) \,,\qquad  q_a\equiv e^{2\pi it_a}\,,
\ee  
any such flux-induced free energy has the following generic instanton expansion:
\be\label{Fjexp}
\cF^{(j)}(t)\ :=\  \cF_{\Gfour_j}(t)\ =\ \sum_{\vec\beta>0} N_{\vec \beta}^{(j)} \,{\rm Li}_2(q^{\vec \beta})\ ,\qquad j=1,...,\tilde h^\ttw_{\ver,U(1)}\,.
\ee
Here  $N_{\vec \beta}^{(j)}$ denote the four-fold generalization  \cite{Klemm:2007in} of the integral  Gopakumar-Vafa
\cite{Gopakumar:1998ii,Gopakumar:1998jq} invariants, and
 the sum is over all effective classes, $\vec\beta = \beta^a C_a$, where $C_a$ denotes some basis of $H_2(Y_4)$.
Furthermore $q^{\vec \beta}\equiv q_1^{\beta^1}...q_{h^{1,1}}^{\beta^{h^{1,1}}} $ encodes the multi-degree of the respective curve wrappings.
In (\ref{Fjexp}) we have omitted possible classical terms which are polynomial in $t_a$. The fundamental correlation functions of the two-dimensional $N=(2,2)$ 
supersymmetric theory are then given by
\cite{Mayr:1996sh,Klemm:1996ts}
\be\label{Ciab}
C_{jab}(t) \ = \ \partial_{t_a}\partial_{t_b}\cF^{(j)}(t).
\ee
Importantly, the  free energy $\cF^{(j)}(t)$ is the 
the A-model mirror of the flux superpotential (\ref{fluxpot}), and can hence be computed by use of mirror symmetry, now applied to Type II strings compactified on
Calabi-Yau four-folds \cite{Greene:1993vm}. 
For the $U(1)$ fluxes $\Gfour_j$ we consider,
the primary physical interpretation of the free energy (\ref{Fjexp}) is therefore as  a superpotential of the two-dimensional Type IIA string 
compactification on the mirror, $X_4$, of $Y_4$,
and {\it not} as a superpotential in four dimensions. The four-dimensional  precursors, or F-theory uplifts, of the potentials $\cF^{(j)}(t)$ are 
the moduli independent Green-Schwarz anomaly cancelling terms, $B\wedge F^{(j)}$, and their supersymmetric partners, the
Fayet-Iliopoulos $D$-terms, $D^{(j)}$. This is why such two-dimensional prepotentials were
dubbed Fayet-Iliopoulos potentials in \cite{Lerche:1997zb}.

The relevance of $\cF^{(j)}(t)$ in the context of obtaining the elliptic genus  is as follows: The family of free energies $\cF^{(j)}(t)$, labelled by $U(1)$ fluxes
$G_j$, is the two-dimensional analog of the familiar free energy, or prepotential, $\cF(t)$ of four-dimensional compatifications with $N=2$ supersymmetry. All these free energies represent BPS saturated threshold corrections (see e.g. \cite{Harvey:1995fq}) whose dependence on continuous moduli arises from  $T^2$ compactifications of chiral theories in two dimensions higher.
In the present context the free energies play the role of partition functions of 
nearly tensionless heterotic
strings that emerge in certain infinite distance limits in the moduli space of F-theory compactifications.  These are exactly the solitonic heterotic strings referred to at the beginning of this section.

For six-dimensional heterotic strings which arise in F-theory compactifications, 
this has been discussed in ref.~\cite{Lee:2018urn,Lee:2018spm}.
When these strings are wrapped on $S^1$ or $T^2$, they
yield BPS particles in five or four dimensions, and their appropriately defined partition function $Z(t)$ can 
be computed via mirror symmetry in terms
of the BPS invariants encoded in $\cF(t)$ \cite{Klemm:1996hh,Minahan:1998vr,Haghighat:2013gba,Haghighat:2014vxa}.
By reinterpreting the modular parameter of $T^2$ in terms
the toroidal world-sheet of a string,  this partition function gains a dual interpretation as the elliptic genus
of a six-dimensional heterotic string compactification.
Note that in the chiral six-dimensional theory, there is neither a prepotential $\cF(t)$, nor a concept of BPS particles, and thus 
the elliptic genus does not encode BPS quantities, but rather a certain sub-spectrum of non-BPS particles.  This has been analyzed in \cite{Lee:2018urn,Lee:2018spm} in the context of Weak Gravity Conjectures. 

Our goal is to apply an analgous logic in two dimensions lower, where we deal with the free energies $\cF^{(j)}(t)$ in two dimensions and
their F-theory uplifts to chiral, anomalous $U(1)$ theories in four dimensions.  
We will find an analogous relation between the invariants 
encoded in the flux-induced free energies $\cF^{(j)}(t)$ and partition functions of certain emergent nearly tensionless 
heterotic strings, which are given by elliptic genera.

\subsection{General properties of the $4d$ elliptic genus}\label{subsec_ellgen}

Let us now elucidate the relation between the free energies ${\cal F}^{(j)}(t)$ and the elliptic genus 
of the four-dimensional heterotic string arising in F-theory in more detail.
The elliptic genus is defined
by taking the string world-sheet  to be a torus $T^2$
 and performing the weighted trace over the Ramond sector of its excitations:
\be \label{Zdef}
Z(\tau, z) = {\rm Tr}_R (-1)^F F q^{H_L} \bar q^{H_R} e^{2 \pi i z Q }  \,.
\ee
Here $q = e^{2\pi i \tau}$ and $\tau$ is the modular parameter of the world-sheet torus $T^2$, while $z$ is an elliptic parameter which takes account of the refinement by weighting the excitations with their $U(1)$ charges;  moreover $Q$ denotes the generator of this symmetry. 
The elliptic genus has an expansion
\be \label{Zexp}
Z(\tau, z) =q^{E_0} \sum_{n \geq 0, r \in \mathbb Z}  N(n,r) q^n \xi^r \,,
\ee
where $\xi =e^{2\pi i z}$. The prefactor, $q^{E_0}$, reflects 
the zero-point energy \cite{DelZotto:2016pvm,Kim:2018gak} and depends on the particular geometry.  

Recall that one of the most important properties of the elliptic genus is its behavior under modular transformations.
Specifically, it is known \cite{Schellekens:1986yi,Schellekens:1986xh} 
that for perturbative chiral heterotic strings in $d=2p+2$ dimensions, which are based on conformal field theory,
the elliptic genus must be a meromorphic modular form of weight $w=-p$, and $E_0=-1$.  

On the other hand, if we consider a heterotic string as a solitonic string within F-theory, it is a priori not clear to what extent this property
should hold if the heterotic string is non-perturbative. We will discuss this issue in the sequel, in particular in Section \ref{subsec_quasimod}.

For the current section, we assume that the elliptic genus as defined
in (\ref{Zdef}) has the usual modular properties as they apply to the perturbative heterotic string. This implies that
in the presence of some $U(1)$ background gauge field,\footnote{For more general, possibly non-abelian gauge fields one encounters certain Weyl invariant generalizations but these are not of present interest.} it turns into a  meromorphic Jacobi form  
$\varphi_{w,m}(\tau,z)$  of weight $w=-p$ and some fugacity index $m$.  
Recall \cite{EichlerZagier,Dabholkar:2012nd} that such Jacobi forms behave under the modular transformations 
as shown in eq.~(\ref{jacobitrApp}).
The fugacity parameter~$z$ plays the role of the $U(1)$ field  strength, and 
the associated fugacity index, $m$, is model dependent.
By standard arguments \cite{Benini:2013xpa,DelZotto:2016pvm,DelZotto:2017mee,DelZotto:2018tcj}, it can be identified with the parameter $m$ that enters
 the 't~Hooft anomaly polynomial $I_4(\Gauge)$ of the string world-sheet theory, which will be given in eq.~(\ref{I4Gpoly}).
This implies that $m$ is always a positive integer in the present context.

For a heterotic string in four dimensions, the elliptic genus  must thus have weight $w=-1$. 
The (potential) pole due to $q^{-1}$ indicates that $Z(\tau, z) $ is (possibly) a meromorphic Jacobi form,  
and modularity suggests that $Z(\tau, z)$ should contain a factor 
$\eta^{-24}(\tau)$, where $\eta(\tau)$ is the Dedekind eta-function; 
physically this reflects, of course, the oscillator partition function in the left-moving sector.
 Since $\eta^{2}(\tau)$ transforms with weight $w=1$ under the modular group, one concludes that 
\be\label{ellgenjaceleven}
Z(\tau, z) = {\eta^{-24}(\tau)\Phi^-_{11,m}(\tau,z)} \,,
\ee
where $\Phi^-_{11,m}(\tau,z)$ is some unspecified weak Jacobi form of indicated weight and index. 
This means that it must be expressable
as a polynomial of the given weight and index in the generators of the ring of weak Jacobi forms, which (for integer $m$) we may take to be \cite{EichlerZagier}
\be\label{Jacringgens}
\cR\ =\ \big\{E_4(\tau), E_6(\tau),  \varphi_{-1,2}(\tau,z), \varphi_{-2,1}(\tau,z),  \varphi_{0,1}(\tau,z) \big\}.
\ee
Here $E_4,E_6$ are the familiar Eisenstein series and  $\varphi_{w,m}$ certain weak Jacobi forms of weight $w$ and index $m$ that are
defined in Appendix~\ref{jacdefs}.

Note that Jacobi forms $\Phi^\pm_{w,m}$
of even or odd weight are symmetric or antisymmetric in $z$, respectively. This is mirrored by their theta-expansion \cite{EichlerZagier,zbMATH02171618},
which can be written as
\bea\label{thetaexp}
\Phi_{w,m}^\pm(\tau,z) &=& \sum_{\ell\in\IZ\, {\rm mod}\,m} h_\ell(\tau) \Theta_{m,\ell}^\pm(\tau,z)\,, \\
 \Theta_{m,\ell}^\pm(\tau,z) &:=& \Theta_{m,\ell}(\tau,z)\pm  \Theta_{m,-\ell}(\tau,z)\,, \nn  \\
  \Theta_{m,\ell}(\tau,z)&:=& \sum_{k\in\IZ}q^{(\ell+2mk)^2/4m}\xi^{\ell+2mk} \,. \nn
 \eea
Here the coefficients, $h_\ell(\tau)$, are unspecified vector-valued modular forms of weight $w-1/2$, whose
precise form is not important to us.

The four-dimensional elliptic genus is therefore proportional to $z$ times an even function of $z$.  
This reflects the fact that it vanishes if there is no anomalous $U(1)$ gauge symmetry present.
Since the ring of weak Jacobi forms, as shown in (\ref{Jacringgens}), has only one generator of odd weight and integer index, 
it follows that
$\Phi^-_{11,m}= \varphi_{-1,2}\Phi^+_{12,m-2}$. Moreover,  the 
constant term of  $\Phi^-_{11,m}(\tau,z)$  vanishes by anti-symmetry.  This suggests the latter to be a cusp form,  and indeed
one can prove (\!\!\cite{EichlerZagier}, p.~110), that this is necessarily the case at least for $m<9$.  Under this condition
the Dedekind $\eta$ function cancels out so that we can simplify (\ref{ellgenjaceleven}) as
\be\label{ellgenjaczero}
Z(\tau,z)\ =\ \varphi_{-1,2}(\tau,z)\ \Phi^+_{0,m-2}(\tau,z)\,.
\ee
It follows that for $m=1$ the elliptic genus vanishes identically and for $m=2$ it is unique, up to normalization.

A related consequence of the theta expansion (\ref{thetaexp}) of an odd weight Jacobi form is that 
all dependence on $\xi$ is in terms of $(\xi^r-\xi^{-r})$, and moreover that all powers for $r=k m$ cancel for $k\in \IZ$.
Therefore there are gaps in the charge spectrum, at least as far as its contribution to the elliptic genus is concerned. This property of the elliptic genus will be derived in more detail in Section \ref{subsec_WGCbounds}.

One of our main tasks in this paper is to determine the elliptic genus (\ref{Zdef}) of certain four-dimensional, chiral $N=1$
 supersymmetric heterotic strings geometrically from F-theory via mirror symmetry. 
As will be explained in more detail in Section \ref{sec_ellgen4foldflux},  
this essentially boils down to equating (\ref{Zexp}) with certain flux-induced free energies (\ref{Fjexp}).  
In this way the modular properties of the  elliptic genus  are inherited from the geometry of  the elliptic
fibration of the compactification manifold,~$Y_4$.

This by itself is analogous to what has been done in six dimensions \cite{Klemm:1996hh,Haghighat:2013gba,Haghighat:2014vxa},
where the modular geometry of elliptic three-folds \cite{Klemm:2012sx,Alim:2012ss,Huang:2015sta}  becomes important.
However, we will see that the situation is much more complicated than in six dimensions. Indeed a crucial
new ingredient is the choice of fluxes, which control the chiral spectrum, supersymmetry breaking, 
and non-perturbative sectors. As it will turn out, only in favorable circumstances (e.g., for the perturbative heterotic string)
will we obtain potentials with `good' modular modular properties. See Section \ref{sec:fluxes} for details. 


At this point we may ask in which way Jacobi forms of weight $w=-1$ can possibly pop out from elliptic four-fold geometries. 
Modular properties of correlators on elliptic four-folds have been initially discussed in ref.~\cite{Haghighat:2015qdq} 
and extended in ref.~\cite{Cota:2017aal}. The upshot there
was that four-point correlators are modular forms of weight $-2$.  Now four-point correlators are not fundamental but rather
factorize into the three-point functions (\ref{Ciab}) as \cite{Mayr:1996sh,Klemm:1996ts}
\be
C_{abcd}(\tau)\ =\ C_{jab}(\tau) (\eta^{-1})^{jk}C_{kcd}(\tau)\ =\  C_{jac}(\tau) (\eta^{-1})^{jk}C_{kbd}(\tau)\ ,
\ee
where $\eta$ denotes the intersection form  (\ref{etaform}) on $H^\ttw_\ver(Y_4)$. Obviously a weight $-2$ modular form can
be obtained by pairing a weight $0$ with a weight $-2$ form. This is what happens for the example discussed
in ref.~\cite{Cota:2017aal}, where it was shown how a modular eigenbasis of $H^\ttw_{\ver,0}(Y_4)$ can be chosen such as to realize the off-diagonal part
of the intersection form (\ref{etaform}).

On the other hand, we have argued that in chiral four-dimensional theores the elliptic genus must have weight equal to $-1$, 
so we need to pair two copies if we want to obtain a total modular weight $-2$. This implies
 that the relevant part of the 4-cohomology must have a non-vanishing intersection pairing between weight $-1$ modular forms,
which corresponds to the block-diagonal part in $\eta$ as shown in (\ref{etaform}). As we will see below,
this is  a feature of precisely the fluxes in $\widetilde H^\ttw_{\ver,U(1)}(Y_4)$ that we consider.

\section{Elliptic genera from four-folds with flux} \label{sec_ellgen4foldflux}

We now explain in more detail the relation between the elliptic genus of a four-dimensional solitonic heterotic string in F-theory, and 
the BPS invariants of the underlying elliptic four-fold $Y_4$ equipped with some chirality generating $U(1)$ flux.
We set the stage in section \ref{subsec_Geomset} by reviewing elements of F-theory of $Y_4$ pertinent to this discussion. The expert reader may wish to jump directly to
section \ref{sec:D3-heterotic}, where we introduce the heterotic string we are interested in.
The relation between the index-like multiplicities of its excitations and flux-dependent BPS invariants will be explained in section \ref{subsec_GW4}.
This discussion will include a new interpretation to the chiral index of massless charged matter in F-theory in terms of certain Gromov-Witten invariants on $Y_4$.

\subsection{Geometric setup and U(1) fluxes} \label{subsec_Geomset}

A four-dimensional, $N=1$ supersymmetric F-theory compactification can be thought of as non-perturbative Type IIB string theory compactified on a positively curved K\"ahler 3-fold, $B_3$, in the presence of 7-branes.
This data is  encoded in the geometry of a Calabi-Yau four-fold $Y_4$  which admits an elliptic fibration
\bea \label{ellfibrationpi}
\pi :\quad \mathbb{E}_\tau \ \rightarrow & \  \ Y_{4} \cr 
& \ \ \downarrow \cr 
& \ \  B_3 \ ,
\eea
and which serves as the compactification space of a formal twelve-dimensional F-theory.
A well-known duality relates the four-dimensional F-theory compactification on $Y_4$ 
to a three-dimensional  theory with $N=2$ supersymmetry, which is obtained by compactifying eleven-dimensional M-theory on the same $Y_4$:
\be \label{Fmduality}
\text{F-theory on} \,\,  \mathbb R^{1,2} \times S_A^1 \times Y_4 \quad  \longleftrightarrow \quad  \text{M-theory on} \, \, \mathbb R^{1,2}  \times Y_4  \,.
\ee
The three-dimensional M-theory reduction is related upon further circle compactification to Type IIA theory on $Y_4$, which is the natural framework for studying mirror symmetry as outlined in the previous section.

Of special importance to us are the abelian gauge symmetries that appear in the three-dimensional M-theory action, and their counterpart in the dual F-theory.\footnote{Many more details can be found e.g in the recent reviews \cite{Weigand:2018rez,Cvetic:2018bni}, to which we also refer for the original references.}
Given any element $w \in H^{1,1}(Y_4)$ we can expand the eleven-dimensional M-theory 3-form $C_3$ as
\be
C_3 =  A_{\rm 3d} \wedge  w + \ldots\,. 
\ee 
This yields a three-dimensional gauge potential $A_{\rm 3d}$, and we will encounter three types of such gauge symmetries.
The first is  the Kaluza-Klein $U(1)_{\rm KK}$ symmetry in the three-dimensional M-theory effective action which arises by reduction of the four-dimensional F-theory along $S_A^1$.
Every elliptic fibration has a zero-section $S_0$, which defines a divisor in $H^{1,1}(Y_4)$, and 
the gauge potential $\tilde A_{\rm 3d} $ obtained from it via
\be \label{KKdef}
C_3 = \tilde A_{\rm 3d} \wedge \tilde S_0 + \ldots \,, \qquad \tilde S_0 = S_0 + \frac{1}{2} \pi^{-1} \bar K \,,
\ee 
describes exactly this KK $U(1)$ \cite{Park:2011ji}. Here $\bar K$ is the anti-canonical class of $B_3$.
This symmetry becomes part of the four-dimensional metric in F-theory.

Second, taking $w = \pi^{-1}(D_k^{\rm b})$, where $D^{\rm b}_k$ form a basis of $H^{1,1}(B_3)$, gives rise to a set of gauge potentials $A^k_{\rm 3d}$ which
map to 2-form gauge potentials in F-theory: The gauge fields 
$A^k_{\rm 3d}$  arise from the expansion of the Type IIB 4-form $C_4 = C_2^k \wedge D^{\rm b}_k$, by reducing  the 2-forms $C_2^k$ on $S^1_A$.

Third, there can be additional abelian gauge symmetries in the four-dimensional F-theory, which are encoded  in extra sections of $Y_4$ besides the zero section $S_0$.\footnote{In addition, non-abelian gauge groups can be broken to their Cartan $U(1)$'s by gauge fluxes, but we do not consider these here.}
For ease of presentation we focus on situations with a single such extra abelian gauge group, associated with
some rational section $S$. This can easily be generalized to more complicated configurations.
A basis of the cohomology group $H^{1,1}(Y_4)$ is then spanned by the independent sections $S_0$, $S$ and the pullback divisors  $\pi^{-1} D^{\rm b}_k$, 
\be
H^{1,1}(Y_4) = \langle S_0, S, \pi^{-1} D^{\rm b}_k  \rangle \,.
\ee
The  extra abelian gauge potential is obtained by expanding the 3-form as $C_3 = A^S_{\rm 3d} \wedge \sigma(S) + \ldots$. Here, the Shioda map $\sigma(S)$ associates to the section $S$ the following linear combination of divisors\footnote{We distinguish the intersection pairings within the rings $H^{*,*}(Y_4)$ and $H^{*,*}(B_3)$ by denoting them by the two symbols, $\circ$ and $\cdot\,$, respectively. Integration of a top form over the space is understood.}
\be
\sigma(S) = S - S_0 - \pi^{-1} \pi_\ast (S \circ S_0) \in H^{1,1}(Y_4) \,.
\ee
The specific properties of the divisor $\sigma(S)$ imply that $A^S_{\rm 3d}$ uplifts, under F/M-theory duality, to an abelian gauge potential in the four-dimensional effective action.  It can be interpreted as a $U(1)$ gauge symmetry 
that arises from a 7-brane which wraps a 4-cycle in $B_3$.  The properties responsible for this are the transversality conditions
\be
\sigma(S) \circ {\bf C} = 0 \,, \qquad \sigma(S) \circ \mathbb E_\tau = 0 \,,
\ee
where the first condition  holds for every curve ${\bf C}$ in the base.
An equivalent way of putting this is that
\be \label{transversality-a}
\sigma(S) \circ S_0 \circ \pi^{-1} D^{\rm b}_i \circ \pi^{-1} D^{\rm b}_j = 0\,,\qquad \quad \sigma(S) \circ \pi^{-1} D^{\rm b}_i    \circ \pi^{-1} D^{\rm b}_j \circ  \pi^{-1} D^{\rm b}_k = 0 \,.
\ee


In presence of an extra section  (assuming that all fibral singularities have been resolved) the fiber $\mathbb E_\tau$ typically splits into two fibral components over various curves on the base $B_3$.
Depending on the details of the fibration, different splitting patterns may occur over different curves $C_\qr$ on $B_3$, 
\be
\pi^{-1}(p)  \rightarrow   {\cal C}^{\rm f}_{\qr} + {\cal C}^{\rm f'}_{\qr} \,, \qquad p \in C_\qr \subset B_3\,,
\ee
with the property that\footnote{The split into fiber components with these properties is always guaranteed at the level of cycles.}
\be
\begin{array}{cccccc}
\sigma(S) \circ {\cal C}^{\rm f}_{\qr} &=& \qr\,, \qquad &  S_0 \circ {\cal C}^{\rm f}_{\qr} &=& 0\,;  \\
\sigma(S) \circ {\cal C}^{\rm f'}_{\qr} &=& -\qr\,, \qquad &  S_0 \circ {\cal C}^{\rm f'}_{\qr} &=& 1 \,.   \\
\end{array}
\ee

The second line follows in fact from the first and the transversality condition $0=\sigma(S) \cdot  \mathbb E_\tau =  \sigma(S) \cdot  \mathbb ({\cal C}^{\rm f}_{\qr} + {\cal C}^{\rm f'}_{\qr})$.
Note that in homology, ${\cal C}^{\rm f}_{\qr} = \qr \, {\cal C}^{\rm f}_{\qr=1}$. 

Therefore,  M2-branes wrapping holomorphic curves in the  class
\be \label{M2class1}
\Gamma_0(n,r):= n \, \mathbb E_\tau + \qr \, {\cal C}^{\rm f}_{\qr=1}
\ee
give rise to chiral multiplets in the three-dimensional effective action, with $U(1)$ charge $\mq =\qr$ and $U(1)_{\rm KK}$ charge~$n$.
They become massless in the F-theory limit where the fiber volume shrinks to zero.
These states represent the familiar KK tower that arises from a massless charged particle in the four-dimensional F-theory effective action, upon circle reduction to M-theory.

An important geometrical quantity in studying the $U(1)$ gauge theory is the height-pairing
\bea \label{heightpairing}
b = - \pi_\ast (\sigma(S) \circ \sigma(S)) \,,
\eea
which is a divisor on $B_3$. In particular, the $U(1)$ gauge coupling is related to the K\"ahler volume of $b$.
Indeed the four-dimensional effective action contains the terms
\be
S = \frac{M^2_{\rm Pl}}{2} \int \sqrt{-g} \, R + \frac{1}{4 g^2_{\rm YM}} \int F_{\mu \nu} F^{\mu \nu}\ ,
\ee
where
\bea \label{Mplgymdef}
M^2_{\rm Pl} = 4 \pi \, {\rm vol}(B_3) \,,\qquad   \frac{1}{g^2_{\rm YM}} = \frac{1}{2 \pi} {\rm vol}({\bf S})\,, \, \qquad \quad {\bf S} = b \,.
\eea
This is in conventions where the Type IIB string length is fixed and given by $\ell_s = 2 \pi \sqrt{\alpha} \equiv 1$.

Gauge fluxes  associated with  $U(1)$ gauge groups have a convenient description in M-theory language in terms of 4-form  fluxes  $\Gfour \in \tilde H^{2,2}_{{\rm vert}, U(1)}(Y_4)$ as
\be\label{FGflux}
G = F \circ \sigma(S) \,, \qquad \quad F = \pi^{-1} (F^{\rm b}) \,,
\ee
for some 2-form $F^{\rm b} \in H^{1,1}(B_3)$. By definition these fluxes lie in the subspace $\widetilde H^\ttw_{\ver,U(1)}(Y_4)$ of $H^{2,2}_{\rm vert}(Y_4)$
which was introduced in (\ref{H22split}).
As a result of (\ref{transversality-a}), they satisfy the orthogonality conditions 
\be \label{transversGflux}
G \circ S_0 \circ \pi^{-1}D_i^{\rm b} = 0 \,,\qquad \quad G \circ \pi^{-1}D_i^{\rm b} \circ \pi^{-1}D_j^{\rm b} = 0 \,.
\ee
Since the 4-forms of type $S_0 \circ \pi^{-1}D_i^{\rm b}$ and $\pi^{-1}D_i^{\rm b} \circ \pi^{-1}D_j^{\rm b}$ span $H^\ttw_{\ver,0}(Y_4)$,
this leads to an intersection metric $\eta$ of the form eq.~(\ref{etaform}).

Vertical gauge fluxes of this sort affect the effective action in several important and interrelated ways:
First, they are responsible for the appearance of net chirality in the spectrum of massless charged four-dimensional $N=1$ chiral multiplets.
Fibering the curve ${\cal C}^{\rm f}_\qr$
over $C_\qr$ gives rise to a matter surface $S_\qr$, whose class likewise lies in $\widetilde H^\ttw_{{\rm vert}, U(1)}$ and which therefore has a non-zero intersection product with the $U(1)$ flux $\Gfour$.
The chiral index of massless matter of charge $\qr$ in the F-theory action is thus given by the overlap \cite{Donagi:2009ra,Braun:2011zm,Marsano:2011hv,Krause:2011xj, Grimm:2011fx}
\be \label{chiqdefinition}
\chi_\qr = \int_{S_\qr} G = \qr \int_{C_\qr} F^{\rm b} \,.
\ee
As a new observation, we will identify in Section \ref{subsec_GW4} this 
quantity as nothing else than the lowest degree, genus-zero Gromov-Witten invariants of the fibral curve,
in the presence of the flux $\Gfour$; see eq.~(\ref{chiisN}).

Due to the induced net chirality, the abelian gauge symmetry generically exhibits both an abelian cubic and a mixed abelian-gravitational  anomaly
at one loop order, which in turn are cancelled by the Green-Schwarz mechanism.
The anomaly cancellation relations involve the height-pairing and take the form \cite{Cvetic:2012xn}
\bea
\sum_\qr   \chi_\qr  \,  \qr^3 &=& 3 \, G \circ \pi^{-1}(b) \circ \pi^{-1}(b) \,, \\
\sum_\qr  \chi_\qr  \,  \qr    &=&   6 \,  G \circ \pi^{-1}(b) \circ \pi^{-1}(\bar K) \,.\nn
\eea
Relatedly,  the $\Gfour$ flux potentially induces a $D$-term in the effective action proportional to
\be \label{xiFth}
\zeta =  J \cdot F^{\rm b} \cdot b \,,
\ee
and this has to vanish in a supersymmetric vacuum with unbroken $U(1)$ gauge symmetry.
More precisely, a solution to $\zeta = 0$ for non-zero base K\"ahler moduli must exist in the closure of the K\"ahler cone of the base $B_3$.
As a result of the $D$-term potential, the gauge field acquires a St\"uckelberg mass, at a scale below which the symmetry is realized as an effective global symmetry in the effective action. It may be further broken to a discrete subgroup, $\mathbb Z_k$, by D3-brane instanton effects \cite{Blumenhagen:2006xt,Ibanez:2006da,BerasaluceGonzalez:2011wy}.  We will return to discussing St\"uckelberg mass terns for the $U(1)$ gauge fields later in Section~\ref{stuck}.

Note that in a globally consistent vacuum the flux background must satisfy the Freed-Witten quantization condition \cite{Witten:1996md}
\be\label{freewit}
\Gfour + \frac{1}{2} c_2(Y_4) \in H^4(Y_4,\mathbb Z)\,,
\ee
and furthermore it is constrained by the D3-brane tadpole condition
\be \label{F-theorytadpole}
\frac{1}{2} \Gfour \circ \Gfour + n_{D3} = \frac{1}{24} \chi(Y_4) \,.
\ee
Here $n_{D3}$ is the number of spacetime-filling D3-branes in F-theory and $\chi(Y_4)$ denotes the Euler characteristic of $Y_4$. Requiring that no anti-D3-branes are present translates into demanding that $n_{D_3} \geq 0$.

\subsection{Quasi-perturbative heterotic strings from wrapped D3-branes}\label{sec:D3-heterotic}

Our goal is to compute the elliptic genus of a specific, well-controlled solitonic string in the extended dimensions $\mathbb R^{1,3}$.
In F-theory, this string is obtained by wrapping a D3-brane along 
 a holomorphic curve $C_\beta$ inside the base $B_3$.
 Such a curve defines a ``base curve'' class 
 \be
 {\bf C}_\beta = S_0 \circ \pi^{-1}(C_\beta) \subset Y_4 \,.
 \ee
 While string-like BPS objects do not exist in a four-dimensional theory with only $N=1$ supersymmetry, 
 the effective theory on the string world-sheet can nonetheless be 
 described  by a chiral 2d $N=(0,2)$ supersymmetric theory. 
 For a single wrapped D3-brane, this world-sheet theory is found \cite{Lawrie:2016axq} by dimensionally reducing the abelian $N=4$ gauge theory theory along the D3-brane on $C_\beta$ with the help of a topological duality twist \cite{Martucci:2014ema} along $C_\beta$.

The zero mode structure of the effective string has been analyzed in detail in \cite{Lawrie:2016axq} and is summarized in Table \ref{tablespec}.
All $N=(0,2)$ supersymmetry multiplets except the ones in the last row originate in the $N=4$ gauge multiplet along the D3-brane. Their
 fermionic fields transform as spinors with respect to the structure group $SO(2)_T$ of the normal bundle in the two extended directions transverse to the string in $\mathbb R^{1,3}$. 
The Fermi multiplets $\lambda_-$ in the last row represent the zero-mode excitations of fundamental 3-7 strings localised at the pointlike intersection of $C_\beta$ with the 7-branes in the F-theory background.
These zero modes are $SO(2)_T$ singlets, but charged under the 7-brane gauge group $\Gauge$, which in our case is taken to be $\Gauge=U(1)$. The 7-brane gauge symmetry hence acts as a global or flavour  symmetry from the perspective of the string world-sheet.

An important quantity of the string is the 't Hooft anomaly polynomial, $I_4$.  It encodes the structure of the two-dimensional
 anomalies of the global symmetries $SO(2)_T$ and $\Gauge=U(1)$, under which the world-sheet fields are charged.
It takes the form \cite{Lawrie:2016axq}
 \be
 I_4  = - \frac{1}{2} {\rm tr} F_T^2 \left( - \frac{1}{4}  \bar K \cdot C_\beta  \right)  -\frac{1}{4} p_1(R) \left({\bar K} \cdot C_\beta \right) + I_4(\Gauge) \,,
 \ee
 where the second term encodes  the gravitational anomaly along the string and $I_4(\Gauge)$ refers to the 't Hooft anomaly for the global $\Gauge=U(1)$ group inherited from the 7-brane gauge symmetry. 
Extrapolating the results  \cite{Weigand:2017gwb} for analogous world-sheet theories from D3-branes wrapped on curves in Calabi-Yau 5-folds, the $U(1)$ anomaly polynomial must take the form
\be \label{I4Gpoly}
I_4(\Gauge) = -  \frac{1}{2}\, m \,  {\rm tr}  F_{U(1)}^2\,,
\ee
with 
 \be
 m = \frac{1}{2}  b \cdot C_\beta  = -\frac{1}{2} \pi_\ast(\sigma(S) \circ \sigma(S)) \cdot C_\beta \,.
\ee
Moreover, the gravitational anomaly is related to the net zero-point energy given by
  \be \label{E0onstring}
E_0 = - \frac{1}{2} C_\beta \cdot \bar K \,,
\ee 
which arises from Casimir forces on the string \cite{DelZotto:2016pvm,Kim:2018gak}.

\begin{table}
  \begin{tabular}{|c|c|c|c|c|} \hline
   \text{Fermions}            & \text{Bosons}       & \text{(0,2)  Multiplet}    &  Multiplicity                  & Multiplicity $C_\beta = C_0$  \\ \hline
                        $\mu_+$ & $\varphi$              &    \text{Chiral}            &  {$h^0(C_\beta, N_{C_\beta/B_3})$} & 2 \\ \hline 
     $ \tilde \psi_+ $          &  $ \sigma $                     & \text{Chiral}                & {$g - 1 + \bar K \cdot C_\beta$}  & 1 \\ \hline
     $  \gamma_+ $          &$  \tau $ & \text{Chiral}  & {$h^0(C_\beta) = 1$} & 1 \\  \hline 
      $ \rho_- $& \text{---} &  \text{Fermi}  & {$h^0(C, N_{C_\beta/B_3}) -  \bar K \cdot C_\beta$}  & 0 \\  \hline
    $ \beta_- $                  & \text{---}               &  \text{Fermi} & {$h^1(C_\beta) = g$} & 0  \\  \hline  \hline
 $\lambda_-$                  &  \text{---} &  \text{Fermi} &  $ 8  \bar K \cdot C_\beta$ & 16 \\ \hline
 \end{tabular}
\caption{2d $(0,2)$ multiplets in the effective theory on $\mathbb R^{1,1} \times C_\beta$,  where $C_\beta$ is
a curve of genus $g$ with normal bundle $N_{C_\beta/B_3}$ inside the base $B_3$ of $Y_4$ (this is based on table 5 in \cite{Lawrie:2016axq}). The subscripts of the fermions denote the chirality along the string world-sheet. \label{tablespec}}
\end{table}

Our prime interest in this work is in a special type of solitonic strings, namely ones which can be interpreted in terms of
 weakly coupled, critical heterotic strings
compactified to four dimensions.
At the level of world-sheet fields, any such string must have the familiar $8$ right-moving scalars plus fermionic partners, $8$ left-moving scalars with no fermionic partners and in addition 16 left-moving fermions.
In view of Table \ref{tablespec}, this spectrum can only be obtained for $C_\beta$ a rational curve with $C_\beta \cdot \bar K = 2$ and 
normal bundle either
 $N_{C_\beta} ={\cal O}_{C_\beta}(1) \oplus  {\cal O}_{C_\beta}(-1)$ or  
  $N_{C_\beta} ={\cal O}_{C_\beta}(0) \oplus  {\cal O}_{C_\beta}(0)$.
 
For the solitonic string that arises from a D3-brane wrapped on $C_\beta$ to be  a critical string, it must contain the massless graviton in its spectrum - 
at least in the asymptotically weakly coupled limit: 
In this regime it defines the duality frame of the fundamental heterotic string.
This condition rules out a curve $C_\beta$ with normal bundle
$N_{C_\beta} ={\cal O}_{C_\beta}(1) \oplus  {\cal O}_{C_\beta}(-1)$:
 Such a $C_\beta$ is rigid at least along a divisor on $B_3$, as indicated by the negative degree of one of the normal bundle summands. The string modes are therefore not free to move in the entire bulk of the compactification, unlike what is required for a string that has the graviton in its spectrum.

This leaves, as the only possibility, a holomorphic curve $C_0$ with 
\be \label{trivialnormalbundle}
N_{C_0/B_3} = {\cal O}_{C_0} \oplus  {\cal O}_{C_0} \,, \qquad \quad C_0 \cdot \bar K = 2 \,.
\ee
 We furthermore demand that it be possible to obtain a parametrically weakly coupled and tensionless critical string,
 by taking the limit
\be \label{criticallimitdef}
{\rm vol}(C_0) \to 0 \qquad \text{for}  \quad {\rm vol}(B_3) \quad \text{finite} \,. 
\ee
These requirements are motivated by the fact that the asymptotically weakly coupled limit for a  critical heterotic string coincides with the limit of vanishing tension in the Einstein frame, while at the same time criticality requires gravity not to be decoupled.\footnote{More precisely, in the heterotic  four-dimensional Einstein frame, the fundamental string tension is given by
$T_{\rm het}  = \frac{2\pi}{\ell^2_{\rm het}} e^{2 \Phi}$\,,
where the dependence on the ten-dimensional dilaton $\Phi$ arises from the Weyl rescaling from string to Einstein frame (cf. the discussion around eq. (\ref{gMNrescaling4d})).
Hence the tensionless limit requires us to take $e^{2 \Phi} \to 0$, while at the same time the volume of the heterotic compactification space $Z_3$ is fixed to keep $M_{\rm Pl}$ finite. As a result, the four-dimensional dilaton
$S_{\rm het}  = e^{-2 \Phi} {\rm vol}(Z_3)/\ell^6_{\rm het} \to \infty$
diverges in the tensionless limit (\ref{criticallimitdef}). \label{footnotehettension}}
The condition (\ref{trivialnormalbundle}) is met, for instance, by any curve with the properties
\be \label{C0realisation}
C_0 = J_0 \cdot J_0 \,,  \quad \text{with} \quad J_0^2 \cdot \bar K  > 0 \,, \quad J_0^3 =0 \,,
\ee
where $J_0$ is a generator of the K\"ahler cone of $B_3$. The  limit (\ref{criticallimitdef}) can furthermore be taken, for such curves, in a certain region in moduli space.
A class of examples of this type is where $B_3$ is a $\mathbb P^1$-fibration over a surface $B_2$ or a blowup thereof, as considered in conventional F-theory - heterotic duality: Here $C_0$ is simply the $\mathbb P^1$ fiber.
This will be detailed in Section \ref{sec_WGC} in the context of the Weak Gravity Conjecture.

Note, however, that the four-dimensional compactification involves $U(1)$ flux data, and the $D$-term constraints from the flux may obstruct taking the limit (\ref{criticallimitdef}). 
The conditions on the flux to allow for such a limit will be discussed in Section \ref{subsec_quasimod}.

To avoid confusion, let us stress that even if the weakly coupled limit (\ref{criticallimitdef}) is possible, this does not yet mean that the heterotic string is fully perturbative: Non-perturbative effects, in particular NS5-branes, can and in general do contribute to
the string dynamics, even at the level of the elliptic genus.
\begin{itemize}
\item
 We call a heterotic string which allows for a limit (\ref{criticallimitdef}) {\it quasi-perturbative}, to the extent that we can take the limit where $S_{\rm het} \to \infty$ (at least away from possible non-perturbative defects such as NS5-branes).
\item
If at the same time the dynamics of the string, as far as its contribution to the elliptic genus is concerned, does not involve any NS5-branes or other non-perturbative elements, we call the string  
 {\it perturbative}. 
 \end{itemize}
\subsection{Elliptic genus and Gromov-Witten invariants of four-folds} \label{subsec_GW4}

Let us now assume that the forementioned conditions for the emergence of a critical, 
perturbative (or quasi-perturbative) four-dimensional heterotic string are fulfilled. In this case we can easily compute the elliptic genus (\ref{Zdef}).
The degeneracies $N(n,r)$ that appear in the expansion (\ref{Zexp})  count the number of left-moving excitations weighted with signs as in (\ref{Zdef}),   at  level $n$ and with $U(1)$ charge $r$. These can be level-matched against oscillator excitations of the right-moving Ramond ground state, to yield physical, albeit non-BPS states at the given excitation and charge levels.

As we now discuss, these degeneracies of  non-BPS states in four dimensions coincide with certain BPS degeneracies of the three dimensional
theory obtained by $S^1_A$  reduction; these correspond, via the M/F duality sketched in 
eq.~(\ref{Fmduality}), to Gopakumar-Vafa invariants of the compactification of M-theory on $Y_4$. 
The logic is analogous to the reasoning underlying the duality between elliptic genera of 6d strings and BPS invariants of M-theory compactified on elliptically fibered three-folds to five dimensions \cite{Klemm:1996hh,Haghighat:2013gba,Haghighat:2014vxa}. 
However, for the time being we are careful to apply it only to the well-controlled quasi-perturbative critical heterotic strings as defined above.
 
More precisely, wrapping the solitonic string $k$ times around $S^1_A$ produces a tower of particles in $d=3$.
Even though both the string and its excitations are not BPS saturated  in the $d=4$, $N=1$ supersymmetric theory, the wrapped string gives rise to BPS states in the $d=3$ theory with $N=2$ supersymmetry.
Since the string is charged with respect to a 2-form in the $d=4$ theory, all particles in the tower carry some charge $k$
with respect to the abelian gauge symmetry to which the 2-form symmetry is reduced after reduction on $S^1_A$.
As long as we are dealing with a weakly coupled critical heterotic string, the usual quantization rules can safely be applied, 
similar to the 6d/5d context \cite{Klemm:1996hh}.
In particular, the left-moving excitation number of the wrapped string is related to the wrapping number $n$ and KK momentum number $w$ along the $S^1_A$ as
\be
n = w \, k \,.
\ee
For the singly wrapped string with $k=1$, the left-moving excitation number of the wrapped string hence equals the KK momentum along the $S^1_A$.
Even though this is a priori a statement about the excitations of the wrapped string, the number of purely left-moving excitations at level $n$ is unchanged in the unwrapped situation.
The problem of counting the left-moving excitations at level $n$ and charge $r$ in four dimensions is therefore 
reduced to counting BPS particles in three dimensions, with KK momentum $w=n$, $U(1)$ charge $\mq=r$, and with charge $k=1$ with respect to the gauge field inherited from the 2-form tensor in four dimensions.
In light of the discussion of Section \ref{subsec_Geomset}, such particles are due to M2-branes wrapping the curve class
 \be \label{C0nrdef}
\Gamma_{C_0}(n,r) = {\bf  C}_0 + n \, \mathbb E_\tau + \qr\, \cC^{\rm f}_{\qr=1} \subset Y_4\,,
 \ee 
where  $\mathbb E_\tau$ is the class of the generic fiber on the 4-fold $Y_4$, while ${\cal C}^{\rm f}_{\qr=1}$ is the fibral curve class of $U(1)$ charge $\qr=1$. 
An M2-brane along this curve carries a total KK charge of 
\be
\tilde S_0 \circ \Gamma_{C_0}(n,r)  = n - 1 = n + E_0 \,.
\ee
Here we are using the $U(1)_{\rm KK}$ generator $\tilde S_0$ as given in (\ref{KKdef}) along with the fact $S_0 \circ {\bf C}_0 = S_0 \circ S_0 \circ \pi^{-1}(C_0) =  - \bar K \cdot C_0$. The shift by the zero-point energy $E_0$   is the same as the prefactor of the elliptic genus (\ref{Zexp}), and is as given in (\ref{E0onstring}).

By the above reasoning, the degeneracies  $N(n,r)$ defined via the  index  (\ref{Zdef}), (\ref{Zexp})  coincide with the
generalization to four-folds \cite{Klemm:2007in}  of the familiar, integral Gopakumar-Vafa invariants \cite{Gopakumar:1998ii,Gopakumar:1998jq}.  They count the BPS invariants of M2-branes wrapped on holomorphic curves in the presence of  4-form flux.  For singly wrapped curves, these generalized
Gopakumar-Vafa invariants are identical to the respective Gromov-Witten invariants.
Therefore, while focusing on $k=1$, our task is to compute the genus-zero Gromov-Witten invariants $N_{C_0; \Gfour}(n,r)$ of the curve (\ref{C0nrdef}) in the presence of $U(1)$ four-flux, $\Gfour$. In terms of these, the degeneracies of the elliptic genus are then simply
\be \label{Nc0nr}
N(n,r)  = N_{C_0; G}(n,r) \,.
\ee

The actual computation proceeds via mirror symmetry, as indicated in Section \ref{subsec_warmup}.
A further circle reduction on $S^1_B$ brings us from the  three-dimensional M-theory to the two-dimensional
Type IIA theory in which this mirror symmetry computation is set.
More explicitly,  let us define an integral basis of $\widetilde H^{2,2}_{{\rm vert}, U(1)}$ and expand the four-flux into it:
 \be
\Gfour = \sum c_j \, \hbas_{j} \,, \qquad \quad \left\{\hbas_{j}: \text{integral basis of } \,  \widetilde H^{2,2}_{{\rm vert}, U(1)} \right\} \,.
 \ee
We can then expand the genus-zero free energy ${\cal F}_{\Gfour}$ into the corresponding flux components as given
in (\ref{Fjexp}). Corresponding to eq.~(\ref{C0nrdef}) we now distinguish the various K\"ahler parameters 
 by defining \be
q = e^{2 \pi i \tau} \,, \qquad \xi = e^{2 \pi i z}
\ee
as the expontentials of the K\"ahler parameter $\tau$  of $\mathbb E_\tau$ and of the K\"ahler parameter $z$ of the fibral curve ${\cal C}^{\rm f}_{\qr=1}$ on $Y_4$,
respectively. We denote the  remaining K\"ahler parameters  associated with generic curves $C_\beta \in H_2(B_3,\mathbb Z)$ by  $t_\beta$, and their
exponentials by $Q_\beta = e^{2 \pi i t_\beta}$. Since we will be interested in an at most linear order in the $Q_\beta$, we write the expansion of
the genus-zero free energy (again omitting classical terms) as follows:
\bea
{\cal F}_{\Gfour}(t)  &=& \sum c_j \, {\cal F}^{(j)}(t)  \,, \\
{\cal F}^{(j)}(t)  &=:& \sum_{C_\beta \in H_2(B_3,\mathbb Z)}  {\cal F}^{(j)}_{C_\beta}(\tau,z) \, Q_\beta + \cO({Q_\beta}^2) \,.
\eea
The Gromov-Witten invariants we are interested in,
\be
N_{C_0; \Gfour}(n,r) = \sum  c_j \, N^{(j)}_{C_0}(n,r) \,,
\ee
are then obtained by linearly combining the integral expansion parameters of
\bea\label{FjC}
{\cal F}^{(j)}_{C_0}(\tau,z) &=& \sum_{n \geq 0, r \in \mathbb Z}  N^{(j)}_{C_0}(n,r) q^n \xi^r \,.
\eea
All this is summarized by identifying the elliptic genus of the nearly tensionless heterotic string associated with the shrinking curve $C_0$, in the presence of flux $\Gfour$, as follows:
\be \label{ZtauszandGW}
Z_{C_0; \Gfour}(\tau,z) = q^{-1} \, {\cal F}_{C_0; \Gfour}(\tau,z)  :=  q^{-1} \sum c_j \,  {\cal F}^{(j)}_{C_0}(\tau,z)   \,.
\ee
This expression for the elliptic genus is determined entirely by the F-theory compactification geometry and the choice of flux,
 and as written no reference is being made as to whether the emerging heterotic string is perturbative or not. 
Thus one may be tempted to
view it as the {\it definition} of a non-perturbative version of the elliptic genus.
This would be similar to the  six-dimensional theories \cite{Klemm:1996hh,Minahan:1998vr,Haghighat:2013gba,Haghighat:2014vxa}, 
where this reasoning has proved to be a valid strategy. In particular, it leads to a (quasi-)modular elliptic genus for the string associated with any heterotic curve $C_0$ with comparable properties \cite{Lee:2018urn}.

In four dimensions,
as we will see in the next section, certain extra conditions need to be
satisfied by the choice of flux in order for $Z_{C_0; \Gfour}(\tau,z)$ to have distinguished properties under modular transformations.

Before delving into this discussion, we conclude this section  by pointing out an interesting interpretation of the genus-zero Gromov-Witten invariants in the context of F-theory. This will also lead to a powerful check of the above assertions.
A geometric definition of the Gromov-Witten invariants on a Calabi-Yau 4-fold $Y_4$ proceeds in terms of  the moduli space  $\overline \cM_{g,s}$ of stable holomorphic maps from a curve class ${\bf C}_\beta \in H_2(Y_4,\mathbb Z)$ to $Y_4$, of genus $g$ and with $s$ marked points.
The mathematical concepts underlying this definition can be found in \cite{Cox:2000vi} and are further  discussed and applied e.g. in \cite{Klemm:2007in,Haghighat:2015qdq,Cota:2017aal}.
In particular, the so-called virtual dimension of $\overline \cM_{g,s}$  is
\be
{\rm dim \, vir}\overline \cM_{g,s} =  (1 - g) + s \,.
\ee
For $g=0$ and $s=1$, ${\rm dim \, vir}\overline \cM_{g,s} =2$, and these are the relevant values for us. 
The genus-zero Gromov-Witten invariants for some ${\bf C}_\beta \in H_2(Y_4,\mathbb Z)$ with respect to some flux $\Gfour\in H^4(Y_4)$  are then defined as the pullback
\be \label{GWdef}
N_{C_\beta; \Gfour} := \int_\mu {\rm ev}^\ast \Gfour\,,
\ee
where $\mu$ is the class in the moduli space $\overline \cM_{0,1}$ associated with ${\bf C}_\beta$ (with one point fixed), and
${\rm ev}^\ast \Gfour$  is the pullback of the 4-form class $\Gfour$ onto $\mu$.

As a simple illustration, consider the fibral curve class $\Gamma_0(n,r) = n \mathbb E_\tau + \qr  \cC^{\rm f}_{\qr=1}$  defined in (\ref{M2class1}), and recall that M2-branes wrapping this class in M-theory produce a KK tower of charged matter fields. These are associated with chiral $N=1$ multiplets in four dimensions.
Since the curve class $\cC^{\rm f}_{\qr=1}$ is a component of 
 the fiber over the curve ${C}_{\qr=1} \subset B_3$, it can  move freely over this curve.
The moduli space of the curve $\Gamma_0(n,r)$ can therefore be identified with the curve ${C}_{\qr=1}$ in the base. To obtain the moduli space $\overline{\cal M}_{0,1}$, we must fix one point on $\Gamma_0(n,r)$. Since the moduli space of a point on a curve is the curve itself,  we end up with a 2-complex dimensional moduli space. Its embedding into $Y_4$ can be identified with the surface $S_{n,\qr}$ obtained by fibering $\Gamma_0(n,r)$ over ${C}_{\qr=1}$.
Then the genus-zero Gromov-Witten invariants (\ref{GWdef}) for $\Gamma_0(n,r)$ reduce to the integral of the 4-form flux $\Gfour$ over this surface,
\be
N_{\Gamma_0(n,r); G} = \int_{S_{n,\qr}} G \,.
\ee
As a consequence of the transversality condition (\ref{transversGflux}), this integral is independent of $n$.
In particular, the chiral index of charged matter fields in four dimensions, (\ref{chiqdefinition}), is nothing but the 
genus-zero Gromov-Witten invariant of $\Gamma_0(n,r)$, evaluated for any $n>0$,
\be \label{chiisN}
\chi_\qr = N_{\Gamma_0(n,r); G} \,.
\ee
Apart from providing a beautiful mathematical interpretation of the expression for the chiral index,
this allows for a non-trivial consistency check of the relation (\ref{ZtauszandGW}). It is based on the following observation:
Consider a situation in which a D3 brane wrapped around $C_0$ indeed corresponds to a critical heterotic string, and assume that there is only one such curve class $C_0$.
The chiral index of the massless charged matter fields in F-theory must then agree with the $n=1$, charge $\qr$ degeneracies of the elliptic genus $Z(q,\tau)$, i.e., with the numbers $N_{C_0; \Gfour}(1, \qr)$. This is because the latter count the index of massless charged excitations in the dual heterotic theory, and the two must match by duality.
This gives the prediction
\be\label{pred}
N_{\Gamma_0(n,r); \Gfour} = N_{\Gamma_0 (1,\qr); \Gfour}  \equiv N_{C_0; \Gfour}(1, \qr)  \,.
\ee
We will confirm this relation in the example studied in Section \ref{sec_Example1}.

\section{Constraints on four-form fluxes for (quasi-)modularity }\label{sec:fluxes}

Even if a curve $C_0$ with the properties (\ref{trivialnormalbundle}) exists that moreover allows for a 
weak coupling limit (\ref{criticallimitdef}) at the level of geometry, additional constraints must be imposed
on the $U(1)$ flux background in order that the elliptic genus is a modular or quasi-modular Jacobi form, as already announced in Section~\ref{subsec_ellgen}.

More precisely, we say that the elliptic genus is {\it modular} if it is of the form (\ref{ellgenjaceleven}) or
(\ref{ellgenjaczero}),  where $\Phi_{w,m}(\tau,z)$ is a meromorphic or weak Jacobi form of weight $w$ and index $m$, respectively.
 This in turn implies
that it can be written as a polynomial in the generators given in eq.~(\ref{Jacringgens}).
We say it is {\it quasi-modular} if the quasi-modular Eisenstein series $E_2(\tau)$ needs to be included as well.

All other cases will be referred to as {\it non-modular}, even though the elliptic genus may still possess more subtle non-trivial transformation properties under the modular group.
Indeed, we do observe hints for this to be case.
A major difference between (quasi-)modular and more general elliptic genera is the appearance of certain gaps in the charge spectrum, as far as its contribution to the index is concerned.
The origin of these gaps, which are not observed in the non-modular case, has been explained on general grounds in section \ref{subsec_ellgen}.

In this section we motivate the constraints on the flux $G\in \widetilde H^\ttw_{\ver,U(1)}(Y_4)$ for obtaining modularity or quasi-modularity of the
elliptic genus. The quasi-modular property will be seen to be intimately tied to the presence of certain non-perturbative branes
in the dual heterotic geometry. The upshot of this discussion is the criterion stated in section \ref{subsec_Generalisation}.

To understand the origin of this criterion, we first have to recall some aspects of the duality between four-dimensional, $N=1$ supersymmetric
compactifications of F-theory and the heterotic string, and especially the role of the NS5-brane in the Green-Schwarz mechanism \cite{Honecker:2006dt,Blumenhagen:2006ux}. The reader familiar with this material is invited to jump directly to section \ref{subsec_quasimod}.

\subsection{Aspects of F-theory - heterotic duality in four dimensions} \label{sec_Aspectshet}

Our starting point is the standard duality between F-theory compactified on the elliptic fibration $Y_4$ over the base $B_3$,
 and the heterotic string compactified on an elliptic Calabi-Yau 3-fold $Z_3$, where $B_3$ and $Z_3$ are related as follows:

\begin{minipage}{5cm}
\bea 
p :\quad C_0 \ \rightarrow & \  \ B_{3} \cr 
& \ \ \downarrow \cr 
& \ \  B_2  \nonumber
\eea
\end{minipage}
\qquad \begin{minipage}{2cm}
$\longleftrightarrow$
\end{minipage}
\begin{minipage}{5cm}
\bea \label{ellfibrationrho}
\rho :\quad  {\mathbb E}_{\tau'}  \rightarrow & \  \ Z_{3} \cr 
& \ \ \downarrow \cr 
& \ \  B_2 \nonumber
\eea
\end{minipage}
\vspace{4mm}

The structure of the elliptic fibration $\rho$ of $Z_3$ is inherited from the elliptic fibration  $\pi$~of~$Y_4$ as defined in eq.~(\ref{ellfibrationpi}).
In particular, if $Y_4$ contains extra rational sections $S_i$ in addition to the zero section~$S_0$, then the same is true for the heterotic fibration $Z_3$ \cite{Cvetic:2015uwu} (see also \cite{Choi:2012pr}). We will denote the zero-section and the extra rational sections on the heterotic side by $S'_0$ and $S_i'$, respectively, or collectively by $S_I'$.

If the rational  fibration $p$ of $B_3$ and the elliptic fibration $\pi$ of $Y_4$ are compatible, $Y_4$ is in itself also an elliptic K3-fibration over $B_2$.
Let us first describe this special situation.


At a microscopic level, the duality relates the solitonic string obtained by wrapping a D3-brane along the fiber $C_0 = \mathbb P^1_f$ to the fundamental heterotic string.
This fits with the fact that the rational fiber $C_0$ satisfies the property (\ref{trivialnormalbundle}). 
Similarly, a D3-brane instanton along a divisor $p^{-1}(C)$, where $C$ is a curve on $B_2$, dualises to a fundamental heterotic world-sheet instanton wrapped on $C$ in the base of $Z_3$.

In addition to the geometric data, both the F-theory and the heterotic duality frames contain gauge backgrounds as crucial ingredients. 
The gauge background on the F-theory side has already been specified in Section \ref{subsec_Geomset} for the type of models studied in this article.
On the heterotic side, the gauge background is encoded in a polystable vector bundle 
\be
W = W^{(1)} \oplus W^{(2)} \,,
\ee
whose structure group $H^{(1)} \times H^{(2)}$ is embedded into the 10d heterotic gauge group $E_8 \times E_8$.
The four-dimensional gauge group, $\Gauge  =\Gauge^{(1)} \times \Gauge^{(2)}$, emerges as the commutant of $H^{(1)} \times H^{(2)}$ inside $E_8 \times E_8$.
Each of the two vector bundles $W^{(i)}$ may in turn be a polystable sum of vector bundles 
\be
W^{(i)} = \oplus V^{(i)}_j \,.
\ee
If $c_1(V_j^{(i)}) \neq 0$, the actual gauge group contains an extra  factor of $U(1)_j^{(i)}$. On the F-theory side,
abelian gauge groups are associated with rational sections and map to certain linear combinations of such heterotic $U(1)_j^{(i)}$.


The spacetime-filling D3-branes that are required to cancel the tadpoles in F-theory
translate into spacetime-filling heterotic NS5-branes, partially wrapping certain holomorphic curves $\gamma_a$ on $Z_3$.
The class of the curves $\gamma_a$ on $Z_3$ is constrained by the Bianchi identity 
\be\label{bianchi}
{\rm ch}_2(W) + c_2({\rm T}Z_3) = \sum_a N_a \gamma_a \,,
\ee
where $N_a$ denote the number of 5-branes in the stacks on $\gamma_a$. 
The identity~\eqref{bianchi} is the heterotic dual of the D3-brane tadpole cancellation condition (\ref{F-theorytadpole}).
In models of the type described so far, where the F-theory fibration $Y_4$ is also elliptically K3-fibered,
the only NS5-branes present on the heterotic side wrap the elliptic fiber, i.e. $\gamma_a = {\mathbb E}_{\tau'}$. This constrains the gauge bundle $W$ accordingly.
More general configurations will be described below.
In the presence of NS5-branes, the heterotic compactification is best thought of as  M-theory on $Z_3 \times S^1/\mathbb Z_2$ \cite{Horava:1995qa}.
The NS5-branes then correspond to M5-branes located at points along the interval $S^1/\mathbb Z_2$ while wrapped around $\gamma_a$.


Heterotic NS5-branes participate in the anomaly cancellation mechanism of the abelian gauge symmetries $U(1)_j^{(i)}$.
This is due to the appearance of two types of Green-Schwarz terms in the effective action, as found in \cite{Honecker:2006dt} and \cite{Blumenhagen:2006ux} for compactifications to six and four dimensions, respectively.
In the context of four-dimensional models, these terms result by dimensional reduction of two interaction terms in the presence of NS5-branes wrapped on $\gamma_a$ \cite{Blumenhagen:2006ux}:
\bea
S^{(1)}_{\rm GS} &=& A^{(1)} \sum_a N_a \int_{\gamma_a}  {\rm B} \wedge ({\rm tr} F_1^2 + {\rm tr} F_2^2 - {\rm tr} R^2) \,, \label{GSheta} \\
S^{(2)}_{\rm GS} &=& A^{(2)} \sum_a N_a \int_{\gamma_a}  {\rm B}^{a} \wedge ({\rm tr} F_1^2 - {\rm tr} F_2^2)\label{GShetb} \,.
\eea
Here ${\rm B}$ denotes the universal heterotic 2-form, ${\rm B}^a$ refers to the chiral tensor field on each 5-brane stack, and $F_1$, $F_2$ are the field strengths of the two $E_8$ factors of the heterotic gauge group. 
The numerical constants $A^{(1)}$ and $A^{(2)}$ play no important role for us.
The additional anomaly cancelling terms contribute St\"uckelberg mass terms of the form $\int_{\mathbb R^{1,3}} {\rm B} \wedge F_i$
 or $\int_{\mathbb R^{1,3}} {\rm B}^a \wedge F_i$ in the four-dimensional effective action, however obviously only so for abelian gauge groups. Furthermore they appear only if the internal part of the gauge field has a non-zero overlap with the curve $\gamma_a$ that is wrapped by the NS5-brane. 

Starting from such a configuration, we can consider blowing up a curve $\Gamma_a$ on $B_2 \subset B_3$ into a divisor $E_a$ of  the blown-up space $\hat B_3$. This means that there exists a contraction map from  $\hat B_3$  onto $B_3$ 
\be
f:  \hat B_3 \rightarrow  B_3 \,,
\ee
which is an isomorphism away from $\Gamma_a$, with the additional property that
\be
 f^{-1}(\Gamma_a) = E_a \,.
\ee
This process destroys the global compatibility of the rational fibration of $B_3$ and of the elliptic fibration of $Y_4$.
As has been studied in detail in \cite{Braun:2018ovc}, the effect on the dual heterotic side is to leave the Calabi-Yau geometry $Z_3$ unchanged, but to include additional NS5-branes along the curve $\Gamma_a$ on $B_2$.

More precisely, the inverse image of each point $Q$ on $\Gamma_a$ is a rational curve $C_a$,
\be
f^{-1}(Q) = C_a \,, \qquad Q \in \Gamma_a \,,
\ee
which we call the fiber of the (exceptional)  blow up divisor $E_a$ on $\hat B_3$.
On $\hat B_3$, the rational curve\footnote{Strictly speaking, we should refer to the rational curve on $\hat B_3$ as $f^{-1}(C_0)$, and similarly for all other curves on $\hat B_3$ which are inherited from $B_3$, but we omit this for notational simplicity. } $C_0$ is still fibered over every point of $B_2$  away from the blow-up locus, where it splits into two components, $C_a$ and $C_a'$.
The solitonic heterotic string obtained by wrapping a D3-brane along $C_0$ on $\hat B_3$ hence splits into two other types of strings as it sweeps out its moduli space, namely strings obtained
from wrapping the D3-brane on either $C_a$ or $C_a'$. Neither of these can be a critical heterotic string. 
Rather, from the properties of $C_a$ and $C_a'$ it follows that they must be non-critical strings, and in the dual heterotic geometry this signals the presence of NS5-branes that wrap the curve $\Gamma_a$.
That is, they can be interpreted as solitonic strings coupling to the chiral tensor fields ${\rm B}^a$ on the NS5-branes, or, in Horava-Witten language,
as M2-branes stretched between the M5-brane and each of the two $E_8$ planes. This is closely related to the six-dimensional phenomenon of fusion, `$E+E=H$,' where a heterotic string
splits into two copies of non-critical E-strings \cite{Haghighat:2014pva}.
Specifically, the K\"ahler volume of the blow-up fiber $C_a$ on $\hat B_3$ maps to the position modulus $\lambda_a$ of these NS5-branes along the Horava-Witten interval.

Above we have focused on a special set of geometries for $B_3$, $\mathbb P^1$-fibrations and their blowups over curves in the base of $B_3$. 
This is sufficient to develop some intuition on which conditions we must impose on the gauge background such that the elliptic genus is (quasi-)modular.
These conditions will momentarily be formulated in full generality.
Nonetheless it is interesting to remark that while in six dimensions, the geometries of the type studied in this section exhaust the list of possible F-theory bases (except for $\mathbb P^2$), 
 the space of K\"ahler three-fold bases $B_3$ for four dimensional F-theory compactifications is considerably richer. 
 This is true even at the level of blowups:
In addition to blowing up a curve on the base of $B_3$, one can blow up the rational fiber itself. This option does not exist for a rationally fibered surface. 
 In Appendix \ref{ex_gen} we will exemplify this construction.

\subsection{Conditions for (quasi-)modularity} \label{subsec_quasimod}

We are now  in a position to determine the criteria on the F-theory $U(1)$ flux, $\Gfour\in \widetilde H^\ttw_{\ver,U(1)}(Y_4)$, to yield a modular, or quasi-modular, elliptic genus for 
a heterotic string that arises from a D3-brane wrapping a curve $ C_0$ of type~(\ref{trivialnormalbundle}). 
\subsubsection{F-theory on $\mathbb P^1$ fibrations}

To gain some intuition, we first assume that the base on the F-theory side is the blowup of a $\mathbb P^1$-fibration over $B_2$ along various curves $\Gamma_a \subset B_2$.
 For notational simplicity, suppose furthermore that the F-theory elliptic fibration has only one extra section $S$, leading to a single abelian gauge group 
 associated with the $U(1)$ flux $\Gfour = F \circ \sigma(S)$ with $F = \pi^{-1} (F^{\rm b})$ for some $F^{\rm b}$ in $H^{1,1}(B_3)$.

According to our discussion of the previous section, this type of geometries has a heterotic dual defined on an elliptic fibration $Z_3$ over the same base
space $B_2$, and we are therefore facing the question which modular properties the elliptic genus for 
the fundamental heterotic string of this theory has. 
From the classic analysis of \cite{Schellekens:1986xh} follows that the elliptic genus of a heterotic string is modular whenever the string 
has a perturbative worldsheet formulation, which is defined entirely at 
the level of conformal field theory. In particular, in the four-dimensional effective theory defined by this heterotic string theory, the only axion participating in the Green-Schwarz anomaly cancellation mechanism is the universal axion, which is dual to the heterotic 2-form field.
Given the close correlation between anomaly cancellation and the modular properties of the elliptic genus revealed in \cite{Schellekens:1986yi,Schellekens:1986xh,Lerche:1987qk}, 
our conjecture is that this is indeed the characteristic property which is responsible for modularity 
 of the string, also in situations where the theory is defined as a geometric compactification rather than directly at the level of conformal field theory.
In other words, we expect that the elliptic genus is fully modular whenever the anomaly cancellation mechanism receives contributions only from the universal axion in the heterotic duality frame, and thus is independent of the axionic partners of the K\"ahler moduli of the base space
$B_2$, as well as of those of the NS5-brane moduli, $\lambda_a$.

Under F-theory-heterotic duality, the axions paired with the K\"ahler moduli of $B_2$ arise from the
expansion of the M-theory 3-form $C_3$ with respect to the divisor classes $D_\alpha  = p^*(C_\alpha)$, where $C_\alpha$ are curves on $B_2$. The F-theory dual of the NS5-brane axions arise by expanding $C_3$ in terms of the blowup divisor classes $E_a$.
The condition for these not to participate in the anomaly cancellation mechanism is equivalent to the condition that their non-axionic partners do not enter the flux-induced D-term (\ref{xiFth}), i.e. 
\begin{equation}
\begin{split} \label{noexceptionalF}
F^{\rm b} \cdot b \cdot p^\ast(C_\alpha) &= 0  \qquad \quad   \forall \quad C_\alpha \in H_2(B_2)  \\
F^{\rm b} \cdot b \cdot E_a &= 0  \qquad \quad \forall  \quad E_a\,.
\end{split}
\end{equation}
Under these conditions, the elliptic genus of the string associated with $C_0$ is expected to be modular, and we refer to $U(1)$ fluxes, $G = F \circ \sigma(S)$, with this property as {\it modular fluxes}.

From experience with the six-dimensional heterotic string we know \cite{Lee:2018urn} that any involvement of the NS5-brane axions in the anomaly-cancellation mechanism
leads to a quasi-modular, as opposed to a modular, elliptic genus. 
This suggests that flux which satisfies
\begin{equation}
\begin{split} \label{noKahlerFtheory}
F^{\rm b} \cdot b \cdot p^\ast(C_\alpha) &= 0  \qquad \quad   \forall \quad C_\alpha \in H_2(B_2)  \\
F^{\rm b} \cdot b \cdot E_a &\neq  0  \qquad \quad \text{for some} \, \,  E_a
\end{split}
\end{equation}
leads only to a quasi-modular elliptic genus. The underlying fluxes are thus called {\it quasi-modular fluxes}.   

Finally, giving up even on this constraint, the elliptic genus will in general have much less distinguished modular properties.
Such generic fluxes will be called {\it non-modular}. They have the property that
\be \label{genericflux}
F^{\rm b} \cdot b \cdot p^\ast(C_\alpha) \neq 0  \qquad \quad   \text{for some} \, \,   p^\ast(C_\alpha)  \,.
\ee

The conditions (\ref{noexceptionalF}) and (\ref{noKahlerFtheory}) can  be equivalently characterised as follows:
If the flux obeys the condition (\ref{noexceptionalF}), then the D-term  (\ref{xiFth})  vanishes identically in the weak coupling limit (\ref{criticallimitdef}), in which ${\rm vol}(C_0) = 0$.
To see this, we expand the K\"ahler form of $B_3$ in terms of the K\"ahler cone generators $J_i$ as $J = t_i J_i$, where each $J_i$ can be written as a linear combination of the basis of divisors 
\be
\{ D_I \}  = \{ D_0 \,, D_\alpha  = p^*(C_\alpha), E_a \} \,.
\ee 
Here $D_0$ a section of the $\mathbb P^1$-fibration $B_3$ and $C_\alpha$ is a basis of $H^2(B_2)$. Altogether the K\"ahler form becomes
\be
J = t_i J_i = t_i \, a_{iJ} \,  D_J \,.
\ee
The flux induced D-term (\ref{xiFth}) follows as
\be
\zeta  = t_i a_{iJ} D_J  \cdot F^{\rm b} \cdot b = t_i \, a_{i0} \, (D_0 \cdot F^{\rm b} \cdot b) \,,
\ee
where we have made use of  (\ref{noexceptionalF}).
To compare this to the volume of $C_0$, note that $p^*(C_\alpha) \cdot C_0 = 0$ because $C_0$ is a fiber of $p^*(C_\alpha)$, and $E_a \cdot C_0 = 0$ because the fiber of the blow-up divisor $E_a$ is contained in $C_0$, while $C_0 \cdot D_0=1$.
As a result,
\be
{\rm vol}(C_0) = C_0 \cdot J = C_0 \cdot t_i a_{iJ} D_J =  t_i \, a_{i0}\,,
\ee
and so 
\be
\zeta|_{{\rm vol}(C_0) = 0} \equiv 0 \quad \text{for flux of type} \, \, (\ref{noexceptionalF}).  \label{noexeptional2}
\ee

By similar reasoning, for a flux obeying (\ref{noKahlerFtheory}), the D-term vanishes identically as a function of the moduli, 
only when we set ${\rm vol}(C_0) = 0$ and also ${\rm vol}(C_a) = 0$
for all fibers $C_a$ of the blow-up divisors $E_a$ over $\Gamma_a$. In order words:
\be
 \zeta|_{ \substack{    {\rm vol}(C_0) = 0 \phantom{\, \forall C_a} \\  {\rm vol}(C_a) = 0 \, \forall C_a} } \equiv 0 \,, \qquad  \text{but} \quad
\zeta|_{{\rm vol}(C_0) = 0} \;\slash\equiv\;  0   \qquad \quad \text{for flux of type} \, \, (\ref{noKahlerFtheory}) \,. \label{noKahlerFtheoryA}
\ee

To avoid potential confusion, we emphasize that the alternative characterization~\eqref{noexeptional2} and~\eqref{noKahlerFtheoryA} of modular and quasi-modular fluxes, respectively, are based only on the functional form of the flux-induced D-term in the K\"ahler moduli. In other words, although ${\rm vol}(C_0)=0$ guarantees that ${\rm vol}(C_a)=0$ for all $C_a$ due to the K\"ahler geometry, the criteria~\eqref{noexeptional2} and~\eqref{noKahlerFtheoryA} should be understood without imposing that the moduli lie in the K\"ahler cone.

\subsubsection{Interpretation via D3-brane/non-critical worldsheet instantons} \label{subsec_instantons}

Before widening the scope of this discussion, let us point out that the conditions (\ref{noexceptionalF}) and (\ref{noKahlerFtheory}) have an interesting interpretation in F-theory in terms of D3-brane instantons wrapping the respective divisors $p^*(C_\alpha)$ and $E_a$.
In presence of gauge flux, $G = F \circ \sigma(S)$, a D3-brane instanton along a general divisor $D_{\rm inst}$ can carry a net $U(1)$ charge \cite{Blumenhagen:2006xt,Ibanez:2006da,Haack:2006cy,Blumenhagen:2009qh}.
At the microscopic level, this is due to charged zero-modes localised on the intersection curve of the D3-brane instanton and the 7-brane divisor. 
For a $U(1)$ gauge group, the role of the 7-brane divisor is played by the height-pairing $b$. The net instanton charge is proportional to the chiral index of charged zero modes on $b \, \cap \,D_{\rm inst}$, which for vanishing instanton flux \cite{Grimm:2011dj} takes the form
\bea
Q_{D_{\rm inst}} = \int_{b \, \cap \, D_{\rm inst}} F^{\rm b}  = b \cdot  D_{\rm inst} \cdot F^{\rm b}  \,.
\eea
If $Q_{D_{\rm inst}} \neq 0$, the instanton generates couplings that violate the $U(1)$ selection rule by an amount of $Q_{D_{\rm inst}}$. 

Such a violation of the $U(1)$ symmetry by instantons along the divisors $p^*(C_\alpha)$ and $E_a$ is particularly drastic in the weak coupling limit: As it turns out, in this limit the volume of these divisors vanishes and the resulting instanton effects are unsuppressed. 
We can therefore rephrase the conditions for (quasi-)modularity of the elliptic genus by saying that this must not happen. More precisely, a flux is quasi-modular if
\bea
&Q_{D_{\rm inst}} = 0 &\qquad \text{for all}~ D_{\rm inst} \in \{ p^*(C_\alpha) \} \,,\label{instquasimod}\\
&Q_{D_{\rm inst}} \neq  0 &\qquad \text{for some}~  D_{\rm inst} \in \{ E_a \}  \,,\nonumber 
\eea
and modular if 
\bea
&Q_{D_{\rm inst}} = 0& \qquad \text{for all}~ D_{\rm inst} \in \{ p^*(C_\alpha), E_a \} \,.  \label{instmod}
\eea

To understand the significance of these two conditions on the dual heterotic side, note that 
the F-theory $U(1)$ symmetry maps, in general, to a linear combination of several abelian gauge group factors 
if the heterotic model involves several sub-bundles $V_j^{(i)}$ with non-zero $c_1(V_j^{(i)})$. All other linear combinations of abelian gauge group factors can only dualize to so-called geometrically massive $U(1)$ factors~\cite{Grimm:2011tb}
on the F-theory side. We can therefore discard these combinations for the purpose of our analysis and 
parameterize the relevant linear combination of abelian gauge factors as
\be
U(1) = \sum_j a_j \, U(1)_j^{(1)} + \sum_j b_j \, U(1)_j^{(2)} \,.
\ee
This $U(1)$ is the commutant within $E_8^{(1)} \times E_8^{(2)}$ of the structure group of the bundle $V$ with
\bea
c_1(V) &=& c_1(V^{(1)}) + c_1(V^{(2)})  \\
&=&  \sum_j a_j \, c_1(V_j^{(1)})+ \sum_j b_j \, c_1(V_j^{(2)}) \,.
\eea
In order for this bundle to map to an F-theory background,  it must satisfy the condition
\be \label{c1VEtauconstr}
 \int_{ {\mathbb E}_{\tau'}} c_1(V) = 0 \,.
 \ee
One way to understand this well-known constraint is that otherwise the NS5-branes along the fiber ${\mathbb E}_{\tau'}$ would participate in the Green-Schwarz cancellation mechanism. This is not possible, however, since the NS5-branes along the elliptic fiber of $Z_3$ dualize to D3-branes. Indeed, on the F-theory side, D3-branes do not participate in the anomaly cancellation mechanism of 7-brane gauge group factors. 
Since by assumption we are dealing with a situation with two independent sections on both the F-theory and the heterotic side, (\ref{c1VEtauconstr}) implies that
 \be \label{c1Vheterotic}
 c_1(V) = x (S' - S_0') + \rho^* c_1(N) \,,
\ee
where $c_1(N) \in H^{1,1}(B_2)$ and $x$ is a suitably quantized parameter. This ansatz is easily seen to obey the constraint (\ref{c1VEtauconstr}) because $\int_{{\mathbb E}_{\tau'}} \rho^* c_1(N) = 0$ and  $\int_{{\mathbb E}_{\tau'}} S' = 1 = \int_{{\mathbb E}_{\tau'}} S_0'$. 

A D3-instanton on $p^\ast(C_\alpha)$ maps to a heterotic world-sheet instanton on $C_\alpha$, which can likewise carry $U(1)$ charge given by
\be
Q_{\rm WS_{inst}} = \int_{C_\alpha} c_1(V) \,.
\ee
We hence conclude that for the elliptic genus to be at least quasi-modular, we must impose that
\be \label{c1VCalphahet}
\int_{C_\alpha} c_1(V) = 0 \qquad \forall \quad C_\alpha \subset B_2 \,.
\ee
This is the heterotic dual of the condition (\ref{instquasimod}).
If  the parameter $x$ in (\ref{c1Vheterotic}) vanishes, this requires $c_1(N) = 0$ and hence $c_1(V)=0$. In such a  case the $U(1)$ gauge theory is non-anomalous and the resulting elliptic genus vanishes identically.
For non-zero $x$, on the other hand, a non-trivial result is possible. 

As reviewed in Section \ref{sec_Aspectshet}, the worldsheet of a heterotic string over a curve $\Gamma_a$ splits into two non-critical E-string worldsheets, 
each of which couples to a different $E^{(i)}_8$ factor.
Correspondingly,
the D3-brane instanton on $p^*(\Gamma_a)$ thus maps to the sum of two {\it non-critical E-string instantons} on the heterotic side. 
The analogue of the modularity condition (\ref{instmod}) is therefore that in addition to (\ref{c1VCalphahet}), each individual E-string instanton over all curves $\Gamma_a$ must be uncharged, rather than only their sum. In other words we must have
\be \label{Gammanoncriticalcharge}
  \int_{\Gamma_a} c_1(V^{(1)}) = 0 = \int_{\Gamma_a}c_1(V^{(2)})    \qquad \forall \quad \Gamma_a \,,
 \ee
in order to maintain modularity. 
Note that the conditions for modularity and quasi-modularity only differ provided the $U(1)$ arises as a non-trivial combination of abelian gauge groups from both $E_8$ factors.

\subsubsection{Generalisation} \label{subsec_Generalisation}

We now propose a generalisation of the two criteria (\ref{instquasimod}) and (\ref{instmod}) to situations where the F-theory base $B_3$ is not simply the blowup of a $\mathbb P^1$-fibration over some base curves $\Gamma_a$.
On  a  general base $B_3$ that contains a curve $C_0$ of type (\ref{trivialnormalbundle}), we can still investigate the criterion (\ref{noexeptional2}) for modular flux, by checking if the flux-induced D-term vanishes identically in the limit of vanishing ${\rm vol}(C_0)$.
Furthermore, the analogue of the blowup fiber classes $C_a$ appearing in (\ref{noKahlerFtheoryA}) are all curves into which $C_0$ can split as it moves along its moduli space in $B_3$.


This motivates the following\\

\noindent{\bf Conjecture} {\it Consider F-theory on an elliptic fibration over base $B_3$ with gauge group $U(1)$,
height-pairing $b$ as in (\ref{heightpairing}) and U(1) flux $\Gfour = F \circ \sigma(S)$.
 Assume that $B_3$ contains a rational curve $C_0$ of type (\ref{trivialnormalbundle}).
 Denote by $C_a$ all curve classes into which $C_0$ can split as it sweeps out its moduli space on $B_3$.
If there exists a geometric limit which takes ${\rm vol}(C_0) \to 0$ while keeping the volume of $B_3$ finite, a D3-brane wrapped on $C_0$ gives rise to a quasi-perturbative, critical heterotic string.
The elliptic genus of this string is \\

\begin{tabular}{lll}
\text{quasi-modular:} \qquad & if  $J \cdot b \cdot F^{\rm b}|_{ \substack{    {\rm vol}(C_0) = 0 \phantom{\, \forall C_a} \\  {\rm vol}(C_a) = 0 \, \forall C_a} } \equiv 0$ \,, \qquad  {but} \quad
$J \cdot b \cdot F^{\rm b}|_{{\rm vol}(C_0) = 0} \;\slash\equiv\;  0$ \vspace{2mm}  \\ 
\text{modular:} \qquad           & if $J \cdot b \cdot F^{\rm b} |_{{\rm vol}(C_0) = 0} \equiv 0$ \,.
\end{tabular}
}

\vspace{2mm}

\section{Example of a (quasi-)modular flux compactification} \label{sec_Example1}

In this section we will present, as an explicit example, a four-dimensional, $N=1$ supersymmetric
F-theory compactification on a definite four-fold $Y_4$,
with a single $U(1)$ gauge group.\footnote{We have computed a number of further examples with analogous results; thus it suffices to present just a representative one here.} 
In particular we will confirm our criteria on the flux to obtain a modular or quasi-modular elliptic genus, as proposed in the previous section.

The elliptic four-fold, $Y_4$, is constructed by fibering elliptic curves over a base three-fold, $B_3$. 
For better readability, we have relegated the definition of $Y_4$ and $B_3$
as well as a detailed presentation of their toric data  to Appendix~\ref{ex_main}. 

The data given there are sufficient for computing, to any given order, the Gromov-Witten invariants, $N_{C_0; \Gfour}(n,r)$, of the curve classes 
(\ref{C0nrdef})
\beq
\Gamma_{C_0}(n,r)=
C_0 + n \, \mathbb E_\tau + \qr\, \cC^{\rm f}_{\qr=1}\,,
\eeq
in the presence of some  $U(1)$ flux $\Gfour$.
Here, the fibral curves classes $\mathbb E_\tau$ and $\cC^{\rm f}_{\qr=1}$ correspond  to the full fiber and the $U(1)$ fibral curve, respectively. They are
uniquely  determined  by their intersection properties
\bea
&S_0 \circ \mathbb E_\tau = 1\,,\quad S \circ \mathbb E_\tau = 1 \quad &\Longrightarrow \quad \sigma(S)\circ \mathbb E_\tau =0 \,, \\ 
&S_0 \circ \cC^{\rm f}_{\qr=1} = 0 \,,\quad S \circ \cC^{\rm f}_{\qr=1} = 1 \quad &\Longrightarrow \quad \sigma(S)\circ \cC^{\rm f}_{\qr=1} =1  \,. \nn
\eea
Written in terms of the Mori cone generators $l^{(1)},..., l^{(5)}$ defined in (\ref{mcm}), the classes $\mathbb E_\tau$ and $\cC^{\rm f}_{\qr=1}$ take the form 
\bea\label{Ef}
\mathbb E_\tau &=&3 l^{(3)} + 2 l^{(5)}  \,,\\
\cC^{\rm f}_{\qr=1} &=& l^{(3)}+l^{(5)}\,.\nn
\eea
On $B_3$ there exists a curve class $C_0=J_0\cdot J_0$ with the properties (\ref{trivialnormalbundle}), which admits a limit (\ref{criticallimitdef}). Its lift ${\bf C}_0 $ to a curve class in $H_2(Y_4, \IZ)$ can be expressed as
\beq
{\bf C}_0 =  l^{(2)} + l^{(4)}+l^{(5)}\,.
\eeq
This, together with~\eqref{Ef}, leads to
\beq
\Gamma_{C_0}(n,r)= l^{(2)} +(3n+r) l^{(3)} + l^{(4)} + (2n+r+1) l^{(5)} \,. 
\eeq
As a last ingredient we need to specify a basis $\Gfour_{j}$ of $U(1)$ fluxes, which we take to be
\beq\label{basisflux}
\Gfour_{j} = \pi^{-1}(D_j) \circ \sigma(S) \,\in \,\widetilde H^\ttw_{\ver,U(1)}(Y_4)\,.
\ee
Here $D_j$ denote the basis elements~\eqref{basischoice} of $H^{1,1}(B_3, \IZ)$.
In this basis, the  intersection   
form on $\widetilde H^\ttw_{\ver,U(1)}(Y_4)$ can be computed as 
\be
\eta\ =\ \left(
 \begin{array}{ccc}
 12 & 0 & -6 \\
 0 & 16 & -18 \\
 -6 & -18 & 12 \\
\end{array}
\right).
\ee
This corresponds to the right-most non-vanishing block in the full intersection metric (\ref{etaform}).

With these data, we can apply standard methods of mirror symmetry~\cite{Hosono:1994ax} for four-folds \cite{Mayr:1996sh,
Klemm:1996ts,
Klemm:2007in,
Haghighat:2015qdq,
Cota:2017aal}
and compute the 
expansion of the generating functions (\ref{FjC}),
\beq
\cF_{C_0}^{(j)}(q,\xi)\equiv \sum_{n \geq 0} \sum_{r} N_{C_0}^{(j)}(n,r) q^n \xi^r \,,
\eeq
to some given order as follows:
 \bea\label{Fjexample}
 \cF_{C_0}^{(1)}(q,\xi)\! \!&=&\! \!
12 q \left(\xi ^{\pm 2}+\xi^{\pm 1} \right) 
+18 q^2\left(51 \xi^{\pm 4}+910 \xi^{\pm 3}+3216 \xi ^{\pm 2}+3894 \xi^{\pm 1} \right) \nn
\\&&
+\,12 q^3 \left(14 \xi ^{\pm 6}+2181 \xi ^{\pm5}+31464 \xi ^{\pm4}+152648 \xi ^{\pm3}+339150 \xi ^{\pm2}+336111 \xi^{\pm1} \right) 
\nn
\\&&
+\,\cO(q^4)  \,,
\nn
\\ 
 \cF_{C_0}^{(2)}(q,\xi)\! \!&=&\! \!
 12 q \left(\xi ^{\pm2}+11 \xi^{\pm1} \right) 
 +6  q^2 \left(51 \xi ^{\pm4}+942 \xi ^{\pm3}+2960 \xi ^{\pm2}+4310 \xi^{\pm1} \right)
 \\&&
+\,12q^3  \left(4 \xi ^{\pm6}+711 \xi ^{\pm5}+10488 \xi ^{\pm4}+51496 \xi ^{\pm3}+110748 \xi ^{\pm2}+114885 \xi^{\pm1} \right) 
\nn
\\&&
+\,\cO(q^4) \,,
\nn
\\ 
 \cF_{C_0}^{(3)}(q,\xi)\! \!&=&\! \!
 48 q  \left(\xi ^{\pm2}+\xi^{\pm1} \right) 
   +18  q^2\left(51 \xi ^{\pm4}+868 \xi ^{\pm3}+3312 \xi ^{\pm2}+3828 \xi^{\pm1} \right)\nn
 \\&&
 +\,12q^3  \left(11 \xi ^{\pm6}+2244 \xi ^{\pm5}+31464 \xi ^{\pm4}+151628 \xi ^{\pm3}+341751 \xi ^{\pm2}+333672 \xi^{\pm1} \right) 
 \nn
\\&&
+\,\cO(q^4)\,. \nn
\eea
The appearance of $\xi^{\pm r}\equiv \xi^r - \xi^{-r}$ reflects that the generating functions are parity-odd in the $U(1)$ fugacity~$z$.

As a first consistency check we confirm  that the pieces with $q$-degree $n=1$ 
correctly reproduce the chiral indices of the zero modes in F-theory with respect to  each of the basis fluxes. 
 Recall from the discussion at the end of Section \ref{subsec_GW4} that these chiral indices are given by the 
  invariants $N^{(j)}_{0}(n,r)$ for the purely fibral
curve classes $n \mathbb E_\tau + r \cC^{\rm f}_{\qr=1}$. We explicitly confirm for $n=1,\ldots, 5$ that the only non-vanishing invariants are
\bea\nn
&&N^{(1)}_{0}(n,\pm 2)=\pm12\,,\quad N^{(1)}_0(n,\pm1)=\pm12 \,, \\
&&N^{(2)}_{0}(n,\pm2)=\pm12\,,\quad N^{(2)}_0(n,\pm1)=\pm132 \,, \\ \nn
&&N^{(3)}_{0}(n,\pm2)=\pm48\,, \quad N^{(3)}_0(n,\pm1)=\pm48 \,.
\eea
These match   the degeneracies at order $q^1$  in~\eqref{Fjexample}, precisely as predicted in~\eqref{pred}.

We now assemble the generating functions $\cF^{(j)}_{C_0}$ for the elliptic genus by specifying the four-flux
in terms of the basis (\ref{basisflux}) as
\beq\label{generalG}
\Gfour= \sum_{j=1}^3 c_j\, \Gfour_{j} \,.
\eeq
The integrality properties of $c_j$ are constrained by the flux quantization condition (\ref{freewit}); in terms of 
a minimal integral basis of $H^4(Y_4)$ the coefficients must be integral or half-integral, but finding such a minimal integral basis is not needed for our discussion. We therefore 
continue with the basis (\ref{basisflux}) and the understanding that the expansion coefficients must be chosen well-quantized.

The generalized elliptic genus, $Z(q, \xi,c)$, associated to the flux $G$ is then 
the corresponding linear combination of the generating functions $\cF_{C_0}^{(i)} (q, \xi)$ in (\ref{Fjexample}):
\beq\label{ZFjexpan}
Z(q, \xi,c_j)\ =\ q^{-1}\sum_{j=1}^3 c_j\, \cF_{C_0}^{(j)}(q, \xi)\ =:\ \frac1{\eta^{24}(q)}\Phi^-_{11}(q,\xi,c_j) \,. 
\eeq
Upon inspection, we do not observe any obvious modular properties 
of $Z(q, \xi,c_j)$ for generic $c_j$, other than
that the expansion coefficients in odd powers of $z$ appear to be quasi-modular forms:
\bea
(\xi \del_\xi)\Phi_{11}^-(q,\xi,c_j)\big|_{\xi=1}&=& 
 \frac{1}{72} {E_4}^3 (48 {c_1}+28 {c_2}+57
   {c_3})+\frac{1}{72} {E_6}^2 (57 {c_1}+7 {c_2}+48 {c_3})
   \nn\\
&&   -
\frac{1}{72} (3 {c_1}+{c_2}+3 {c_3})E_2\left(  {E_2}{E_4}^2 -34  {E_4}
   {E_6} \right),
\\
(\xi \del_\xi )^3\Phi_{11}^-(q,\xi,c_j)\big|_{\xi=1}&=& 
\frac{1}{72} {E_2} \left( {E_4}^3 (219 {c_1}+89
   {c_2}+246 {c_3})+ {E_6}^2 (195 {c_1}+49 {c_2}+168
   {c_3})\right) 
   \nn\\
 &&   -
\frac{1}{36}(3 {c_1}+{c_2}+3 {c_3}) \left( {E_2}^3 {E_4}^2  
+ 33 {E_2}^2 {E_4} {E_6}+
   35 {E_4}^2 {E_6} 
      \right),
   \nn
\eea
and so on. Despite this property, $\Phi_{11}^-(\tau,z,c_j)$ cannot be expanded  -  for general $c_j$ -  in terms of the generators (\ref{Jacringgens}) of weak Jacobi forms,
even if the quasi-modular form $E_2(\tau)$ is included.  Equivalently, it does not have a theta expansion (\ref{thetaexp}) of definite index $m$. 
Functions whose expansion is (quasi)-modular order-by-order in a formal power series in $z$, but which
 fail to have standard periodicity properties in $z$ are called 
Jacobi-like forms \cite{Dabholkar:2012nd,KanekoZagier,ChoieLee}, and this is what we seem to encounter here.\footnote{In fact \cite{ChoieLee},
Jacobi-like forms are a natural next step in the sequence of liftings: modular $\rightarrow$ quasi-modular $\rightarrow$ Jacobi-like.
This suggests that good modular properties might be recovered once we attribute suitable transformation properties to the fluxes as well.}
Indeed we have argued in Section \ref{subsec_quasimod} that generic fluxes would be in tension with standard modular or quasi-modular transformation properties, and this observation confirms our expectations.
On the other hand, if we restrict the flux such as to obey the constraints 
(\ref{noexceptionalF}) and (\ref{noKahlerFtheory}), we will momentarily show  that $Z(q, \xi,c_j)$ has the expected, well-defined modular
or quasi-modular properties. 

To this end, we first determine the divisors of $B_3$ of the form $p^\ast(C_\alpha)$ and the exceptional divisors $E_a$.
In the geometry under consideration, these sets of divisors are generated by
\bea
\langle p^\ast(C_\alpha) \rangle &=& \langle d_{y_1} \rangle  \equiv  \langle j_1 \rangle \,,
\\
\langle E_a \rangle &=& \langle  d_e \rangle  \equiv \langle j_1 + j_3 - j_2 \rangle \,.
\eea
Here $d_{y_1}$ and $d_e$ refer to the toric divisors in  Table \ref{tb:GLSM_B3}, and $j_i$ are the K\"ahler cone generators~\eqref{kcgen_B3} of the base, with triple intersection numbers~\eqref{inter_B3}.

To check for quasi-modular fluxes, we require eq.~(\ref{noKahlerFtheory}) to hold, i.e.
\be \label{quasimodular_ex}
0 \stackrel{!}{=} d_{y_1} \cdot F^{\rm b} \cdot b =  \sum_{j=1}^3 c_j  d_{y_1} \cdot D_j \cdot b 
 = 6c_1 + 2 c_2 + 6c_3 \,.
\ee



Before identifying the elliptic genus on this sub-family of fluxes as a modular or quasi-modular Jacobi form,
we first need to determine the fugacity index $m$ from the geometry of $Y_4$. 
In the basis~\eqref{basischoice}, the divisor of the height pairing takes the form
\bea
b &=& 6 \bar K - 2 \beta \\ \label{hp_ex}
 &=& (12-2{\rm x}) D_1 + (24 - 2{\rm y}) D_2 + (30-2{\rm z}) D_3 \,,\nn
\eea
with $({\rm x}, {\rm y}, {\rm z}) = (2, 2, 4)$ in this example.
The $U(1)$ fugacity index of the heterotic elliptic genus is therefore computed as
\beq
m= \frac{1}{2} b \cdot C_0 = 6- \rm x = 4 \,,
\eeq
where we used $C_0=J_0\cdot J_0$ and $J_0=j_1$ as per Appendix~\ref{ex_main}.

After eliminating $c_3$ via (\ref{quasimodular_ex}), it is now easy to verify that the elliptic genus~\eqref{ZFjexpan} is indeed quasi-modular of weight
 $w=-1$ and index $m=4$. By matching a sufficient number of Gromov-Witten invariants to a general ansatz in terms of the generating functions (\ref{Jacringgens}) augmented by $E_2(\tau)$, we find that the denominator
 $\eta^{24}$ cancels out as expected, and obtain the following
 quasi-modular expansion into weak Jacobi forms:
\bea\label{quasimodular_Z}
Z(q, \xi,c_1,c_2, c_3=-c_1-\frac{c_2}{3}) &=& q^{-1}  \big(c_1\, \cF^{(1)}(q, \xi) + c_2\, \cF^{(2)}(q, \xi) - (c_1+\frac{c_2}{3})\, \cF^{(3)}(q, \xi)\big)  \nn\\ 
&=& \varphi _{-1,2} 
\left(
\frac{3}{4}\left(c_2-c_1\right) {\varphi _{0,1}}^2
+ \frac{1}{12} E_4 \left(27 c_1+13 c_2\right) {\varphi _{-2,1}}^2 \right.
\nn \\
&& \qquad \,  \left. -\frac16  E_2 \left(9 c_1+11 c_2\right) \varphi _{-2,1} \varphi _{0,1}
\right). 
\eea 
In addition we can impose  the extra condition~\eqref{noexceptionalF} to  decouple the NS5-brane axions from the Green-Schwarz mechanism,
\beq\label{modular_ex}
0 \stackrel{!}{=} d_e \cdot F ^{\rm b} \cdot b  = c_1 + 3 c_2 - 2 c_3 \,.
\eeq 
When combined with~\eqref{quasimodular_ex}, this yields
\beq
9c_1 + 11 c_2 = 0 \,.
\eeq
Precisely this combination makes the quasi-modular piece proportional to $E_2$ disappear, 
 rendering the elliptic genus into a genuine weak Jacobi form  of weight
 $w=-1$ and index $m=4$:
\beq\label{modular_Z}
Z(q,\xi,c_1 = -\frac{11}{9} c_2, c_2, c_3 = \frac{8}{9} c_2) = \frac{5}{3}  c_2\,\varphi_{-1,2}\,( {\varphi_{0,1}}^2- E_4 {\varphi_{-2,1}}^2) \,.
\eeq
The first terms of its expansion are
\bea
Z(q,\xi,c_1 = -\frac{11}{9} c_2, c_2, c_3 = \frac{8}{9} c_2) &=& 
40c_2 \left(4 \xi^{\pm1} +\xi ^{\pm2}\right)+480c_2\, q \left(3 \xi^{\pm1} -\xi ^{\pm3}\right) \\
&+& 40c_2\, q^2 \left(204 \xi^{\pm1} +3 \xi ^{\pm2}-88 \xi ^{\pm3}+12 \xi ^{\pm5}-\xi ^{\pm6}\right) +\cO(q^3).  \nonumber 
\eea
Here, as anticipated purely based on modularity in Section \ref{subsec_ellgen}, we  observe a gap at charge level ${\mq}=\pm m=\pm 4$.  

\begin{figure}[t!]
\centering
\includegraphics[width=12cm] {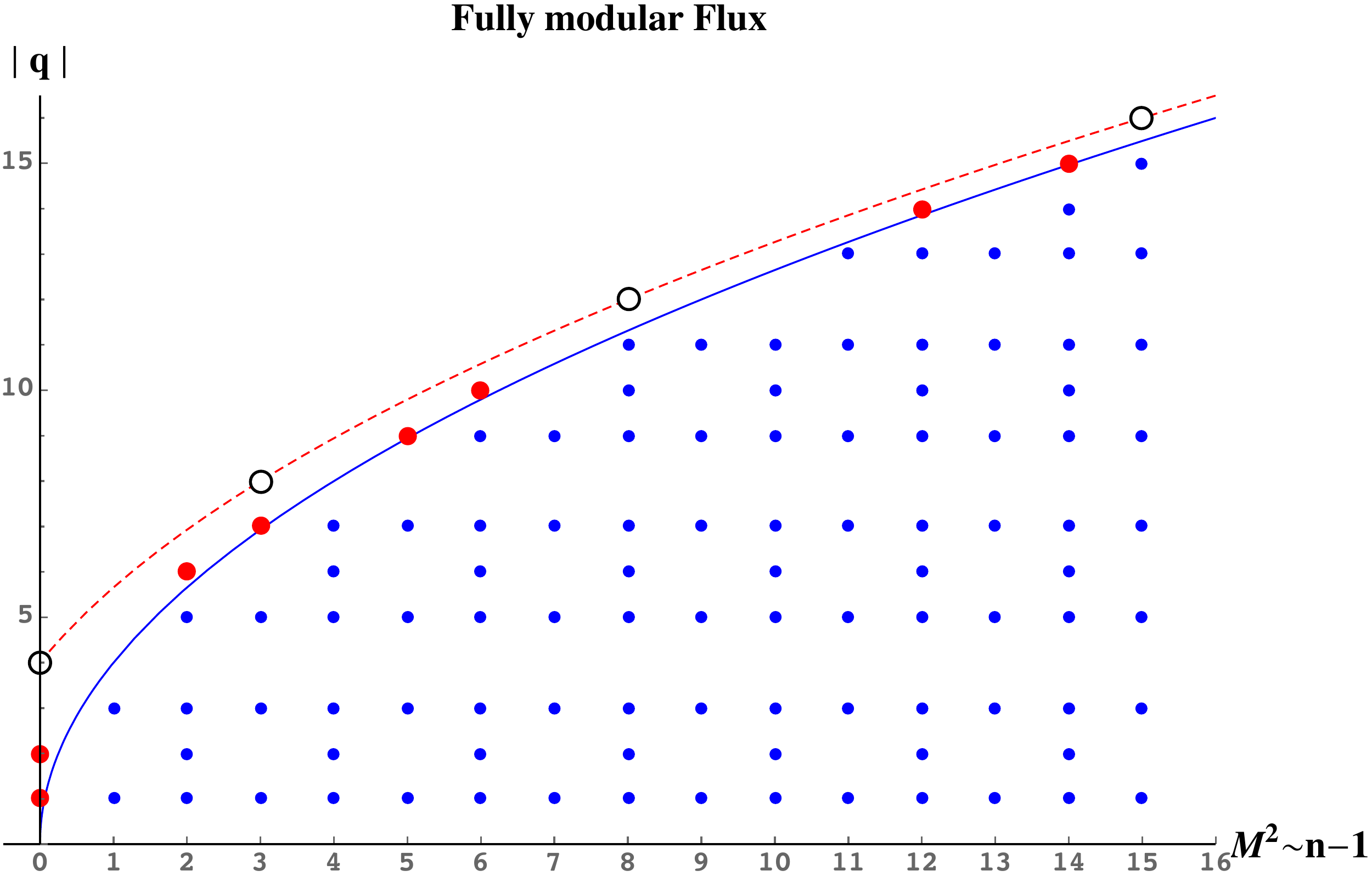}
\caption{Shown is the charge-mass spectrum encoded in the fully modular elliptic genus~(\ref{modular_Z}).
The fat red dots indicate super-extremal states, which lie above the
solid blue line encoding the charge-to-mass ratio of classical extremal black holes as discussed in Section \ref{subsec_WGCbounds}.
Note the gaps at charges ${\mq}=4\IZ$; in particular the open dots,
which would correspond to the possible maximal super-extremal states and which would form a charge sublattice by themselves,
are not populated. Qualitatively this picture does not only apply to the present example, but to the generic situation with modular fluxes.
We will come back to it in the next section in the context of Weak Gravity Conjectures.
 }

\label{f:modularflux}
\end{figure}

The charge pattern encoded in (\ref{modular_Z}) over an extended range of excitation levels $n$ is depicted in Figure~\ref{f:modularflux}. We clearly see charge gaps at ${\mq}=mk$, $k\in\IZ$, 
which follow already on general grounds from the theta expansion (\ref{thetaexp})  of the elliptic genus.
Note that for a perturbative heterotic string we can identify the excitation level $n-1$ with the mass $M^2$ of the physical states associated with the excitations, more precisely $M^2/(8 \pi T) = n-1$, where $T$ denotes the string tension.
The solid blue curve is defined by ${\mq}^2/(M^2/8 \pi T)=4m$ and
characterizes the charge-to-mass ratio of extremal dilatonic black holes which play a role in the context of the Weak Gravity Conjecture. This will be explained in Section \ref{subsec_WGCbounds}. The
open dots correspond to the possible, but  not populated maximally super-extremal states which would lie on the dashed red line that corresponds to
the leading orbit of $\ell=0,m$ in the theta expansion.

\begin{figure}[t!]
\centering
\includegraphics[width=17cm] {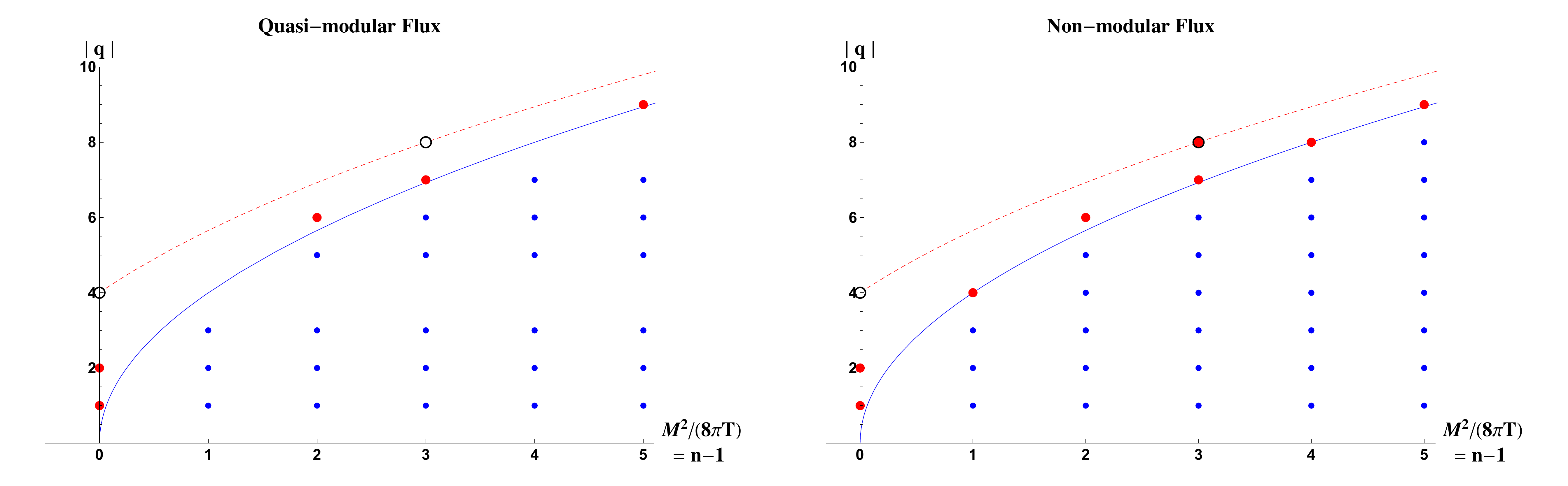}
\caption{Same as Figure 1, except for quasi-modular (\ref{quasimodular_Z}), and generic non-modular  background fluxes, for which there are fewer and fewer cancellations in the spectrum. 
Note that in the last case, some super-extremal states do appear to form a sublattice, ${\mq}=2k$. 
However, for other models this is generically not the case. }
\label{f:lessmodularflux}
\end{figure}

We also have exhibited, as red fat dots, the existence of super-extremal states, whose charges
just barely lie above the blue curve. 
There are also other gaps in the part of spectrum encoded in the elliptic genus, for example at $(M^2/8 \pi T,q)=(5,6)$ in Figure~\ref{f:modularflux}.
It turns out that these intermediate locations become populated for quasi-modular fluxes leading to
 (\ref{quasimodular_Z}).  Moreover, all gaps are fully populated
for the non-modular elliptic genus (\ref{ZFjexpan})  for a generic choice for the $c_j$. This is what we have shown in Figure~\ref{f:lessmodularflux}. Thus, morally speaking, the better behaved under modular transformations 
the elliptic genus becomes upon judiciously choosing the flux configuration, the more non-trivial cancellations occur.

At this point one may worry whether the criterion of completeness of the charge lattice \cite{Polchinski:2003bq,Banks:2010zn}, which is part of the lore of quantum gravity conjectures,
may be violated due to the gaps. And relatedly, whether the sublattice conjecture of the super-extremal states \cite{Heidenreich:2016aqi} is violated too,
since in general the charges of the super-extremal states do not appear to form a sublattice.
We just note here the obvious fact that the absence of contributions of certain charges to the elliptic genus does not necessarily imply that
the states do not exist, and refer to the concluding Section \ref{sec_conc} for a further discussion.

\section{Weak Gravity Conjectures in four dimensions} \label{sec_WGC}


We are now in the position to quantitatively address the Weak Gravity Conjectures for F-theory, when compactified 
on an elliptically fibered four-fold $Y_4$ to four dimensions, additionally equipped with activated gauge fluxes.
We assume that an unbroken gauge group $\Gauge$ arises from a stack of 7-branes along  some divisor ${\bf S}$ on the base, $B_3$, of the fibration.
As reviewed in Section \ref{subsec_Geomset}, if $\Gauge= U(1)$, this divisor is the height-pairing associated with the $U(1)$ symmetry, whereas for non-abelian gauge groups ${\bf S}$ is a component of the discriminant locus of the elliptic fibration $Y_4$. 


We will analyze  the Weak Gravity Conjecture in the 
 weak coupling limit of the gauge theory in the presence of gravity, i.e. in the limit
\be \label{limitvolSinfty}
 \frac{1}{g^2_{\rm YM}} \sim  {\rm vol}({\bf S}) \to \infty \,, \qquad \quad M^2_{\rm Pl}  \sim {\rm vol}(B_3) \quad {\rm finite} \,.
\ee
 To this end, we will show in Section~\ref{subsec_geometry1} quite generally  that whenever a limit of the form (\ref{limitvolSinfty})
can be taken,
there exists a curve $C_0$ in $B_3$  whose classical volume  goes to zero.
The string associated with a D3-brane wrapped on $C_0$ hence becomes tensionless in the limit (\ref{limitvolSinfty}).

The crucial question is then which precise type of tensionless string will arise, and this depends on the specific properties the shrinking curve $C_0$.
The analogous problem in six dimensions allowed for a complete classification of the possibilities \cite{Lee:2018urn,Lee:2018spm}: In this case, the curve $C_0$ on the base $B_2$ of the elliptic threefold $Y_3$ is a rational curve with trivial normal bundle, i.e. $C_0 \cdot C_0 =0$.
A D3-brane wrapped on any such $C_0$ gives rise to an asymptotically tensionless, weakly coupled heterotic string. Its excitations, as characterized by the elliptic genus, contain states satisfying the Weak Gravity Conjecture \cite{ArkaniHamed:2006dz} in its sublattice version \cite{Heidenreich:2016aqi}. 
A complementary recent study of infinite distance and weak coupling limits in the K\"ahler moduli space of Calabi-Yau three-folds, with or without elliptic fibration, has furthermore been performed in \cite{Corvilain:2018lgw}.

As we will see, for F-theory on an elliptic four-fold fibration, $Y_4$, there is a considerably richer set of possibilities to realize the limit (\ref{limitvolSinfty}) in the K\"ahler moduli space of the base $B_3$. 
We will identify two types of weak coupling limits: The first type is similar to the situation in six dimensions in that $C_0$ is guaranteed to be a rational curve with trivial normal bundle, i.e. of type (\ref{trivialnormalbundle}). It therefore leads again to an asymptotically tensionless, weakly coupled heterotic string.
Combining this with the results of the previous sections, we will show in Section~\ref{subsec_WGCbounds} that its spectrum contains states that satisfy the 
original Weak Gravity Conjecture.  On the other hand, we show that the related sub-lattice conjecture, which posits that the super-extremal states should include a charge sublattice, is in general not satisfied, at least at the level of the index.

The second type of weak coupling limit, by contrast, corresponds to a qualitatively new situation with no analogue in six dimensions: Here $C_0$ may give rise to a {\it non-critical} string because $C_0$ is not necessarily of the form (\ref{trivialnormalbundle}). In this situation an explicit analysis of the string spectrum remains 
a particularly exciting challenge for future work.

\subsection{Geometry of the weak coupling limit} \label{subsec_geometry1}

To describe the limit (\ref{limitvolSinfty}), consider the most general ansatz for the K\"ahler form of $B_3$,
\beq
J = \sum_{i \in \cI} t_i J_i \,,
\eeq
in terms of the generators $J_i$ of the K\"ahler cone, with $t_i \geq 0$ for all $i\in \cI$.
The K\"ahler cone generators have the important property that their mutual intersection numbers are non-negative,
\be \label{JiJjJk}
d_{ijk}:=J_i \cdot J_j \cdot J_k \geq 0 \,.
\ee
 Since the K\"ahler cone generators are nef, they are guaranteed to be pseudo-effective, i.e. they lie in the closure of the cone of effective divisors.
Throughout this paper we are making the technical assumption{\footnote{
The effectiveness of $J_i$ follows if the cone of effective divisors is closed. The latter may fail as illustrated in an example by Mumford (see e.g. \cite{Lazarsfeld});
however this geometry is a 
surface with $b_1 \neq 0$ and can hence not serve as an F-theory base.  While we are not aware of any example with a non-closed cone of effective divisors which can be the base of an elliptically fibered Calabi-Yau manifold, it would be interesting to investigate this possibility further.}
that  
all generators $J_i$ are in fact {\it effective} and that the generators are furthermore {\it irreducible}.

To take ${\rm vol}({\bf S}) \to \infty$, at least one of the generators, called $J_0$, must have a coefficient $t$ which scales to infinity. We can hence split the index set $\cI$ for the K\"ahler cone generators into
\beq 
\cI = \cI_0 \cup \cI_1 \cup \cI_2 \cup \cI_3\,,
\eeq
and write
\be \label{Jansatzgen}
J = t J_0 +  \sum_{\nu \in \cI_1} a'_\nu J_\nu +  \sum_{\mu \in \cI_2} b'_\mu J_\mu +   \sum_{r \in \cI_3} c_r J_r \,.
\ee
Here $J_0$ is the only K\"ahler cone generator of type $\cI_0$. The remaining generators are divided into sets with the respective properties
\bea
&&J_0^2 \cdot J_\nu \neq 0 \,,  \quad \nu \in \cI_1\,, \label{Jnugenerators} \\
&&J_0^2 \cdot J_\mu = 0  \,, \quad \mu \in \cI_2 \,,\quad \text{and} \,  \text{there exists} \, \, \mu' \in \cI_2: J_0 \cdot J_{\mu} \cdot J_{\mu'}  \neq 0 \,,  \label{Jmugenerators} \\
&& J_0^2 \cdot  J_r = 0 \,,\quad r \in \cI_3 \,,\quad \,  \text{and for all}~ i \in \cI_2 \cup \cI_3: \, \,  J_0 \cdot  J_{r} \cdot  J_{i}= 0 \,. \label{Jr1} 
\eea
In order for 
\be
{\rm vol}(B_3) = \frac{1}{6} J^3
\ee
 to remain finite in the limit $t \to \infty$, the generator $J_0$ must obey
 \be\label{J03=0}
J_0^3 = 0 \,.
\ee
This is because (\ref{JiJjJk}) makes cancellations among the contributions from the different K\"ahler moduli impossible.
If no such K\"ahler cone generator $J_0$ exists, the base $B_3$ does not admit the limit (\ref{limitvolSinfty}).
An example of such a geometry without a weak coupling limit is $B_3 = \mathbb P^3$.

The two qualitatively different cases to consider are now distinguished as follows:
\bea
&\rm{Type \, \, A:} \qquad \quad &J_0 \cdot J_0 \neq 0 \\
&\rm{Type \, \, B:} \qquad \quad &J_0 \cdot J_0 = 0  \,.
\eea

Depending on the geometry of the base space $B_3$ and the choice of the divisor ${\bf S}$ on it, 
the weak coupling limit (\ref{limitvolSinfty}) might be realizable 
both for an ansatz (\ref{Jansatzgen}) of Type A or of Type B,
each for a different choice of K\"ahler cone generator $J_0$ in (\ref{Jansatzgen}). An example of this phenomenon is discussed in detail in Appendix \ref{App_Illustration}.
For other pairs $(B_3, {\bf S})$, the ansatz   (\ref{Jansatzgen}) may always be of the same Type A or Type B. 
Of course, as noted already, $(B_3, {\bf S})$ might also not admit a weak coupling limit (\ref{limitvolSinfty}) at all.

 \vskip .5cm
\noindent
We will now turn to discussing the two types of weak coupling limits in more detail.

\subsubsection*{\bf Type A: $J_0 \cdot J_0 \neq 0$}

In Appendix \ref{pf_case1} we prove the following properties of geometries for
 which the weak coupling limit (\ref{limitvolSinfty}) is realized as in (\ref{Jansatzgen}) with $J_0 \cdot J_0 \neq 0$:
\begin{enumerate}
\item
The index set $\cI_2$ of K\"ahler cone generators is empty, and the weak coupling limit must take the form
\beq \label{Jsplit}
J = t J_0 + \sum_{\nu \in \cI_1} \frac{a_\nu}{t^2} J_\nu +  \sum_{r \in \cI_3} c_r J_r \,,
\eeq 
where  the K\"ahler parameters $a_\nu$ and $\frac{c_r}{t}$ must be finite as $t \to \infty$ in order for the volume of $B_3$ to remain finite. 

\item
The non-trivial cycle 
\be \label{C0definitionWGC}
C_0 = J_0 \cdot J_0
\ee
on $B_3$ is a rational holomorphic curve with trivial normal bundle $N_{C_0/B_3} = {\cal O}_{B_3} \oplus {\cal O}_{B_3}$.
The proof of this makes use of Mori's cone theorem together with the fact that $C_0$ intersects the divisor ${\bf S}$ non-trivially,
\be
m:= \frac{1}{2} C_0 \cdot {\bf S} = \frac{1}{2} J_0 \cdot J_0 \cdot {\bf S} \neq 0 \,.
\ee
The fact that $m \neq 0$ follows from the fact that ${\rm vol}({\bf S}) \to \infty$ as $t \to \infty$.
Triviality of the normal bundle is a consequence of $J_0^3=0$, which is a defining property of the K\"ahler cone generator $J_0$ responsible for the presence of gravity in the weak coupling limit.

\item
In the weak coupling limit (\ref{limitvolSinfty}), the classical volume of $C_0$ shrinks as
\be
{\rm vol}(C_0)  \to \frac{a_\nu}{t^2} d_{0 0 \nu} \qquad \quad {\rm for} \, \, t \to \infty \,.
\ee
A D3-brane on $C_0$ hence gives rise to an asymptotically tensionless, weakly coupled heterotic string as  $t \to \infty$.

Furthermore,
\beq \label{volSvolC0}
{\rm vol}({\bf S}) \, {\rm vol}(C_0)   =  2m \,  {\rm vol}(B_3)  \qquad \quad {\rm as} \, \, t \to \infty  \,.
\eeq 
This  will play a crucial role in our quantitative proof  of the Weak Gravity Conjecture in Section~\ref{subsec_WGCbounds}.
\end{enumerate}

The above results are a direct generalization of the findings for the weak coupling limit in six-dimensional F-theory models \cite{Lee:2018urn}.
In that case, the curve $C_0$ supporting a heterotic string can easily be proven to be the {\it only} curve with this property which shrinks in the limit.
This reflects the fact that there exists a unique heterotic duality frame in which the dynamics of the weak coupling limit is described. 
On the other hand, for four-dimensional compactifications with 3-fold base spaces, $B_3$, proving the analogous statement would be more involved because of the considerably richer structure of the Mori cone
of K\"ahler 3-folds. We are not aware of a counter-example to the conjecture that also in this case, for limits of Type A, the shrinking heterotic curve $C_0$ is unique, and in fact illustrate this 
for the example of Section \ref{sec_Example1} in Appendix \ref{ex_main}. 

All statements so far are in the realm of classical geometry.
In the context of nearly tensionless strings in four dimensions with $N=1$ supersymmetry, one might worry  that quantum corrections modify these findings because 
the quantum volume of a curve need in general not vanish even if the classical volume does \cite{Mayr:1996sh}.
However, the classical zero-volume limit of $C_0$ coincides with the limit where the dual heterotic string is weakly coupled, i.e.  $S_{\rm het} \to \infty$ with $S_{\rm het}$ the heterotic dilaton (see footnote \ref{footnotehettension}).
This limit is not expected to be obstructed by quantum corrections. We therefore advocate that the statements under point 3 above survive potential quantum corrections, at least to leading order.

\subsubsection*{\bf Type B: $J_0 \cdot J_0 = 0$}

Suppose, by contrast, that the weak coupling  limit is realized by a K\"ahler form (\ref{Jansatzgen})  in a setting where the K\"ahler cone generator $J_0$ obeys $J_0 \cdot J_0 = 0$.
As an immediate consequence, the cycle (\ref{C0definitionWGC}) ceases to exist. While one can still show that there is a holomorphic curve $C_0$ which shrinks in the weak coupling limit,
it is no longer guaranteed that it is a rational curve with trivial normal bundle. Rather its normal bundle in $B_3$ is of the form ${\cal O}_{C_0} \oplus {\cal O}_{C_0}(d)$ for some $d \geq 0$.
More precisely, we will show in Appendix~\ref{pf_case2} that for $J_0 \cdot J_0 = 0$, the weak coupling limit has the following properties:
\begin{enumerate}
\item
There are no generators of type $\cI_1$ since $J_0 \cdot J_0 = 0$, and furthermore one finds that there cannot be any generators of type $\cI_3$. The limit (\ref{Jansatzgen}) hence takes the form
\be \label{TypeIIlimitansatz}
J = t J_0 + b'_\mu J_\mu \,, \qquad \quad \mu \in \cI_2 \,,
\ee
for suitably constrained parameters $b'_\mu$.
\item
The fact that ${\rm vol}({\bf S}) \to \infty$ implies that there must exist some K\"ahler cone generators $J_{\mu_0}$, $J_{\nu_0}$  with $\mu_0, \nu_0 \in \cI_2$ such that $d_{0 \mu_0 \nu_0} \neq 0$ and 
\bea
b'_{\nu_0} &=& b_{\nu_0} \, t^{-1 + a_{\nu_0}}  \\
b'_{\mu_0} &=& b_{\mu_0} \, t^{- a_{\nu_0} - \Delta_{\mu_0}} 
\eea
for $a_{\nu_0} >0$ and $\Delta_{\mu_0} \geq 0$. The parameters $b_{\nu_0}$ and $ b_{\mu_0}$ stay finite in the limit $t \to \infty$.
\item
The cycle $C_{\nu_0}:= J_0 \cdot J_{\nu_0}$ is a non-trivial holomorphic curve which classically shrinks in the limit as
\be
{\rm vol}(C_{\nu_0}) \, \to \,  d_{0 \nu_0 \mu} b'_{\mu} \lesssim {\cal O}(t^{- a_{\nu_0}}) \qquad {\rm as} \quad  t \to \infty \,.
\ee

\item
The normal bundle of the shrinking curve $C_{\nu_0}$  takes the form
 \be
  N_{C_{\nu_0}/B_3} = {\cal O}_{C_{\nu_0}} \oplus  {\cal O}_{C_{\nu_0}}(d_{\nu_0}) 
  \ee 
with
\be
 d_{\nu_0} := d_{0 \nu_0 \nu_0} \geq 0 \,.
 \ee
If $d_{\nu_0} = 0$, then a D3-brane along $C_{\nu_0}$ gives rise to an asymptotically tensionless critical heterotic string.
An example of this type is discussed in Appendix \ref{TypeIIExample-App}, where we also verify the analogue of relation (\ref{volSvolC0}) for $C_{\nu_0}$.

For non-zero $d_{\nu_0}$, the effective string from a D3-brane on $C_{\nu_0}$ is a more exotic, non-critical string.
An example where in fact no rational curve class with trivial normal bundle shrinks in the limit (\ref{Jansatzgen})  with $t \to \infty$ is discussed in
 Appendix \ref{App_Illustration}. The Weak Gravity Conjecture strongly suggests that a tower of massless states arises also from such strings in the tensionless limit, but a quantitative proof of this conjecture
 is beyond the scope of this article.

\end{enumerate}

Two more comments are in order:
{ First, unlike for limits of Type A, we do know of examples where several heterotic curves shrink simultaneously in the weak coupling limit.
This is exemplified in Appendix \ref{TypeIIExample-App} and represents a striking phenomenon from the perspective of string duality. In such exotic situations, the dual theory is in 
a self-dual regime where a `fundamental' and a `solitonic' heterotic string become equally important. 
Clearly, this impedes a conclusive quantitative analysis of the spectrum. As stressed already, such complications are a novelty in four dimensions with no analogue in six-dimensional models.
}

Second, since for $d_{\nu_0} >0$ the curve $C_{\nu_0}$ does not give rise to a critical heterotic  string, the question of quantum corrections modifying its tension becomes even more pressing.
From the perspective of the Swampland Distance conjecture, the appearance of a tensionless string in the limit of $g_{\rm YM} \to 0$ is a natural expectation: Such a tensionless string would in particular 
explain the breakdown of the effective field theory in the infinite distance limit where a global symmetry emerges in presence of gravity.
We take this as suggestive that the limit of vanishing volume of $C_{\nu_0}$ is indeed attained even after taking quantum corrections into account.
Note that $C_{\nu_0}$ always has non-negative normal bundle. The string it supports is therefore fundamentally different from what we would call the four-dimensional analogue of E-strings and their cousins in six dimensions: The latter are inherently strongly coupled and become tensionless at finite distance points in moduli space. It is not implausible that the structure of quantum corrections distinguishes between curves of positive and negative normal bundle, and in particular the former case of curves is more protected.

\subsection{Compatibility of the weak coupling limit with D-terms}\label{sec:compatibility}

In four-dimensional compactifications with $N=1$ supersymmetry, the scalar potential can in principle obstruct a  weak coupling limit of the form (\ref{limitvolSinfty}) even if it is attainable at the level of pure geometry.
F-term obstructions  have recently been discussed from the perspective of various quantum gravity conjectures in \cite{Gonzalo:2018guu}.
In our context, even if we ignore the possibility of a non-perturbatively generated superpotential, the  D-term potential (\ref{xiFth}) induced by gauge fluxes may not be compatible with the weak coupling limit.

From the general analysis in Section \ref{subsec_quasimod}, we deduce, on the other hand, that both quasi-modular and modular fluxes are automatically compatible with a vanishing D-term, when taking the limit (\ref{limitvolSinfty}) and hence ${\rm vol}(C_0) \to 0$.\footnote{The discussion of (quasi-) modular fluxes is of course only sensible in limits of Type A, or of Type B for an asymptotically tensionless rational curve $C_0$ of trivial normal bundle.}
Indeed, for modular fluxes the induced D-term vanishes identically for ${\rm vol}(C_0) \to 0$, and the induced scalar potential dynamically drives the system to the weak coupling limit. 
When the flux is only quasi-modular, corresponding to (\ref{noKahlerFtheory}), the D-term depends not only on ${\rm vol}(C_0)$, but also on the K\"ahler moduli of the fibers $C_a$ of the blow-up divisors; however, to stay within the closure of the K\"ahler cone of $B_3$, we must also take their volume to zero whenever ${\rm vol}(C_0) \to 0$, since the blow-up fibers lie inside the fiber $C_0$ of $B_3$. In this sense, the D-term constraint is again automatically satisfied in the limit ${\rm vol}(C_0) \to 0$.

Beyond the realm of (quasi-)modularity, the question of whether or not the D-term is compatible with taking ${\rm vol}(C_0) \to 0$ depends on the concrete choice of fluxes. As long as the D-term can be set to zero for a non-zero K\"ahler form on the boundary of the K\"ahler cone, the weak coupling limit is compatible with the choice of flux. 
For fluxes other than the (quasi-) modular ones discussed above, this requires $h^{1,1}(B_3) \geq 3$ because it rests on a cancellation of the contributions of two or more K\"ahler moduli different from ${\rm vol}(C_0)$ and ${\rm vol}(C_a)$.

It is understood from now on that we restrict ourselves to gauge fluxes whose associated D-term is indeed compatible with taking the weak coupling limit. According to the above considerations, the most {\it generic} such fluxes will not lead to a quasi-modular or modular elliptic genus for the asymptotically tensionless heterotic string, but may nonetheless define a bona fide weak coupling limit in F-theory.

\subsection{WGC bounds: elliptic genus versus supergravity} \label{subsec_WGCbounds}

We now provide a quantitative proof of the Weak Gravity Conjecture in the weak coupling limit (\ref{limitvolSinfty}),
under the assumption that a single curve $C_0$ with the properties (\ref{trivialnormalbundle}) and (\ref{volSvolC0}) exists
in the geometry. As one of our main results we will exemplify that a tower of super-extremal states exists, even though it does in general not form a charge sublattice.

As an initial step, we assume that the gauge flux leads to a modular or quasi-modular elliptic genus.
Despite some important differences, the proof of the Weak Gravity Conjecture is analogous to the analysis in six dimensions \cite{Lee:2018urn,Lee:2018spm},
and rests on the properties of the elliptic genus as a Jacobi form, in connection with a shift of the vacuum energy.
Let us first recapitulate the situation in six dimensions. Here the elliptic genus has the generic form 
$Z(\tau,z)=\eta^{-24}(\tau)\Phi^+_{10,m}(\tau,z)$, where $\Phi^+_{10,m}(\tau,z)$ is some weak Jacobi form of weight 
$w=10$. It is necessarily
symmetric under $z\to-z$.  Thus there can be a non-vanishing
constant term and in fact there will be always one, namely corresponding to the (unphysical) tachyon of the heterotic string.  
Therefore the term with $\ell=0$ in the theta expansion (\ref{thetaexp}) of $\Phi^+_{10,m}$,
\be
h_0(\tau)\Theta^+_{0,m}(\tau,z) = ({\rm const}.+\cO(q))\sum_{k\in\IZ} q^{(2mk)^2/4m}\xi^{2mk} \,,
\ee
is always present and encodes a lattice orbit of states with $\mq_k=2mk$ and $n_k=\mq_k^2/4m$, which trivially satisfy
\be\label{maxsuper}
\mq_k^2=4mn_k\,.
\ee
These  ``maximally super-extremal" states are always populated and automatically satisfy the condition 
$
\mq_k^2  >  4m(n_k-1)\,,
$
which lies at the heart of the proof of the Weak Gravity Conjecture.
Indeed, after identifying the excitation level
with the squared mass, $n_k-1 \sim {M^2_{n_k}}$ (including the shift by the vacuum energy),
this eventually translates into the bound $\mq_k^2 g^2_{\rm YM}> {M^2_{n_k}}/M^{4}_{\rm Pl}$,
which is what the Weak Gravity Conjecture postulates.

The situation in four dimensions is different in that the elliptic genus  has  weight $w=-1$ and thus is of the form
$Z(\tau,z)=\eta^{-24}(\tau)\Phi^-_{11,m}(\tau,z)$. 
 As we pointed out in Section~\ref{subsec_ellgen}, this means that $\Phi^-_{11,m}(\tau,z)$ is odd
under $z\to-z$ and so cannot have a constant term. Therefore  $Z(\tau,z)$ is not a meromorphic
but actually a holomorphic Jacobi form.
Moreover,  the terms with $\ell=0,m$ cancel in the theta expansion  (\ref{thetaexp}).
This implies that the previously leading terms in the  theta expansion are absent,  and that the
corresponding orbit of maximally super-extremal states is missing in the elliptic genus
(signified by the open dots in Figure~\ref{f:modularflux} for our
example). 

On the other hand, massless physical states must necessarily show up at order $q^1$ in the expansion of $\Phi^-_{11,m}(q,\xi)$.
Therefore, for each pair of massless states with charges $ \mq=\pm\ell$,
the theta-function expansion (\ref{thetaexp}) must contain non-vanishing terms of the form
\bea
h_\ell(\tau)\Theta^-_{\ell,m}(\tau,z) \ &=&\ \left(q^{1-\ell^2/4m}\ ({\rm const}.+\cO(q))\right)\sum_{k\in\IZ} q^{(2mk+\ell)^2/4m}\xi^{2mk+\ell} - (\ell \to -\ell)  \nonumber\\
 &\sim & q^1(\xi^\ell-\xi^{-\ell})+\dots\,.
\eea
The theta function then implies that for each such massless pair there must exist lattice orbits of states with 
\be\label{nolatt}
\mq_k=2mk\pm\ell\,,  \qquad n_k=k(mk\pm\ell)+1 \,,\qquad k\in\IZ\,,
\ee
which therefore obey
\be\label{notmaximal}
{\mq_k}^2 = 4m(n_k-1)+ \ell^2\,.
\ee
These states are less super-extremal than the ones in (\ref{maxsuper}) which arise from the $\ell=0$ orbit, which is why
we have called the latter maximally super-extremal. 
Since these do not show up in the elliptic genus in four dimensions, we need to test the Weak Gravity Bound
for the states with $\ell\not=0$ that satisfy the weaker equation~(\ref{notmaximal}).
Despite the shift $n_k\to n_k-1$, which a priori tends to create tension with the conjectured bound, the day is saved by
the extra contribution of $\ell^2$; in effect the modification is to trade the loss of the offset by $1$ against a gain of $\ell^2/4m$,
which is still positive for $\ell\not=0$.

Note that from equation~(\ref{nolatt}) it is evident that the super-extremal states do not form a sublattice of the charge lattice,
rather they lie on sublattices $\Delta\mq=2km$ shifted by $\pm\ell$.  This has been visualized for our example
in Figure~\ref{f:modularflux}, where we have depicted some of these super-extremal states by fat red dots.  They belong to the four lattice orbits
that originate from the massless states with $\mq=\pm1,\pm2$.

Moreover, as pointed out before, Figure~\ref{f:modularflux} shows charge gaps at $\mq_k=mk$ \footnote{Figure~\ref{f:modularflux} 
displays a further $\IZ_2$ sub-structure of gaps, which arises in an analogous way and which is tied
to the fact that $m=4$ is not prime in our example.}, which generally follow from the cancellation of the theta functions for
$\ell=0,m$. One can understand this physically by describing the lattice sector of the world-sheet theory in terms of a free boson, by writing
 the $U(1)$ current as $J_{U(1)}=-i\sqrt{2m}\partial H$. 
Then the lattice charge sector labelled by $\ell$ is generated by a vertex operator given by $V_\ell=e^{i\ell H/\sqrt{2m}}$
(the physical vertex operators contain in general additional holomorphic and anti-holomorphic pieces).
On the other hand, lattice shifts are generated by conserved, 
holomorphic symmetry currents, $V_{\pm2m}$, of conformal dimensions $(m,0)$ and charges $\mq=\pm 2m$. 
These may be viewed as generating a spectral flow
symmetry that links together pairs of states, $V_{\pm m}$, which contribute with opposite signs to the elliptic genus. 

Up to now, we have assumed that the $U(1)$ flux is chosen such that it gives rise to a fully modular or quasi-modular elliptic genus,
in line with our considerations of Section~\ref{sec:fluxes}.
As we have been arguing, more generic choices of $U(1)$ fluxes do not lead 
to (quasi-)modular Jacobi forms. Note that such fluxes satisfy the $D$-term conditions at best on a subspace of the closure of the K\"ahler moduli space, even in the weak coupling regime.
For such fluxes the elliptic genus generically 
does not enjoy a theta expansion of the form discussed above.
As a result, the above charge gaps do not need to occur. Indeed this is what we have demonstrated in our example in Section~\ref{sec_Example1}.  We have found that
the charge spectrum is nevertheless bounded by the curve (\ref{maxsuper}) and just differs from the (quasi-)modular spectra in that all the gaps become populated.
We do not have a precise understanding of this general situation, given that we have much less a rigid mathematical structure at our disposal and are lacking a closed expression for the elliptic genus.
However in view of our observations, it seems reasonable to assume that this pattern holds in general, which would allow us to extend the discussion of the Weak Gravity Conjecture also to cover these more general situations.

In order to complete the proof of the WGC, we now
put the dimensionful quantities back in order to obtain expressions that are directly relevant for physics.  Dropping
the subscript $k$,  we thus obtain
\be
\mq^2 \, g^2_{\rm YM} = (4 \,  m \, (n-1) +\ell^2)\, g^2_{\rm YM} = 4 \, m \, g^2_{\rm YM} \left(\frac{M_n^2}{8 \pi  \, T} + \frac{\ell^2}{4m}\right) \,,
\ee
where we have traded the excitation number $n$ for the spacetime  mass $M_n$ via the standard relation
\be
M_n^2 =  8 \pi \, T (n-1) \,,\qquad \quad T = 2 \pi \,  {\rm vol}(C_0) \,.
\ee
This relation is valid for the critical string associated with the curve $C_0$ at least in the weak coupling limit.
As a consequence of the offset by $\ell^2/4m$, we have
\be \label{inequality1}
\mq^2 \, g^2_{\rm YM}  \geq \frac{4 \, m \, M_n^2}{8 \pi \, {\rm vol}({\bf S}) {\rm vol}(C_0)   } = \frac{M_n^2}{4 \pi \, {\rm vol}(B_3)} = \frac{M_n^2}{M^2_{\rm Pl}} \,,
\ee
where we have used (\ref{Mplgymdef}).
The second step crucially depends on the relation (\ref{volSvolC0}).

The remaining task in testing the Weak Gravity Conjecture is to compare this relation to the precise bound 
that takes the contribution of the scalar fields into account \cite{Heidenreich:2015nta,Palti:2017elp}.
The requirement is that a set of particles must be super-extremal with respect to all non-BPS extremal charged black holes in the same theory,
that is
\be
 g^2_{\rm YM}    \frac{\frak{q}^2  }{M^2}       \stackrel{!}{\geq}       g^2_{\rm YM}    \frac{Q^2_{\rm BH} \, }{M_{\rm ADM}^2}   \,   .  
\ee
As discussed in detail for six-dimensional theories \cite{Lee:2018urn}, the relevant black holes in the weak coupling limit (\ref{limitvolSinfty}) are extremal dilatonic Reissner-Nordstrom black holes \cite{Heidenreich:2015nta}, and the same logic applies to the  situation in $d=4$.

It is in fact illuminating to perform the computation for an arbitrary number $d$ of spacetime dimensions.
Consider therefore  $d$-dimensional Einstein-Maxwell-dilaton theory with action
\be \label{Sdcannorm}
S= \int_{\mathbb R^{1,d-1}} \frac{M^{d-2}_{\rm Pl}}{2} \left( \sqrt{-g} R -  d \phi  \wedge \ast d \phi \right)  - \frac{1}{2 g^2_{\rm YM}} e^{\alpha \phi} F \wedge \ast F \,,
\ee
where $\alpha \in \mathbb R$ determines the coupling of the massless dilaton $\phi$ to the gauge kinetic term.
The charge-to-mass ratio of an extremal dilatonic Reissner-Nordstrom black hole solution in this theory is \cite{Gibbons:1987ps,Heidenreich:2015nta}
\be \label{chargetomassBH}
\frac{Q^2_{\rm BH} \,  g^2_{\rm YM}}{M_{\rm ADM}^2} = \frac{1}{M^{d-2}_{\rm Pl}} \left(  \frac{d-3}{d-2}  + \frac{\alpha^2}{4}\right) \,.
\ee
The first, $d$-dependent term is the purely gravitational contribution, while the second term is due to the scalar charge of the black hole with respect to the massless dilaton $\phi$.

In the present setup, the gauge coupling scales with the volume modulus ${\rm vol}(C_0)$ in the F-theory duality frame. The situation is, however, most simply described directly in the dual  heterotic frame, which is the physically relevant duality frame 
in the asymptotic weak coupling limit (\ref{limitvolSinfty}). 
In the heterotic frame, the inverse gauge coupling square is proportional to the dilaton field, which in $d=4$ dimensions takes the form 
\be
S^{d=4}_{\rm het} = e^{ - 2 \Phi} {\rm vol}( B_3)\,,
\ee
where $\Phi$ is the ten dimensional string frame dilaton. The weak coupling limit hence translates into the limit 
\be
S_{\rm het} \to \infty \,, \qquad M_{\rm Pl} \, \,  \text{finite} \,.
\ee

The relevant pieces of the dynamics follow by straightforward dimensional reduction of the 10d heterotic string frame action
\be
S_{\rm 10d} =  \int_{\mathbb R^{1,9}} \frac{1}{2 \kappa^2_{10}}  \left(  \sqrt{- g}  R + 4 \,  d \Phi \wedge \ast d \Phi  \right) -   \frac{1}{2 g^2_{\rm YM,10}}   e^{-2 \Phi} F \wedge \ast F \,.
\ee
The important information for us is the precise normalization of the kinetic term of the dilaton $\Phi$ and how it enters the exponential in the gauge kinetic term. 
After dimensional reduction to $d$ dimensions all terms pick up a factor of ${\rm vol}(Z_{5-d/2})$, which for the present purpose we can simply normalize to one.
The relevant step consists in the usual Weyl rescaling of the $d$-dimensional metric
\be \label{gMNrescaling4d}
g_{MN} \, \,  \to \, \,  e^{\frac{4}{d-2} \Phi} \, g_{MN}  \,, \qquad \quad M,N = 0,1,\ldots, d-1\,, 
\ee
to remove the $\Phi$-dependence of the Einstein-Hilbert term.
After this rescaling we obtain the action
\be \label{action-4da}
S = \int_{\mathbb R^{1,d-1}}  \frac{M^2_{\rm Pl}}{2} \sqrt{-g } \left( R - \frac{4}{d-2} d \Phi \wedge \ast d \Phi \right) - \frac{1}{2 g^2_{\rm YM}}  e^{- \frac{4}{d-2} \Phi} F \wedge \ast F \,.
\ee
Comparison with the canonically normalised action (\ref{Sdcannorm}) identifies
\be
\phi = \sqrt{\frac{4}{d-2}} \Phi \,, \qquad \quad \alpha = - \sqrt{ \frac{4}{d-2}}  \,.
\ee
In particular, the charge-to-mass ratio (\ref{chargetomassBH}) of an extremal dilatonic charged black hole is independent of the number of spacetime dimensions, and universally given by
\be
\frac{Q^2_{\rm BH} \,  g^2_{\rm YM}}{M_{\rm ADM}^2} = \frac{1}{M^{d-2}_{\rm Pl}} \,.
\ee
This is a result of the interplay between the purely gravitational and the dilaton-dependent \cite{Heidenreich:2015nta} contribution in (\ref{chargetomassBH}).
The latter can equivalently be interpreted  in terms of the Yukawa interactions between the charged particles in the weak coupling limit \cite{Palti:2017elp}, as further discussed also in \cite{Lee:2018spm}.

To conclude, the tower of charged states satisfying the inequality (\ref{inequality1}) are in precise agreement with the super-extremality condition, 
as required by the Weak Gravity Conjecture. 
Furthermore, the tower of super-extremal states does in general not form a charge sublattice.
To the best of our understanding, such a sublattice of super-extremal states is a {\it sufficient} condition \cite{Heidenreich:2015nta,Heidenreich:2016aqi} for the Weak Gravity Conjecture to be stable under dimensional reduction, and moreover realized in 
a number of string theoretic models such as \cite{Heidenreich:2016aqi,Montero:2016tif,Lee:2018urn}. That the sublattice version of the conjecture might in principle be allowed to fail (as long as a suitable tower of super-extremal states exists) has been pointed in \cite{Andriolo:2018lvp}, based on consistency arguments for the effective field theory.
Our four-dimensional $N=1$ supersymmetric setup provides explicit examples for this, at least at the level of the index for the unique heterotic string becoming tensionless in the weak coupling limit.

\subsection{St\"uckelberg masses and the Weak Gravity Conjecture}\label{stuck}

We now interpret our results on the Weak Gravity Conjecture in four dimensions  in light of the phenomenon that 
in the presence of gauge fluxes, the $U(1)$ gauge boson generically acquires a St\"uckelberg mass.   
As is well known, this is a consequence of the Green-Schwarz anomaly cancelling term, which is given by $\int B\wedge F$ in four dimensions and necessarily appears 
\cite{Lerche:1987sg,Lerche:1987qk} whenever the elliptic genus is nonzero. 
More generally it arises from the flux-induced gauging of the shift symmetry of a certain linear combination of axions.
That is, if we denote the specific linear combination of axions participating in the St\"uckelberg mechanism by $a$, the four-dimensional effective action contains a term\footnote{To read off the physical St\"uckelberg mass, we have rescaled the gauge potential compared to the action (\ref{action-4da}) such as to arrive at a canonically normalised gauge kinetic term.}
\be \label{axionkinetic}
S = \int_{\mathbb R^{1,3}} \frac{1}{2} F \wedge \ast F +    \frac{1}{2} f_a^2 (d a - g_{\rm YM} A) \wedge \ast (d a - g_{\rm YM} A) + \ldots \,,
\ee
from which one deduces 
\be
M^2_{\rm St} =  g^2_{\rm YM} f_a^2  \,.
\ee
At energy scales below $M_{\rm St}$, the abelian gauge symmetry is broken to a global $U(1)$ symmetry, which by 
itself is in general further
 broken by instanton effects \cite{Blumenhagen:2006xt,Ibanez:2006da,BerasaluceGonzalez:2011wy}  to a discrete symmetry.
 Interesting consequences of the Weak Gravity Conjecture for St\"uckelberg massive $U(1)$s have been pointed out in \cite{Reece:2018zvv}. 

For generic values of $g_{\rm YM}$ and $f_a$ of order one, the mass of the $U(1)$ gauge field
sits near the compactification scale $M_{\rm KK}$ of the theory. However, in the weak coupling limit we consider,
the St\"uckelberg mass term tends to zero {\it unless} at the same time the decay constant $f_a$ for the St\"uckelberg axion diverges such as to compensate for the vanishing of $g^2_{\rm YM}$.
To determine the parametric behaviour of $M^2_{\rm St}$ in the weak coupling limit, note that in the Type IIB/F-theory frame, the St\"uckelberg axion is a linear combination of the axionic partners of the K\"ahler moduli of the base $B_3$.
At general points in K\"ahler moduli space, the mass is given by (up to numbers of order one 
which we neglect, see e.g.  \cite{Buican:2006sn,Plauschinn:2008yd,Conlon:2008wa})
\bea \label{MStexplicit}
M^2_{\rm St} = g^2_{\rm YM} \, G^{\alpha \beta}  \Pi_\alpha \Pi_\beta \,.
\eea
Here $G^{\alpha \beta}$ is the inverse of the metric 
\be \label{Galphabeta}
G_{\alpha \beta} = - \frac{ {3}/{2}}{{\cal K}} {\cal K}_{\alpha \beta}  + \left(\frac{ {3}/{2}}{{\cal K}} \right)^2 {\cal K}_\alpha {\cal K}_\beta \,
\ee
on the K\"ahler moduli space of Type IIB compactified on $B_3$ \cite{Grimm:2004uq,Jockers:2004yj}.
It is defined in terms of the K\"ahler metric $J = v^\alpha \omega_\alpha$ in some basis, $ \omega_\alpha$, of $H^{1,1}(B_3)$, with intersection numbers ${\cal K}_{\alpha \beta \gamma} = \omega_\alpha \cdot \omega_\beta \cdot \omega_\gamma$ and 
\bea
{\cal K}_{\alpha \beta} &=& {\cal K}_{\alpha \beta \gamma} v^\gamma \,,\\
 {\cal K}_{\alpha } &=&  {\cal K}_{\alpha \beta \gamma} v^\beta v^\gamma \,, \\
{\cal K} &=&  {\cal K}_{\alpha \beta \gamma} v^\alpha v^\beta v^\gamma \equiv 6 {\rm vol}(B_3) \,.
\eea
The quantity 
\be
\Pi_\alpha = \omega_\alpha \cdot {\bf S} \cdot F^{\rm b} \,,
\ee
on the other hand, is purely topological. Here
 $F^{\rm b} \in H^{1,1}(B_3)$ defines the $U(1)$ four-flux  $\Gfour = \pi^{-1}(F^{\rm b}) \circ \sigma(S)$, as before.
 
Even without entering a detailed analysis of the K\"ahler metric $G_{\alpha \beta}$, we can deduce the following parametric behavior of $M^2_{\rm St}$ in the weak coupling limit: \\

\begin{tabular}{lll}
{$G$ modular}: \qquad &     $M^2_{\rm St} \sim \frac{1}{t^6} M^2_{\rm KK}$    \qquad &  \vspace{2mm} \\
{$G$ quasi-modular}: \qquad &     $M^2_{\rm St} \sim \frac{1}{t^3} M^2_{\rm KK}$ \, \,      \qquad & \qquad \quad \quad  \text{as} \quad $g^2_{\rm YM} \sim \frac{1}{t^2}$ for $t \to \infty$  \,.\vspace{2mm} \\
{$G$ generic}: \qquad &$M^2_{\rm St} \sim M^2_{\rm KK}$ \qquad &  \\
\end{tabular}
\vspace{3mm}

To see this, let us identify the basis $\omega_\alpha$ with the generators of the K\"ahler cone (or a subset thereof if the K\"ahler cone is non-simplicial).
Assume first that the flux $G$ is chosen such as to give rise to a modular elliptic genus. According to the criterion (\ref{noexceptionalF}) this means that the only non-vanishing contribution to 
$\Pi_\alpha$ is from the overlap with the direction in K\"ahler moduli space that is dual to the heterotic dilaton, $S_{\rm het}$.
The St\"uckelberg axion $a$, which is its axionic partner, is hence dual to the universal heterotic 2-form field, ${\rm B}_2$, $d{\rm B}_2 = f_a^2 \ast da$.
The kinetic term of ${\rm B}_2$ in the four-dimensional Einstein frame
 follows by dimensional reduction of the ten-dimensional  kinetic term in the string frame:
\be
\frac{1}{4 \kappa^2_{10}} \int_{\mathbb R^{1,9}} e^{-2\Phi} d{\rm B}_2 \wedge \ast d{\rm B}_2 \qquad     \Longrightarrow     \qquad    \frac{ {\rm vol}(B_3) }{4 \kappa^2_{10}} \int_{\mathbb R^{1,3}} e^{-4 \Phi}  d{\rm B}_2 \wedge \ast d{\rm B}_2 \,.
\ee
The extra scale factor of $e^{-2 \Phi}$ appears in going from the four-dimensional string to Einstein frame via the Weyl rescaling (\ref{gMNrescaling4d}).
In view of (\ref{action-4da}), we can furthermore identify the dilaton with the K\"ahler parameter that implements the weak coupling limit,
\be
t \sim e^{-\Phi} \,,
\ee
and conclude $f_a^2 \sim \frac{1}{t^4}$.\footnote{Note that the kinetic term of the 2-form dual to an axion with kinetic term (\ref{axionkinetic}) scales as $\frac{1}{f_a^2} \int_{\mathbb R^{1,3}}d{\rm B}_2 \wedge \ast d{\rm B}_2$.} This combines with $g^2_{\rm YM} \sim \frac{1}{t^2}$ to the parametric behaviour of  $M^2_{\rm St} \sim \frac{1}{t^6} M^2_{\rm KK}$.


For quasi-modular fluxes, the St\"uckelberg axion is a linear combination of the universal axion and the axions dual to the chiral 2-forms ${\rm B}^a_2$ associated with the heterotic NS5-branes.
The kinetic term of ${\rm B}^a_2$ in the four-dimensional Einstein frame follows from the pseudo-action of the NS5-brane wrapped along a curve $\Gamma^{\rm H}_a$, where the index ${\rm H}$  reminds us that this curve is viewed as a curve on the base of the heterotic elliptically fibered three-fold $Z_3$. Up to numerical prefactors, and taking into account again the rescaling to four-dimensional Einstein frame, this gives: 
\be \label{NSaxkin}
 \int_{\mathbb R^{1,5}} d{\rm B}^a_2 \wedge \ast d{\rm B}^a_2 \qquad     \Longrightarrow     \qquad   {\rm vol}(\Gamma^{\rm H}_a)  \int_{\mathbb R^{1,3}} e^{-2 \Phi}  d{\rm B}^a_2 \wedge \ast d{\rm B}^a_2 \,.
\ee
Note that unlike for the universal axion, the six-dimensional kinetic term of ${\rm B}^a_2$ carries no prefactor of $e^{-2\Phi}$. 
The volume of the  curve $\Gamma^{\rm H}_a$ generically scales like 
\be \label{volGamma}
 {\rm vol}(\Gamma^{\rm H}_a) \sim \frac{1}{t} \,.  
\ee
To see this, consider the corresponding curve in the dual F-theory geometry $B_3$, which we denote by $\Gamma^{\rm F}_a$ for clarity.
A D3-brane instanton on $p^*(\Gamma^{\rm F}_a) \subset B_3$ on the F-theory side maps to a heterotic worldsheet instanton on $\Gamma^{\rm H}_a$.
Both instanton suppression factors must coincide. On the F-theory side, $S_{\rm D3} \sim {\rm vol}(p^*(\Gamma^{\rm F}_a)) \sim t \times \frac{1}{t^2} = \frac{1}{t}$, where the first factor of $t$ is the volume of the curve $\Gamma^{\rm F}_a$ and the second factor comes from the fiber $C_0$. The scaling of ${\rm vol}(\Gamma_a^{\rm F}) \sim t$ in turn follows from the scaling of the K\"ahler cone generators in the weak coupling limit.
On the other hand, the worldsheet instanton action on the heterotic side scales like  ${\rm vol}(\Gamma^{\rm H}_a)$, leading to (\ref{volGamma}).\footnote{Note that there are no further factors of $e^{2\Phi}$ in the heterotic worldsheet instanton action even in the four-dimensional heterotic string frame, since going from the four-dimensional string to Einstein frame leaves the metric on $Z_3$ unchanged.}
Combining (\ref{volGamma}) with (\ref{NSaxkin}) implies $f^2_a \sim \frac{1}{t}$ for the axions dual to ${\rm B}^a_2$, which dominates over the contribution from the universal axion and results in the following scaling:~$M^2_{\rm St}\sim \frac{1}{t^3} M^2_{\rm KK}$.

The remaining case corresponds to  generic, i.e. non-modular fluxes, for which also the K\"ahler moduli axions in the dual heterotic frame may mix into the St\"uckelberg axion.
It is now easier to analyze the system in the Type IIB frame. The generic behaviour of $G_{\alpha \beta}$ can be deduced by inspection of (\ref{Galphabeta}) for the weak coupling limit under consideration. 
The leading 
component of its inverse $G^{\alpha \beta}$ turns out to scale as $t^2$.
For generic $\Pi_\alpha$, this leading term in the axion decay constant together with  $g_{\rm YM}^2 \sim 1/t^2$  implies that $M^2_{\rm St} \sim \frac{1}{t^2} \times t^2 M^2_{\rm KK} \sim M^2_{\rm KK} $, which is in marked contrast as compared to the (quasi-)modular fluxes.

This behaviour is indeed explicity confirmed by direct evaluation of (\ref{MStexplicit}) for the three different types of fluxes, in particular
 for the specific model presented in Section \ref{sec_Example1}.

The suppression pattern of the  St\"uckelberg mass for quasi-modular versus generic fluxes resonates perfectly with the results of Section \ref{subsec_instantons}: Whenever the St\"uckelberg mass is parametrically vanishing in the weak coupling limit because the fluxes obey the (quasi-)modularity constraints,
all unsuppressed instanton effects respect the $U(1)$ symmetry, as expected for an intact gauge symmetry.

The parametric behaviour of $M_{\rm St}$ is particularly interesting in the context of the Weak Gravity Conjecture.
For the special case of (quasi-)modular fluxes, we see that the St\"uckelberg mass is parametrically suppressed with respect to the compactification scale $M_{\rm KK}$, at least in the weak coupling regime $t \to \infty$ where we are testing the Weak Gravity Conjecture. In this sense, there exists a broad energy range $M_{\rm St}\sim \frac{1}{t^s} M_{\rm KK} \leq E \leq M_{\rm KK}$ (with $s=3$ or $s=3/2$ for modular or quasi-modular fluxes) where the abelian symmetry effectively acts as a gauge symmetry, and hence the Weak Gravity Conjecture must clearly hold. This is the type of fluxes for which we are explicitly observing a tower of super-extremal states with the expected properties, albeit not a charge sublattice.

On the other hand, for more generic non-modular fluxes, 
the $U(1)$ symmetry behaves like a global symmetry at all energies below $M_{\rm KK}$, even in the limit where $g_{\rm YM} \to 0$.
One might wonder if in such a situation the super-extremality bound is actually required to be satisfied in the four-dimensional effective theory.
Our take on this interesting question is that the $U(1)$ symmetry should be viewed as part of the gauge symmetry in the ultra-violet, in particular at all energies above $M_{\rm KK}$. In this regime, the theory becomes higher-dimensional by definition, and the effective gauge symmetry may enhance from $U(1)$ to a bigger gauge group. Nonetheless the theory must obey the higher-dimensional super-extremality condition with respect to this abelian subgroup of the full gauge symmetry. 
As we have discussed in the previous section, the quantitative bound is the same across all dimensions. In this sense, the observed super-extremality may be seen as the four-dimensional shadow of the higher dimensional Weak Gravity Conjecture.

For non-modular fluxes there is no closed expression for the elliptic genus as a weak Jacobi form, but in all examples we have studied the predictions of the Weak Gravity Conjecture
are satisfied for a tower of states. More precisely, this statement holds by  extrapolation of the low-excitation spectrum which we are explicitly computing via mirror symmetry.
It would be extremely interesting to understand
if there is any deeper meaning to the appearance of gaps in the tower of super-extremal states for (quasi-) modular fluxes,
 in view of the special scaling behaviour of the St\"uckelberg mass as compared to more generic fluxes.

\section{Conclusions and Outlook}\label{sec_conc}

In this work we have launched a quantitative analysis of the Weak Gravity Conjecture in string compactifications to four dimensions with only $N=1$ supersymmetry.
Despite major differences compared to
 six-dimensional compactifications with eight supercharges \cite{Lee:2018urn,Lee:2018spm}, 
we have again been able to confirm, under certain assumptions,
 the predictions of the Weak Gravity Conjecture combined with the Swampland Distance Conjecture:
 Modulo some caveats that were discussed in Section \ref{subsec_geometry1}, in the vicinity of a weak coupling point of a gauge theory coupled to gravity (here for gauge group $U(1)$), a tower of charged states becomes light. It contains, as a subset, a tower of states whose charge-to-mass ratio exceeds that of certain charged dilatonic extremal black holes. 
 
We have reached this conclusion in the framework of F-theory compactified on Calabi-Yau four-folds $Y_4$ with 
specific four-form fluxes activated. 
The tower of states forms a subset of the excitations of a solitonic string that arises from a D3-brane wrapping a distinguished, 
shrinking curve $C_0$ on $Y_4$.
In favorable situations this string is identified as a critical heterotic string, which is not necessarily perturbative, and its elliptic genus provides valuable information about its charged spectrum.

At a technical level, what makes the non-perturbative
computation of the elliptic genus possible is the duality with M-theory/Type IIA theory on the same elliptic fibration. This is similar to the situation in six dimensions \cite{Klemm:1996hh,Haghighat:2013gba,Haghighat:2014vxa}.
The elliptic genus can be inferred from the genus-zero free energy of the topological string on $Y_4$, in the presence of the  four-flux.
This free energy in turn is computable via standard methods of mirror symmetry on four-folds \cite{Mayr:1996sh,
Klemm:1996ts,
Klemm:2007in,
Haghighat:2015qdq,
Cota:2017aal}.

The new feature that we had to take into account in this work is the flux background, which lies in a particular component of the primary vertical  subspace of $H^{2,2}(Y_4)$. The fluxes contained in this subspace lift to chirality generating $U(1)$ gauge fluxes in F-theory.
Their intersection form is in agreement with the fact that the induced genus-zero free energy is, under suitable conditions, a quasi-modular Jacobi form
 of weight $w=-1$, in contradistinction with  different types of vertical fluxes which do not lift to gauge fluxes in F-theory \cite{Haghighat:2015qdq,Cota:2017aal}. 
 
More precisely,  we have found that only a subset of the a priori possible gauge fluxes leads to a quasi-modular or fully modular elliptic genus. 
 The quasi-modular and the modular fluxes are in turn distinguished by whether or not NS5-branes \cite{Honecker:2006dt,Blumenhagen:2006ux} (in the dual heterotic frame) contribute to anomaly cancellation. 
In this sense, we have provided a non-perturbative generalisation of the classical, conformal field theoretical elliptic genus. 
The behaviour for more general fluxes is a particularly exciting direction for future research: The elliptic genus is still anti-symmetric in the $U(1)$ field strength fugacity, albeit not a standard Jacobi form, and it is tempting to investigate to what extent it obeys distinguished arithmetic properties.

To be more specific, let us recapitulate the meaning of the charge spectrum as encoded in the elliptic genus. In contrast to its reductions to three or two dimensions,
this spectrum is not BPS saturated. Rather it consists of level-matched pairs of the left-moving tower of charged states, as encoded in the elliptic genus, with right-moving oscillator excitations of the Ramond ground states. In the parity-odd sector of the partition function, 
these states necessarily cancel out at massive levels due to world-sheet supersymmetry.
Nevertheless, individually these massive states must exist, at least at weak coupling where the description in terms of conformal field theory
is accurate. Note that these states are protected, at tree level,
 against deformations by moduli, due to holomorphic factorization, modularity and their pairing with oscillator excitations.
This observation is sufficient for proving, at tree level,  the Weak Gravity Conjecture in the presence of $U(1)$ flux, which posits that at least {\it some} states with a charge-to-mass ratio larger than that of certain extremal black holes exist.

Moreover, we have found that
if the elliptic genus is modular or quasi-modular, it contains gaps in the charge spectrum (see Figure~\ref{f:modularflux}).
This issue is  more drastic for a non-chiral theory where the elliptic genus vanishes identically, 
and even in an otherwise chiral four-dimensional F-theory compactification it
necessarily vanishes if the  $U(1)$ four-flux vanishes. The question is whether this jeopardizes some of the
Quantum Gravity Conjectures, in particular the Completeness Conjecture \cite{Polchinski:2003bq,Banks:2010zn} or the sublattice Weak Gravity Conjecture \cite{Heidenreich:2016aqi}.\footnote{At a heuristic level we could say that the gaps occur only for specific choices of flux while they disappear for generic fluxes, 
so computing the elliptic genus for a generic background samples 
the maximal set of {\it possible} string excitations at given charge and mass level.}
%
All we can safely say is that these states are not visible
in the parity odd, RR sector of the theory, though they may well appear in the other sectors. Note that those other sectors are not protected 
by holomorphic factorization and modularity,
and so may in general depend on moduli already at tree level.  In particular the masses of some of these states
may become arbitrarily large in certain regions of the moduli space, so that they effectively decouple from the theory.
Thus, for quantities that are not encoded in the elliptic genus, it is unclear to us how to draw 
conclusions about the Weak Gravity Conjecture
for $N=1$ supersymmetric theories; it would be interesting to explore the physical implications further, in particular also in light of refs.~\cite{Andriolo:2018lvp,Gonzalo:2018guu}.

A significant distinction between theories with (quasi-)modular fluxes and generic non-modular fluxes concerns the St\"uckelberg mass that the
$U(1)$ gauge field necessarily acquires whenever the elliptic genus does not vanish. In the former case, the mass is parametrically suppressed
in the weak coupling regime, and we can consider the gauge field as being massless as far as the WGC is concerned. On the other hand,
for non-(quasi-)modular fluxes, the St\"uckelberg mass is generically not suppressed with respect to the compactification scale, even in the weak coupling limit.
A deeper understanding of this phenomenon, possibly in conjunction with the presence or absence of the gaps in the spectrum observed for (quasi-)modular fluxes, is clearly desirable.

As a cautionary remark, 
we are certainly aware that since we have only four supercharges at our disposal, our analysis is necessarily less robust as compared to the previously studied models with more supersymmetries. A priori, many of our statements are strictly valid only in the limit of weak coupling. On the other
hand, we expect the elliptic genus, as geometrically determined via mirror symmetry from F-theory,
 to be robust beyond perturbation theory. This should hold at least as far as charged state counting is concerned, even though these states are not BPS in four dimensions. 
 An indication for this to be true is that the quasi-modular elliptic genus we found confirms the expected predictions,  despite the fact that it does
 not correspond to a fully perturbative heterotic string.
This is not to say that numerical charge-to-mass ratios will not be renormalized as soon as we leave the 
regime of weak coupling.
Subleading corrections both to the stringy analysis and to the black hole solution are expected to modify our classical results for the super-extremality condition and its realisation on a tower of states,
but an order-one deviation seems implausible. 

Another new feature is that the three-fold base $B_3$ of  the  elliptic four-fold $Y_4$ enjoys a considerably richer K\"ahler geometry, as compared to
 the base $B_2$ of an elliptic three-fold $Y_3$ that is relevant for six-dimensional compactifications.
For the latter, the behaviour near the weak coupling region of the F-theory compactification can be determined in full generality \cite{Lee:2018urn,Lee:2018spm}. 
In particular, the weak coupling limit is governed by a single K\"ahler modulus becoming infinite.

By contrast, for the elliptic four-folds under consideration with three-dimensional  bases $B_3$, 
we have identified two different classes of weak coupling limits. While in the first class the appearance of a heterotic string of the type discussed above is ensured, the second type of limit 
may only give rise to an asymptotically tensionless non-critical string from a D3-brane wrapping a vanishing rational curve of positive normal bundle.
It would be extremely interesting to understand the charge spectrum of this type of strings,
 at a level comparable to the analysis performed in this paper for the heterotic string.

\subsection*{Acknowledgements}

We thank Eran Palti and Fabian R\"uhle for helpful discussions.

\appendix

\section{Jacobi forms}\label{jacdefs}

For reference we briefly list a few properties and definitions of certain weak Jacobi forms. More details can be found for instance in
\cite{EichlerZagier,Dabholkar:2012nd}. Recall that by definition
a Jacobi form of weight $w$ and index $m$ is a holomorphic function from $\mathbb H \times  \mathbb C $   to $\mathbb C$,
which  transforms under the modular group as follows:
\bea\label{jacobitrApp}
\varphi_{w,m}  \left(\frac{a \tau + b}{c \tau +d}, \frac{z}{c \tau +d} \right) &=& (c \tau+d)^w e^{2\pi  i  \frac{m c}{c\tau +d}  z^2}    \varphi_{w,m}(\tau,z)\ ,
\\
\varphi_{w,m}\left( \tau , z + \lambda \tau + \mu \right) &=& e^{-2 \pi i   m (z^2 \tau  + 2  \lambda  z )  }   \varphi_{w, m} (\tau,  z)\,,
\quad \lambda, \mu \in \mathbb Z \,.\nn
\eea
For integer $m$, the ring of such Jacobi forms is generated by the elements of $\cR$ shown in eq.~(\ref{Jacringgens}).
Explicitly we have in terms of the standard theta and eta-functions:
\bea 
\varphi_{0,1}(\tau, z)\! &=&\! 4 \!\left(  \frac{\vartheta_2(\tau,z)^2}{\vartheta_2(\tau,0)^2}  + \frac{\vartheta_3(\tau,z)^2}{\vartheta_3(\tau,0)^2}  + \frac{\vartheta_4(\tau,z)^2}{\vartheta_4(\tau,0)^2}  \right) ,
\nn\\
 \varphi_{-1,2}(\tau, z) \! &=&\! \frac{\vartheta_1(\tau,2z)}{\eta^3(\tau)},
 \\
 \varphi_{-2,1}(\tau, z) \! &=&\! \frac{\vartheta_1(\tau,z)^2}{\eta^6(\tau)}.
\nn
\eea
These generators are not free but satisfy the following relation:
\be
432\varphi_{-1,2}^2 = {\varphi_{-2,1}}\left( \varphi_{0,1}^3-3 E_4  \varphi_{-2,1}^2\varphi_{0,1}+2 E_6  \varphi_{-2,1}^3\right).
\ee

\section{Details of the four-fold $Y_4$ discussed in Section~\ref{sec_Example1}}\label{ex_main}

We start by introducing the geometry of the three-fold base, $B_3$, of the elliptic four-fold, $Y_4$, in question.
Let us first consider a $\IP^1$-fibered three-fold $\mathbb H_t$ over $\IP^2$,
\beq\label{def_Ht}
\mathbb H_t = \IP(\cO_{\IP^2} \oplus \cO_{\IP^2}(t)) \,,\quad t \in \IZ \,,
\eeq
where the parameter $t$ characterises how the $\IP^1$ fibration is twisted. The two-dimensional cohomology, $H^{1,1}(\mathbb H_t)$, is spanned by the pull-back $H$ of the hyperplane class in $\IP^2$ and the section $S_\infty$ of the rational fibration `at infinity,' with the intersection property
\beq\label{Sinftyprop}
S_\infty \cdot S_\infty = -\,t\, S_\infty \cdot H \,.
\eeq
Note that $\mathbb H_t$ can be thought of as a three-fold generalisation of the Hirzebruch surface $\mathbb F_a$, where $H$ is the analogue of the fibral class, and $S_\infty$ that of the section class. Altogether~\eqref{Sinftyprop} leads to the intersection polynomial 
\beq
I({\mathbb H_t}) = t^2\, S^3_{\infty} - t\, S^2_{\infty}H + S_\infty H^2 \,,
\eeq
and its first Chern class is given by
\beq
c_1(\mathbb H_t) = 2 S_\infty + (3+t) H\,.
\eeq

In order to fully explore the different types of fluxes discussed in Section~\ref{sec:fluxes},  we specialise to a base $B_3$ which is the blowup of $\mathbb H_{t=1}$ along a hyperplane in the $\IP^2$, i.e., along $S_\infty \cdot H$. To be specific, the toric coordinates of $B_3$ can be described as in Table~\ref{tb:GLSM_B3}, in terms of the charges of a gauged linear sigma model (GLSM). The first two and the next three columns in the table represent the fiber and the base coordinates of the $\mathbb H_1$, respectively, while the last column is introduced as the blow-up divisor. 

\begin{table}
\begin{center}
\begin{tabular}{|c ||cccccc|}
\hline 
& $\nu_{x_0}$ & $\nu_{x_1}$ & $\nu_{y_0}$ & $\nu_{y_1}$ & $\nu_{y_2}$ & $\nu_e$ \\ \hline\hline
$U(1)_{S_{\infty}}$ &1&1&$0$&0&0&0 \\ \hline
$U(1)_{H}$ &0&$t=1$&1&1&1& 0 \\ \hline
$U(1)_{E}$ &$-1$&$0$&$-1$&0&0& 1 \\ \hline
\end{tabular}\end{center}
\caption{GLSM charges of the toric coordinates of the internal manifold $B_3$, obtained by blowing up the $\IP^1$-fibered three-fold $\mathbb H_{t=1}$ along a hyperplane curve in its $\IP^2$ base.} 
\label{tb:GLSM_B3}
\end{table}

To characterise the emergent weakly-coupled heterotic string, it is necessary to analyse the topological and geometrical properties of $B_3$ itself. 
Given the toric description in Table~\ref{tb:GLSM_B3}, this proceeds via appropriate combinatorial computations, e.g., 
by making use of PALP~\cite{Braun:2012vh} and SAGE~\cite{sagemath}. The result of such an analysis is as follows. 
Firstly, the Mori cone $\bold{M}(B_3)$ is spanned by three generators ${\ell}^{(i)}$, which we describe in terms of their intersection numbers with the $6$ base toric divisors ${d}_\rho:=\{\nu_\rho=0\}$, for $\rho = x_0$, $x_1$, $y_0$, $y_1$, $y_2$, and $e$, as follows:
\beq
\begin{array}{lclrrrrrrr}
\ell^{(1)}&=&(&-2,& 0,& 0,& 1,& 1,& 1)\,, \\ 
\ell^{(2)}&=&(&1,& 0,& 1,& 0,& 0,& -1)\,, \\ 
\ell^{(3)}&=&(&0,& 1,& -1,& 0,& 0,& 1)\,.
\end{array}
\eeq
Then,  in turn, (the closure of) the K\"ahler cone ${\bold{K}(B_3)}$ is spanned by the three divisor classes, 
\bea\label{kcgen_B3}\nn 
j_1 &=& d_{y_1} \,,\\ \label{j123}
j_2 &=& d_{x_1} +d_{y_1} -d_{e}  \,,\\ \nn
j_3 &=& d_{x_1} \,, 
\eea
where $\int_{\ell^{(a)}} j_b = \delta^a_b$ for $a,b=1, 2, 3$. Note that $d_{y_1}$ is the pullback of a curve class from $B_2$ to $B_3$.
Other relevant topological properties are the triple intersection numbers, which we list in terms of an intersection polynomial as follows:
\beq\label{inter_B3}
I(B_3) = j_1^2 j_2 + j_1^2 j_3 +  2 j_1 j_2^2 + 2 j_1 j_2 j_3 + j_1 j_3^2 + 4 j_2^3 + 4 j_2^2 j_3 + 2 j_2 j_3^2 + j_3^3 \,.
\eeq

As discussed in Section~\ref{subsec_geometry1}, a weakly-coupled, nearly tensionless heterotic string is obtained by 
wrapping a $D3$-brane on a curve $C_0\subset B_3$ that shrinks asymptotically to zero volume. 
Recall that one of the various important properties~\eqref{C0definitionWGC} of $C_0$ is that it takes the form\footnote{Furthermore, one can also show that \eqref{shrinking} is the only curve that can lead to a weakly-coupled tensionless heterotic string. This can be seen, for example, by writing a most general curve class as a non-negative linear combination $C=\sum_{i=1}^3 c_i \ell^{(i)}$ of the Mori cone generators and demanding that $\bar K \cdot C = c_1+c_2+c_3=2$. For $C$ to shrink in the global limit with respect to~\eqref{J0}, one immediately learns that $c_1=0$. Then, the only possible choices for the other coefficients are $(c_2,c_3)=(2,0)$, $(1,1)$, and $(0,2)$. However, the first and the third are shrinkable at a finite distance in the moduli space with $J=j_1 + \epsilon j_2 +  j_3$ and $j_1+j_2 + \epsilon j_3$ (as $\epsilon \to 0^+$), respectively, both of which leave ${\rm vol}(B_3)$ finite, and hence cannot lead to a heterotic string. The second choice on the other hand can easily be seen to correspond to~\eqref{J0}.}
\beq\label{shrinking}
C_0 = J_0 \cdot J_0 \,,
\eeq
where $J_0$ is one of the K\"ahler cone generators with $J_0 \cdot J_0 \neq 0$ and $J_0^3=0$. From the intersection polynomial~\eqref{inter_B3}, we immediately identify it as
\beq\label{J0}
J_0 = j_1 (=d_{y_1}) \,,
\eeq
which leads to global limits of Type A defined in Section~\ref{subsec_geometry1}. Note that this is the only choice for $J_0$ and in particular that global limits of Type B are not available for the present base three-fold $B_3$, given that the other two K\"ahler cone generators have non-trivial triple self-intersections. 

Having defined the base $B_3$ of the elliptic fibration, we now turn to the geometry of $Y_4$ itself.
To realise a $U(1)$ gauge symmetry, we will take the elliptic fiber over $B_3$ to be a general hypersurface of degree $4$ in ${\rm Bl_1} \IP^2_{112}$. Fibrations of this type have a rank--$1$ Mordell-Weil group of rational sections~\cite{Morrison:2012ei} and are obtained as a hypersurface of the form
\beq\label{PMP}
P_{\rm MP} \equiv sw^2 + b_0s^2u^2w + b_1suvw + b_2v^2w + c_0s^3u^4 + c_1s^2u^3v + c_2su^2v^2 + c_3uv^3 = 0 \,.
\eeq
Here, $u$, $v$, $w$ and $s$ are the four homogeneous coordinates of the fibral ambient space ${\rm Bl}_1\IP^2_{112}$, while the coefficients $b_i$ and $c_i$ are sections of line bundles on the base $B_3$, whose classes are described in terms of the anti-canonical class $\bar K$
and a class $\beta\in H^2(B_3, \IZ)$ as follows:
\begin{center}
  \begin{tabular}{| l || c | c| c | c | c | c | c | c | }
      \hline
 Coefficients in $P_{\rm MP}$ & $b_0$ & $b_1$ & $b_2$ & $c_0$ & $c_1$ & $c_2$ & $c_3$ & $c_4$ \\
    \hline 
                & & & & & & & & 
    \\[-1.1em] 
 Classes in $H^{2}(B_3, \IZ)$ & $\beta$ & $\bar K$ & $2\bar K - \beta$ & $2\beta$ & $\bar K + \beta$ & $2\bar K$ & $3\bar K - \beta$ & $4 \bar K - 2 \beta$ \\
    \hline
  \end{tabular}
\end{center}
Given the three-fold base $B_3$ discussed above, one can therefore specify a model via a choice of $\beta$, which necessarily has to lie in the range
\beq
0 \leq \beta \leq 2 \bar K \,.
\eeq
The inequalities mean that the difference of two classes is an effective class. Denoting by $\cL_u$ and $\cL_s$ the line bundles for which
\beq
u \in H^0(Y_4, \mathcal L_u)\,,\quad s\in H^0(Y_4, \mathcal L_s) \,,
\eeq
we see that due to~\eqref{PMP} the other two fibral ambient coordinates must be a section of the following line bundles
\beq
v\in H^0(Y_4, \mathcal L_u \otimes \mathcal L_s \otimes \cO(\beta-\bar K))\,,\quad w\in H^0(Y_4, \mathcal L_u^2 \otimes \mathcal L_s \otimes \cO(\beta))\,. 
\eeq
The toric coordinates of $Y_4$ can thus be described as in Table~\ref{tb:GLSM_Y4}, where $\beta$ has been parameterized as 
\beq
\beta = {\rm x} D_1 + {\rm y} D_2 + {\rm z} D_3 \,.
\eeq
This refers to a basis of $H^{1,1}(B_3, \IZ)$ defined as follows:
\beq\label{basischoice}
D_1=d_{x_0} \,,\quad D_2=d_{y_0}\,,\quad D_3=d_{e}\,.
\eeq
The model analyzed in  Section \ref{sec_Example1} is then specified by the choice
\beq\label{choice}
{\rm x} = 2 \,,\quad {\rm y}=2 \,, \quad {\rm z}=4\,. 
\eeq

\begin{table}
\begin{center}
\begin{tabular}{|c ||cccccc|cccc|}
\hline 
& $\nu_{x_0}$ & $\nu_{x_1}$ & $\nu_{y_0}$ & $\nu_{y_1}$ & $\nu_{y_2}$ & $\nu_e$ & $\nu_u$ & $\nu_v$ & $\nu_w$ & $\nu_s$ \\ \hline\hline
$U(1)_{S_{\infty}-E}$ &1&1&$0$&0&0&0& 0 & ${\rm x} - 2$ & $\rm x$ & 0  \\ \hline
$U(1)_{H-E}$ &0&$1$&1&1&1& 0 & 0 & ${\rm y}-4$ & $\rm y$ & 0 \\ \hline
$U(1)_{E}$ &$0$&$2$&$0$&1&1& 1 & 0 & ${\rm z}-5$ & $\rm z$ & 0 \\ \hline
$U(1)_{U}$ &0&0&0&0&0&0 & 1 & 2 & 1 &0 \\ \hline
$U(1)_{S}$ &0&0&0&0&0&0& 0 & 1 & 1 & 1 \\ \hline
\end{tabular}\end{center}
\caption{GLSM charges of the toric coordinates of the four-fold $Y_4$ that is an elliptic fibration over the base three-fold $B_3$, which by itself is described in Table~\ref{tb:GLSM_B3}.}\label{tb:GLSM_Y4}
\end{table}

Given the toric data in Table~\ref{tb:GLSM_Y4}, we can perform similar computations as we did for the three-fold base $B_3$ before. This yields
 $18$ triangulations of the lattice polytope and one of them turns out to be compatible with a flat elliptic fibration.\footnote{A criterion for a flat elliptic fibration is that a $2\times 6$ sub-block of zeros exists in the $5\times 10$ Mori cone matrix. This singles out the phase whose properties are listed below.}
The Mori cone $\bold M(Y_4)$ associated with this phase is generated by the five curves
\be\label{mcm}
\begin{array}{lcrrrrrrrrrrr}
l^{(1)}&=&(& -2, & 0, & 0, & 1, & 1, & 1, & 0, & -1, & 0, & 0) \,,\\
l^{(2)}&=&(& 1, & 0, & 1, & 0, & 0, & -1, & 0, & -1, & 0, & 0) \,,\\
l^{(3)}&=&(& 0, & 0, & 0, & 0, & 0, & 0, & 1, & 0, & 1, & -1) \,,\\
l^{(4)}&=&(& 0, & 1, & -1, & 0, & 0, & 1, & -1, & 0, & 0, & 0)\,, \\
l^{(5)}&=&(& 0, & 0, & 0, & 0, & 0, & 0, & -1, & 1, & 0, & 2)\,. \\
\end{array}
\ee
Here, each of the generators is described in terms of its intersection numbers with the $10$ toric divisors  ${d}_\rho:=\{\nu_\rho=0\}$, for $\rho = x_0, \cdots, s$, 
where we keep the same ordering as in Table~\ref{tb:GLSM_Y4}. The (closure of the) K\"ahler cone $\bold K(Y_4)$ is then spanned by the divisor classes,
\bea
J_1 &=& d_{y_1}\,, \nn\\
J_2 &=& d_{x_1} +d_{y_1} -d_{e} \,, \nn\\
J_3 &=& d_w\,, \\
J_4 &=& d_{x_1} \,,\nn \\
J_5 &=& d_w-d_{x_1}-d_u\nn \,,  
\eea
where $\int_{l^{(a)}} J_b = \delta^a_b$ for $a,b=1, \dots, 5$. Note that three of the generators $J_1$, $J_2$, and $J_4$ of $\bold{K}(Y_4)$ are respectively pull-backs of the generators $j_1$, $j_2$, and $j_3$ in~\eqref{j123} of $\bold K(B_3)$, where, by slight abuse of notation, the base toric divisors and their pull-backs are denoted by the same symbols. 
Finally, the intersection polynomial is given as
\bea
I(Y_4) &=& 10 J_1{}^2 J_3{}^2+46 J_2{}^2 J_3{}^2+262 J_3{}^4+16 J_3{}^2 J_4{}^2+4 J_1{}^2 J_5{}^2+20 J_2{}^2 J_5{}^2+134 J_3{}^2 J_5{}^2
\nn\\&&
 +\,8 J_4{}^2 J_5{}^2+52 J_5{}^4+57 J_3{}^3
   J_1+16 J_5{}^3 J_1+117 J_3{}^3 J_2+34 J_5{}^3 J_2+23 J_3{}^2 J_1 J_2
  \\&&
   +\,10 J_5{}^2 J_1 J_2+12 J_2{}^3 J_3+3 J_4{}^3 J_3+91 J_5{}^3 J_3+6 J_2{}^2 J_1 J_3+3 J_4{}^2 J_1
   J_3+28 J_5{}^2 J_1 J_3  
 \nn  \\&&
   +\,3 J_1{}^2 J_2 J_3+6 J_4{}^2 J_2 J_3+59 J_5{}^2 J_2 J_3+72 J_3{}^3 J_4+26 J_5{}^3 J_4+16 J_3{}^2 J_1 J_4+8 J_5{}^2 J_1 J_4
 \nn       \\&&
      +\,32 J_3{}^2 J_2 J_4+16
   J_5{}^2 J_2 J_4+3 J_1{}^2 J_3 J_4+12 J_2{}^2 J_3 J_4+43 J_5{}^2 J_3 J_4+6 J_1 J_2 J_3 J_4+8 J_2{}^3 J_5
\nn      \\&&
      +\,190 J_3{}^3 J_5+2 J_4{}^3 J_5+4 J_2{}^2 J_1 J_5+41 J_3{}^2 J_1
   J_5+2 J_4{}^2 J_1 J_5+2 J_1{}^2 J_2 J_5+85 J_3{}^2 J_2 J_5
 \nn     \\&&
      +\,4 J_4{}^2 J_2 J_5+7 J_1{}^2 J_3 J_5+34 J_2{}^2 J_3 J_5+13 J_4{}^2 J_3 J_5+17 J_1 J_2 J_3 J_5+2 J_1{}^2 J_4
   J_5+8 J_2{}^2 J_4 J_5
\nn      \\&&
      +\,56 J_3{}^2 J_4 J_5+4 J_1 J_2 J_4 J_5+13 J_1 J_3 J_4 J_5+26 J_2 J_3 J_4 J_5 \,.
      \nn
\eea
From this the intersection ideal, or principal part of the Picard-Fuchs operators, can be read off as ($\theta_a\equiv z_a\partial_a$)
\allowdisplaybreaks
\begin{align}
\nn & 
3 \theta_4^2 - \theta_1 \theta_5 - \theta_2 \theta_5 - \theta_4 \theta_5 + \theta_5^2=0\,,\\\nn &  
3 \theta_2 \theta_4 - 2 \theta_1 \theta_5 - 2 \theta_2 \theta_5 - 2 \theta_4 \theta_5 + 2 \theta_5^2=0\,,\\\nn &  
3 \theta_1 \theta_4 - \theta_1 \theta_5 - \theta_2 \theta_5 - \theta_4 \theta_5 + \theta_5^2=0\,,\\\nn &  
\theta_3^2 - \theta_3 \theta_4 - \theta_3 \theta_5=0\,,\\\nn &  
\theta_1 \theta_3 + \theta_2 \theta_3 - 2 \theta_1 \theta_5 - 2 \theta_2 \theta_5 - \theta_3 \theta_5 + 
 2 \theta_5^2=0\,,\\\nn &  2 \theta_1 \theta_2 - \theta_2^2=0\,,\\\nn &  
48 \theta_3 \theta_4 \theta_5 - 8 \theta_1 \theta_5^2 - 8 \theta_2 \theta_5^2 - 30 \theta_3 \theta_5^2 - 
 65 \theta_4 \theta_5^2 + 53 \theta_5^3=0\,,\\\nn &  
96 \theta_2 \theta_3 \theta_5 - 32 \theta_1 \theta_5^2 - 176 \theta_2 \theta_5^2 - 
 66 \theta_3 \theta_5^2 + \theta_4 \theta_5^2 + 131 \theta_5^3=0\,,\\ &  
3 \theta_2^2 \theta_5 - 4 \theta_1 \theta_5^2 - 6 \theta_2 \theta_5^2 + 4 \theta_5^3=0\,,\\\nn &  
3 \theta_1^2 \theta_5 - 4 \theta_1 \theta_5^2 + \theta_5^3=0\,,\\\nn &  
144 \theta_2^2 \theta_3 - 416 \theta_1 \theta_5^2 - 560 \theta_2 \theta_5^2 - 66 \theta_3 \theta_5^2 +
  \theta_4 \theta_5^2 + 515 \theta_5^3=0\,,\\\nn &  
9 \theta_2^3 - 16 \theta_1 \theta_5^2 - 16 \theta_2 \theta_5^2 - 4 \theta_4 \theta_5^2 + 
 16 \theta_5^3=0\,,\\\nn &  \theta_1^3=0\,,\\\nn &  2 \theta_4 \theta_5^3 - \theta_5^4=0\,,\\\nn &  4 \theta_3 \theta_5^3 - 7 \theta_5^4=0\,,\\\nn &  
26 \theta_2 \theta_5^3 - 17 \theta_5^4=0\,,\\\nn &  13 \theta_1 \theta_5^3 - 4 \theta_5^4=0\,,\\\nn &  \theta_5^5=0\,.
\end{align}

\section{Weak coupling limits of Type A for another example base $B_3$}\label{ex_gen}

In this appendix we will analyze global limits of vanishing gauge coupling of Type A, as defined in Section~\ref{subsec_geometry1}, for F-theory on a specific base three-fold,  $B_3$. This will serve as an illustrative example for the base three-folds whose K\"ahler cone has a more general structure than that of the main example in Section~\ref{sec_Example1}. More specifically, for the global limits of Type A explored here, both the $\cI_1$-type and the $\cI_3$-type $(1,1)$-forms, as defined in~\eqref{Jnugenerators} and~\eqref{Jr1}, respectively, will be present amongst the cone generators. In particular, we will verify that the asymptotic expression~\eqref{volSvolC0} indeed applies to this example as well.

Like we did in Appendix~\ref{ex_main}, let us first consider a $\IP^1$-fibered three-fold $\mathbb H_{t=1}$ over $\IP^2$, with twist $t=1$ in~\eqref{def_Ht}.
Its $H^{1,1}$ cohomology is spanned by $H$ and $S_{\infty}$ which have the intersection property~\eqref{Sinftyprop}. The base space, $B_3$, of our interest is then constructed by blowing $\mathbb H_{t=1}$ up along the $\IP^1$ fiber, i.e., along $H \cdot H$. Specifically, the toric coordinates of $B_3$ can be described as in Table~\ref{tb:GLSM_B3_gen}; the first two and the next three columns in the table represent the fiber and the base coordinates of $\mathbb H_1$, respectively, while the last column is introduced as the blow-up divisor. 

\begin{table}
\begin{center}
\begin{tabular}{|c ||cccccc|}
\hline 
& $\nu_{x_0}$ & $\nu_{x_1}$ & $\nu_{y_0}$ & $\nu_{y_1}$ & $\nu_{y_2}$ & $\nu_e$ \\ \hline\hline
$U(1)_{S_{\infty}}$ &1&1&$0$&0&0&0 \\ \hline
$U(1)_{H}$ &0&$1$&1&1&1& 0 \\ \hline
$U(1)_{E}$ &$0$&$0$&$-1$&$-1$&0& 1 \\ \hline
\end{tabular}\end{center}
\caption{GLSM charges of the toric coordinates of the base $B_3$, obtained by blowing up the $\IP^1$-fibered three-fold $\mathbb H_{1}$ along its fiber $\IP^1$.} 
\label{tb:GLSM_B3_gen}
\end{table}

Given this toric description, the relevant topological and geometrical properties of $B_3$ can be easily determined. The Mori cone has three generators, which we describe in terms of their intersections  with the $6$ base toric divisors ${d}_\rho:=\{\nu_\rho=0\}$, for $\rho = x_0$, $x_1$, $y_0$, $y_1$, $y_2$, and $e$, as follows:
\beq\label{mcgen_B3_gen}
\begin{array}{lclrrrrrrr}
\ell^{(1)}&=&(&-1,& 0,& 0,& 0,& 1,& 1)\,, \\ 
\ell^{(2)}&=&(&0,& 0,& 1,& 1,& 0,& -1)\,, \\ 
\ell^{(3)}&=&(&1,& 1,& 0,& 0,& 0,& 0)\,.
\end{array}
\eeq
The (closure of the) K\"ahler cone has the three generators, 
\bea\label{kcgen_B3_gen}\nn 
j_1 &=& d_{y_2} \,,\\ \label{j123_gen}
j_2 &=& d_{y_0}  \,,\\ \nn
j_3 &=& d_{x_1} \,, 
\eea
which are dual to the Mori cone generators, $\ell^{(i)}$.
Finally, the intersection polynomial is given by
\beq\label{inter_B3_gen}
I(B_3) = j_1^2 j_3 +   j_1 j_2 j_3 + j_1 j_3^2 +  j_2 j_3^2 + j_3^3 \,. 
\eeq
Note that two of the generators, $j_1$ and $j_2$, have a vanishing triple self-intersection, 
\beq
j_1^3 = 0 = j_2^3 \,.
\eeq
On the other hand, $j_2 \cdot j_2$ turns out to be a trivial element of $H_2(B_3, \IZ)$ as $j_2 \cdot j_2 \cdot j_i =0$ for all $i=1,2,3$. Therefore, the global limits of Type A are obtained via the identification 
\beq
J_0 = j_1 \,.
\eeq
Associated to this is the curve $C_0 = J_0 \cdot J_0$ whose shrinking leads to the tensionless heterotic string.

Furthermore, from the intersection polynomial~\eqref{inter_B3_gen} we immediately learn that $j_2$ and $j_3$ are generators of type $\cI_3$  and of type $\cI_1$, respectively. 
Therefore, the global limit~\eqref{Jsplit} of the base geometry should take the form 
\bea\nn 
J &=& t_1 j_1 + t_2 j_2 + t_3 j_3  \\ \label{global_kf_ex}
&=& t j_1 + \frac{a}{t^2} j_3 + c j_2 \,,
\eea
where $t_1 = t$ is the large parameter that determines the global limit, and $t_3 = \frac{a}{t^2}$ is a small parameter for which $a$ is of order one, while $t_2 = c$ is yet to be constrained further. 

From the Mori cone data shown in Table~\eqref{mcgen_B3_gen}, the anti-canonical class can be determined as 
\beq
\bar K= j_1 + j_2 + 2 j_3\,, 
\eeq
and for appropriate choices of $\beta={\rm x} j_1 + {\rm y} j_2 + {\rm z} j_3$ we can specify  a family of elliptic fibrations for F-theory models.
For these, the divisor of the height pairing  takes the form
\beq
b= 6 \bar K - 2 \beta = (6- 2{\rm x}) j_1 + (6-2{\rm y}) j_2 + (12- 2 {\rm z}) j_3 \,.
\eeq
The $U(1)$ fugacity index $m$ is then computed as
\beq
m=\frac12 C_0 \cdot b = 6 - {\rm z} \,,
\eeq
where ${\rm z}$ needs to satisfy ${\rm z} \leq 4$ so that $m \geq 2$.
This follows if we require an F-theory background that is compatible with anomaly cancellation.  

In terms of the K\"ahler form~\eqref{global_kf_ex} we then proceed to compute the volumes of $B_3$, $b$ and $C_0$ in turn. First, the volume of the three-fold $B_3$,
\beq
{\rm vol}(B_3) = \frac{1}{6} J^3 = \frac{a}{2} + \frac{ac}{t} + \frac{a^2 c}{2t^4} + \frac{a^2}{2t^3} + \frac{a^3}{6t^6} \,,
\eeq
must remain finite in the global limit. Given that $a$ is a finite parameter, $c$ must thus be constrained so that $\alpha := \frac{c}{t}$ be finite in the limit. The volume is then further simplified as
\beq
{\rm vol}(B_3) ~\to~ (\frac12+\alpha)a \,,
\eeq
with $0 \leq \alpha < \infty$. 
Next, the volumes of $b$ and $C_0$ are computed, respectively, as 
\bea \nn
{\rm vol}(b) &=& m (1+2\alpha) t^2  + \cdots\,,  \\ \nn
{\rm vol}(C_0) &=& \frac{a}{t^2} \,, 
\eea
where the ellipsis indicates subleading pieces that vanish in the limit. 
Therefore, in the global limit, the product of the two volumes obey
\beq
{\rm vol}(b) \,{\rm vol}(C_0) = ma(1+ 2\alpha)= 2m\, {\rm vol}(B_3) \,,
\eeq
again in precise agreement with~\eqref{volSvolC0}. 

It can also be shown easily that no other curve class on $B_3$ contains a rational curve with trivial normal bundle that shrinks in the limit~\eqref{global_kf_ex}. Therefore, the asymptotically tensionless heterotic string associated with the $C_0$ is unique and leads to a well-defined effective theory in the weakly-coupled heterotic duality frame.

\section{General properties of weak coupling limits}  \label{app_GlobalLimit}

As discussed in Section \ref{subsec_geometry1},  for every weak coupling limit (\ref{limitvolSinfty}) the K\"ahler form of the base space $B_3$ must take the
general form
\beq \label{JAnsatzgen-App}
J = t J_0 + \sum_{\nu \in \cI_1} a'_\nu J_\nu + \sum_{\mu\in \cI_2} b'_\mu J_\mu + \sum_{r \in \cI_3} c_r J_r \,,
\eeq
where $t$ is the large parameter that contributes an infinite amount to the volume of $\bold S$. 

In this appendix, we prove some general properties of the global limit that we have announced in Section~\ref{subsec_geometry1}.
Our proof relies on the technical assumption that the K\"ahler cone of $B_3$ is closed, i.e., that the K\"ahler cone generators are not merely pseudo-effective, but effective, as well as that they are irreducible.

As a preparation let us recall the following simple
\begin{Lemma} \label{LemmaCS}
On a K\"ahler surface ${\Sigma}$, suppose two non-trivial cycle classes $V, V' \in H^{1,1}(\Sigma)$ satisfy
\bea
V \cdot V &=& 0\,,\\
V' \cdot V &=& 0 \,.
\eea 
Then
\be
V' \cdot V' \leq 0 \quad  {\text and} \quad V' \cdot V' = 0 \quad \text{iff} \quad  V = \alpha V' \,.
\ee
\end{Lemma}
{\bf Proof}: Introduce an orthogonal basis $\{w_0, w_i\}$ of $H^{1,1}(\Sigma)$ for which only $\omega_0 \cdot \omega_0 = 1$ and $\omega_i \cdot \omega_j = -\delta_{ij}$ are non-zero.
Such a basis always exists because the intersection form has signature $(1, h^{2}(\Sigma)-1)$.
Expand $V = v_0 \omega_0 + v_i \omega_i$ and $V' = v'_0 \omega_0 + v'_i \omega_i$. Then by the assumed properties of $V$ and  $V'$,
\be
 v_0^2 = v_i v_i \, \qquad  v_0 v_0' = v_i v'_i 
\ee
with $v_0 \neq 0$ because $V$ is non-trivial.
By the Cauchy-Schwarz inequality therefore
\be
V' \cdot V'  = v'_0 v'_0 - v'_i v'_i  = \frac{(v_i v_i')^2 }{v^2_0} - v'_i v'_i  \leq  \frac{(v_i v_i)(v'_i v_i') }{v^2_0} - v'_i v'_i  =0
\ee
and equality holds if and only if $v_i = \alpha v_i'$ for some $\alpha$ and therefore, since  $ v_0^2 = v_i v_i$, also $V = \alpha V'$.

We will furthermore make use of
\begin{Lemma} \label{KahlerLemma}
Any pair of {\it distinct} divisors $D$ and $D'$ that lie in the closure of the K\"ahler cone of a K\"ahler surface $\Sigma$ must have a non-trivial intersection. 
\end{Lemma}
{\bf Proof}: 
In the same orthogonal basis $\{w_0, w_i\}$ of $H^{1,1}(\Sigma)$
 we expand $D = a_0 \omega_0 + a_i \omega_i$  and $D'=a'_0 \omega_0 + a'_i \omega_i$. Since $D$ and $D'$ are in the closure of the K\"ahler cone of $\Sigma$,
\bea
0 &\leq&  D \cdot D = a_0^2 - \sum_i a_i^2  \,, \label{DD'inequ1} \\
0 &\leq&  D' \cdot D' = (a'_0)^2 - \sum_i (a'_i)^2  \,,\label{DD'inequ2} \\
0 &\leq& D \cdot D' = a_0 a'_0 - \sum_i a_i a_i' \label{DD'inequ} \,.
\eea
By the Cauchy-Schwarz inequality we have
\beq \label{CSDD'}
(\sum_i a_i {a}_i')^2 \leq (\sum_i a_i^2) (\sum_i {a}_i'^2) \leq (a_0 a_0')^2  \,,
\eeq
where the last inequality uses (\ref{DD'inequ1}) and (\ref{DD'inequ2}).
The inequality (\ref{DD'inequ})  can hence be saturated only for $D = \beta D'$ (and in addition $D^2 = 0$, $D'^2=0$).
Therefore, two {\it distinct} K\"ahler cone generators $D$, $D'$ must intersect.

 \vskip .5cm
\noindent
We now discuss the two different types of the global limits in turn.

\subsection{Type A: $J_0 \cdot J_0 \neq 0$}\label{pf_case1}

We now discuss in detail the global limit for the situation where $J_0 \cdot J_0 \neq 0$.

\begin{proposition} \label{Propo1}
To take the weak coupling limit (\ref{limitvolSinfty}), the K\"ahler form on $B_3$ must be of the form    (\ref{JAnsatzgen-App}), which for $J_0 \cdot J_0 \neq 0$  reduces further to
\be \label{JTypeIApp}
J = t J_0 + \sum_{\nu \in \cI_1} \frac{a_\nu}{t^2} J_\nu +  \sum_{r \in \cI_3} c_r J_r \,.
\ee
\end{proposition}

{\bf Proof}: 
It has already been argued that the K\"ahler form must have an expansion as shown in  (\ref{JAnsatzgen-App}).
Furthermore we now show that $\cI_2$ is empty if $J_0 \cdot J_0 \neq 0$, i.e. there are no K\"ahler generators with the property (\ref{Jmugenerators}):
Let $S_0 \in B_3$ be the irreducible subvariety of $B_3$ with the class $\left[ S_0 \right] = J_0$.
Its existence is guaranteed by our technical assumptions on the K\"ahler cone.
 Let 
 \be
 C_{i, 0} = J_i|_{S_0} \in H_2(S_0)
 \ee
 be the two-cycle class obtained by restricting the K\"ahler cone generator $J_i$ to $S_0$. Then, $C_{0,0}$ and $C_{\mu,0}$ are non-trivial since $J_0 \cdot J_0 \neq 0$ and $J_0 \cdot J_\mu \neq 0$ for $\mu \in \cI_2$. 
 Furthermore by (\ref{J03=0}) and (\ref{Jmugenerators}) we have\footnote{We  use the same symbol for the intersection form on $S_0$ and $B_3$ and trust that this does not lead to any confusion.}
\bea
C_{0,0} \cdot C_{0,0} &=& J_0 \cdot J_0 \cdot J_0 = 0\,, \\
C_{\mu, 0} \cdot C_{0,0} &=& J_0 \cdot J_0 \cdot J_\mu = 0 \,.
\eea
Lemma \ref{LemmaCS} then implies that
\beq
C_{\mu, 0} \cdot C_{\mu, 0} \leq 0 \,,
\eeq
where the equality can only hold if $C_{\mu,0}$ is proportional to $C_0$. On the other hand, because of (\ref{JiJjJk}
\be
C_{\mu,0} \cdot C_{\mu, 0} = J_0 \cdot J_\mu \cdot J_\mu   \geq 0
\ee
 the equality must hold and consequently 
\beq
C_{\mu, 0} = n_\mu C_{0,0} \,.
\eeq
However, this indicates for all $\mu' \in \cI_2$ that
\beq
J_0 \cdot J_\mu \cdot J_{\mu'} = C_{\mu, 0} \cdot C_{\mu',0} = n_\mu n_{\mu'} C_{0,0} \cdot C_{0,0} = 0 \,,
\eeq
which contradicts the assumption that
 there exists $\mu' \in \cI_2$ with a nontrivial triple intersection $J_0 \cdot J_\mu \cdot J_{\mu'}$, i.e. that there are no generators of type $\cI_2$.
Hence, in order for ${\rm vol}(B_3)$ to remain  finite in the limit,
the ansatz (\ref{Jansatzgen}) for the asymptotic K\"ahler form reduces to  (\ref{JTypeIApp}),
where $a'_\nu = \frac{a_\nu}{t^2}$ for some finite $a_\nu$.

\begin{proposition} \label{Prop2}
In the weak coupling limit (\ref{JTypeIApp}) with $t \to \infty$,
\bea
{\rm vol} ({\bf S}) &=&(\frac12 t^2 J_0 \cdot J_0 + \frac12 c_r c_s J_r\cdot J_s  + t c_r J_0 \cdot J_r + a_\nu \frac{c_r}{t^2} J_\nu \cdot J_r ) \cdot {\bf S}  + \cdots \, \label{volS1}, \\
{\rm vol} (B_3) &=& \frac16 c_r c_s c_t d_{rst} + \frac12 a_\nu d_{00\nu} + a_\nu \frac{c_r}{t} d_{0\nu r} + \frac12 a_\nu \frac{c_r c_s}{t^2} d_{\nu r s}+\frac12 a_\nu a_{\nu'} \frac{c_r}{t} d_{r \nu \nu'} + \cdots \label{volB1} \,
\eea
where
\bea
 J_0 \cdot J_r &=& n_r J_0 \cdot J_0 \quad  \text{for} \quad  n_r >0 \label{JrJ0nonzero} \\
J_{r} \cdot J_s &=& n_{rs} J_0 \cdot J_0  \quad  \text{for}   \quad n_{rs} \geq0   \label{JrJsnonzero} 
\eea
and
\be \label{drstzero}
d_{rst} = 0   \qquad \forall r,s,t  \, \in \cI_3.
\ee
Finiteness of $B_3$ then requires that 
\be \label{anucrfinite}
\forall r \in \cI_3: \, \, \frac{c_r}{t} \quad   \text{finite  as} \quad  t \to \infty \,,
\ee
and moreover demanding ${\rm vol}({\bf S}) \to \infty$ for $t \to \infty$ implies
\be \label{J0J0Snonzero}
J_0 \cdot J_0 \cdot {\bf S} >  0 \,.
\ee
\end{proposition}

{\bf Proof}: The expressions (\ref{volS1}) and  (\ref{volB1}) follow by expanding  ${\rm vol}({\bf S}) = \frac{1}{2} J \cdot J \cdot {\bf S}$ and ${\rm vol}(B_3) = \frac{1}{6} J \cdot J \cdot J$, and keeping only those terms which do not obviously vanish in the limit $t \to \infty$.

\begin{itemize}
\item
To show (\ref{JrJ0nonzero}) observe first that $J_0 \cdot J_r$ is non-trivial for every $r$. This can be seen by considering the very ample divisor $\cS$ with class 
\beq
\left[\cS\right]=m(J_0 + \sum_{\nu} J_\nu + \sum_r J_r) \,,
\eeq
for an appropriate $m \in \IZ$. 
Indeed, the expression in the brackets is a K\"ahler form and hence ample, and therefore for $m$ sufficiently large $\left[\cS\right]$ is very ample.
By Bertini's theorem (see e.g.~\cite{Lazarsfeld}), $\cS$ is irreducible and connected. Furthermore, the restrictions 
\be \label{cC0App}
\cC_0 = J_0 |_{\cS} \,, \qquad \quad  \cC_r = J_r|_{\cS}
\ee
are non-trivial classes on ${\cS}$. This is because $\left[\cS\right]$ is in the interior of the K\"ahler cone of $B_3$ and hence (\ref{cC0App}) represents non-vanishing volumes of the effective divisors $J_0$ and $J_r$ with respect to  $\left[\cS\right]$. 
Furthermore, Lefschetz's hyperplane theorem guarantees that $\cC_0$ and $\cC_r$ are distinct classes on $\left[\cS\right]$. They lie in the closure of the K\"ahler cone of $\cS$ because $J_0$ and $J_r$ are in the closure of the K\"ahler cone of the ambient space $B_3$.
By Lemma \ref{KahlerLemma}, $\cC_0$ and $\cC_r$ must therefore intersect on $\cS$, i.e.
\be
 0 \neq  \cC_0 \cdot \cC_r = \left[\cS\right] \cdot J_0 \cdot J_r \,.
\ee
It follows that $J_0 \cdot J_r$ must be non-trivial on $B_3$.

To arrive at (\ref{JrJ0nonzero}), we consider again the surface $S_0 = [J_0]$ and the cycles 
\bea
C_{0,0} &=& J_0 |_{S_0} \\
C_{r,0} &=& J_r |_{S_0} \,.
\eea
Since $J_0 \dot J_0 \neq 0$ and $J_0 \cdot J_r \neq 0$, both are non-trivial cycles. 
A similar application of 
Lemma \ref{LemmaCS} as in the proof of Proposition \ref{Propo1} yields that $C_{r,0} = n_r C_{0,0}$ with $n_r \neq 0$ and therefore $n_r > 0$.

\item

Likewise we can show that $J_{rs} = n_{rs} J_0 \cdot J_0$ for $n_{rs} \geq0$. To see this, we first observe that 
\be
J_r \cdot J_s = n_{rs}' J_0 \cdot J_s
\ee 
by applying Lemma \ref{LemmaCS}  to the two-cycles $C_{0,s}$ and $C_{r,s}$ in $S_s = [J_s]$, where $C_{0,s}$ is non-trivial again thanks to (\ref{JrJ0nonzero}). 
Therefore, 
\beq
J_r \cdot J_s = n'_{rs} J_0 \cdot J_s = n'_{rs} n_s J_0 \cdot J_0 =: n_{rs} J_0 \cdot J_0 \,.
\eeq

This implies that
\be
d_{r s t} = J_r \cdot J_s \cdot J_t = n_{rs} J_0 \cdot J_0 \cdot J_t = 0
\ee
by the defining  property (\ref{Jr1}) of the generators $J_r$.

\item

Finally we  prove (\ref{anucrfinite}). Consider the third term in the expression (\ref{volB1}) for ${\rm vol}(B_3)$. In order that ${\rm vol}(B_3)$ remain finite in the limit, this expression must be finite.
Now, we have for the intersection number
\be
d_{0\nu r} = J_0 \cdot J_r \cdot J_\nu = n_r \, J_0 \cdot J_0 \cdot J_\nu = n_r \, d_{00\nu} \geq 1,
\ee
because, by definition of $J_\nu$, we know that $J_0 \cdot J_0 \cdot J_\nu \neq 0$ for all $\nu$.
Hence in the above expression for ${\rm vol}(B_3)$, there must exist a contribution not less than $\frac{a_\nu c_r}{t}$, which must be finite. Since $a_\nu$ is finite by itself and since for at least one $\nu$, the parameter $a_\nu$ is of order 1 so that ${\rm vol}(B_3)$  is not zero, we conclude that for all $r$ also $c_r/t$ must be finite.
\end{itemize}

These results indeed imply (\ref{J0J0Snonzero}):
In the expression (\ref{volS1}) the last term vanishes in the limit $t \to \infty$ because $\frac{c_r}{t}$ is finite.
The other remaining terms are all proportional to $J_0 \cdot J_0 \cdot{\bf S}$. In order to obtain ${\rm vol}(\bf S) \to \infty$ as $t \to \infty$, we therefore
have $J_0 \cdot J_0 \cdot{\bf S} \neq 0$.

\begin{proposition} \label{PropC0exists}
The cycle  $C_0:=J_0 \cdot J_0$ is a rational curve with a trivial normal bundle. 
\end{proposition}

{\bf Proof}: 
If $C_0$ is an effective curve, it is clear that the normal bundle is trivial by adjunction because $J_0 \cdot J_0 \cdot J_0 = 0$.
To establish that it is a rational curve, we
make use of the fact that $J_0 \cdot J_0 \cdot{\bf S} > 0$, as shown above,
as well as 
Mori's Cone Theorem, which holds for any smooth projective variety: There are countably many rational curves $C_i$ on $B_3$, satisfying $0<\bar K \cdot C_i \leq {\rm dim}(B_3) +1 = 4$, such that the closure of the Mori cone has the structure
\beq\label{dec}
\overline{NE(B_3)} = \overline{NE(B_3)}_{\bar K \leq 0} + \sum \IR_{\geq0} C_i \,.
\eeq
Note first that $J_0^2 \in \overline{NE(B_3)}$. This is the case because $J_0$ is in the closure of the cone of effective divisors, which implies that $J_0^2$ must lie in the closure of the Mori cone. 

If $C_0$ belongs to the second part of the decomposition~\eqref{dec} then it is effective. In this case we may apply the Adjunction formula to compute its Euler number,
\beq
2-2g = \int_{B_3} (\bar K - 2 J_0) J_0^2 = \bar K \cdot C_0  > 0 \,, 
\eeq
where the last inequality follows 
from the assumption that $C_0$ lies in the second part of (\ref{dec}). 
We thus have $g=0$ with $\bar K\cdot C_0 = 2$. 

Suppose now that $C_0$ belongs to the first part of the decomposition~\eqref{dec}. If the gauge symmetry supported on ${\bf S}$ is non-abelian, we have
\beq
12 \bar K = {\bf S} + {\bf S}' \,,
\eeq
where ${\bf S}'$ is the remainder of the residual piece of the discriminant, which also has to be effective. Upon intersecting with $C_0$, we obtain
\beq
12 \bar K \cdot C_0 = {\bf S} \cdot J_0^2 + {\bf S}' \cdot J_0^2 \,, 
\eeq
where both terms ${\bf S} \cdot J_0^2$ and ${\bf S}' \cdot J_0^2$ are non-negative. Since we have shown above that ${\bf S} \cdot J_0^2 >0$, we conclude 
\beq
\bar K \cdot C_0 > 0\,,
\eeq
contradicting that $C_0$ belongs to the first part of~\eqref{dec}.  
When ${\bf S}$ supports a $U(1)$ gauge group in the sense that it is the height-pairing divisor of a section, it has already been conjectured in \cite{Lee:2018urn} that there exists an integer $n >0$ such that 
${\bf S} < n \, \bar K$. Morally speaking, the height-pairing is a linear combination of 7-brane divisors (in Type IIB language), whose total class is constrained by the 7-brane tadpole condition, and the O7-plane class is
proportional to $\bar K$. The relation ${\bf S} < n \, \bar K$ is indeed satisfied for all classes of $U(1)$ models in the literature that  we know of.

\begin{proposition}
In the weak coupling limit, we have
\beq
{\rm vol}({\bf S}) \, {\rm vol}(C_0)   
= 2m \, {\rm vol}(B_3) \qquad \quad {\rm as} \quad t \to \infty\,,
\eeq
where 
\be
m = \frac{1}{2} C_0 \cdot {\bf S} > 0 \,.
\ee
\end{proposition}
{\bf Proof}: Let us expand the surface ${\bf S}$ as
\be
{\bf S} := h_0 J_0 + h_\nu J_\nu + h_r J_r \,,
\ee
where the parameters are not necessarily positive.

We can now express the volumes of $C_0$, ${\bf S}$ and $B_3$, up to terms which vanish as $t \to \infty$,
as 
\bea
{\rm vol} (C_0) &=& \frac{a_\nu}{t^2} d_{00\nu} \,,\\ 
{\rm vol} ({\bf S}) &=&(\frac12 t^2 J_0 \cdot J_0 + \frac12 c_r c_s J_r\cdot J_s  + t c_r J_0 \cdot J_r + a_\nu \frac{c_r}{t^2} J_\nu \cdot J_r ) \cdot {\bf S}  + \cdots  \\
&=& (\frac12 t^2 h_\nu d_{00\nu} + \frac12 h_\nu c_r c_s d_{rs\nu} + t h_\nu c_r d_{0r\nu} ) + \cdots \\ 
&=& h_\nu d_{00\nu} (\frac12 t^2 + \frac12 c_r c_s n_{rs} + t n_r c_r) + \cdots \,,\\
{\rm vol} (B_3) &=& \frac16 c_r c_s c_t d_{rst} + \frac12 a_\nu d_{00\nu} + a_\nu \frac{c_r}{t} d_{0\nu r} + \frac12 a_\nu \frac{c_r c_s}{t^2} d_{\nu r s}+\frac12 a_\nu a_{\nu'} \frac{c_r}{t^4} d_{r \nu \nu'} + \cdots  \\ 
&=& \frac12 a_\nu d_{00\nu} + a_\nu \frac{c_r}{t} d_{0 \nu r} + \frac12 a_\nu \frac{c_r c_s}{t^2} d_{\nu r s} + \cdots \\ 
&=& a_\nu d_{00\nu} (\frac 12 + \frac{n_r c_r}{t} + \frac 12 \frac{n_{rs} c_r c_s }{t^2}) + \cdots \,, 
\eea
Here we have made use of eqs.~(\ref{JrJ0nonzero}),  (\ref{JrJsnonzero}) and (\ref{drstzero}).
Furthermore, 
\beq
m= \frac12 {\bf S} \cdot J_0 \cdot J_0 = \frac12 h_\nu d_{00\nu}  \,. 
\eeq
Altogether, we conclude that for $t \to \infty$:
\beq
{\rm vol}({\bf S}){\rm vol}(C_0)   =  (\sum_{\nu} h_\nu d_{00\nu}) (\sum_\nu a_\nu d_{00\nu}) (\frac12  + \frac12 \frac{c_r c_s n_{rs}}{t^2} + \frac{n_r c_r}{t}) = 2m {\rm vol}(B_3) \,.
\eeq

\subsection{Type B: $J_0 \cdot J_0 = 0$}\label{pf_case2}

If $J_0 \cdot J_0 = 0$, the weak coupling limit takes a rather different form. Specifically we have the following

\begin{proposition}

There are no generators of type $\cI_3$ and in the global limit the K\"ahler form looks
\be \label{TypeIIlimitansatz}
J = t J_0 + b'_\mu J_\mu \,, \qquad \quad \mu \in \cI_2 \,, \
\ee
for suitably constrained parameters $b'_\mu$.
In particular there exists some $\nu_0$ such that
\be \label{b'nu0-App}
b'_{\nu_0} = b_{\nu_0} \, t^{-1 + a_{\nu_0}} \qquad \quad {\rm for} \quad a_{\nu_0} > 0 \,
\ee
with $b_{\nu_0}$ finite as $t \to \infty$ and
\be \label{bmu-App}
\forall \mu \quad {\rm with} \quad d_{0 \mu \nu_0} \neq 0: \, \, b'_{\mu}  = b_{\mu}\,  t^{-a_{\nu_0} - \Delta_{\mu}}   \qquad \quad \Delta_{\mu} \geq 0 
\ee
for $b_{\mu_0}$ finite as $t \to \infty$.
\end{proposition}
{\bf Proof}: To see that no generators of type  $\cI_3$ can exist, we first note that these generators would have to satisfy $J_0 \cdot J_r \neq 0$: This follows as in the proof of Proposition \ref{Prop2} by restricting $J_0$ and $J_r$ to the very ample divisor $[{\cal S}] = m (J_0 + \sum_\mu J_\mu + \sum_r J_r)$, with $\mu \in \cI_2$ and $r \in \cI_3$ (assuming the latter exist).
On the other hand, by definition of $\cI_3$,
\bea
&& J_0 \cdot  J_r \cdot J_0 = 0 \,,\\
&& J_0 \cdot  J_r \cdot J_\mu = 0 \qquad \forall \,  \mu \in  \cI_2 \,,\\
&& J_0 \cdot  J_r \cdot J_s = 0 \qquad \forall s \in  \cI_3 \,.
\eea
Since in the present situation there are no generators of type $\cI_1$, the class   $J_0 \cdot  J_r$ now vanishes: $J_0 \cdot  J_r =0$. This contradiction can only be resolved by concluding that no generators of $ \cI_3$ exist.

We now expand ${\bf S} = h_0 J_0 + h_\mu J_\mu$ for $\mu \in \cI_2$ and observe that
\bea
{\rm vol}({\bf S}) &=& \frac{1}{2} h_\mu b'_\nu b'_\rho d_{\mu \nu \rho} + t h_\mu b'_\nu d_{0 \mu \nu} \,,\\
{\rm vol}(B_3) &=& \frac{1}{2} t b'_\mu b'_\nu d_{0 \mu \nu} + \frac{1}{6} b'_\mu b'_\nu b'_\rho d_{\mu \nu \rho} \,.
\eea
By assumption, in the weak coupling limit where $t \to \infty$,   ${\rm vol}({\bf S})$ receives an infinite contribution from the terms involving a factor of $J_0$ - otherwise we relabel the K\"ahler cone generators accordingly.
This means that there must exist some subset of  $\tilde \nu_0 \in \cI_2$ such that
\be \label{0nu0S}
J_0 \cdot J_{\tilde \nu_0} \cdot {\bf S} > 0 \,,
\ee
and moreover a further subset $\{\nu_0\}\subset \{\tilde\nu_0\}$ exists that leads to an infinite contribution to $ {\rm vol}({\bf S})$:
\be
   t \, b'_{\nu_0}  J_0\cdot J_{\nu_0} \cdot  {\bf S} \to \infty   \qquad \quad {\rm as} \quad t \to \infty \,.
\ee
This guarantees that for this subset, $\nu_0$, the K\"ahler parameters scale as in  (\ref{b'nu0-App}).
At the same time, ${\rm vol}(B_3)$ contains the term
\be
d_{0 \mu \nu_0} b'_{\mu} \, b'_{\nu_0} \, t = d_{0 \mu \nu_0} b'_{\mu} \, b_{\nu_0} \, t^{a_{\nu_0}}\,.
\ee
Thus, in order for ${\rm vol}(B_3)$ to remain finite, we must impose eq.~(\ref{bmu-App}).

\begin{proposition}
There exists a rational curve $C_{\nu_0} = J_0 \cdot J_{\nu_0}$ with normal bundle 
 \be \label{normalCnu0}
  N_{C_{\nu_0}/B_3} = {\cal O}_{C_{\nu_0}} \oplus  {\cal O}_{C_{\nu_0}}(d_{\nu_0}) \,,\qquad \quad d_{\nu_0} := d_{0 \nu_0 \nu_0} \geq 0 \,,
  \ee 
whose volume vanishes in the weak coupling limit.
\end{proposition}
{\bf Proof}: 
If the cycle $C_{\nu_0} = J_0 \cdot J_{\nu_0}$ is a holomorphic curve, its normal bundle must be of the form (\ref{normalCnu0}) because $J_0 \cdot J_0 =0$.
The proof that $C_{\nu_0}$ is a holomorphic rational curve is analogous to the proof of Proposition \ref{PropC0exists}: This time we use Mori's cone theorem together with the fact that $C_{\nu_0} \cdot {\bf S} > 0$ as established in (\ref{0nu0S}).
By the same reasoning as in Proposition \ref{PropC0exists}, we conclude that $C_{\nu_0} \cdot {\bar K}$ cannot vanish and hence by Mori's cone theorem $C_{\nu_0}$ is a rational curve.
Note that the adjunction formula now gives
\be
2 = 2-2g =  (\bar K -  J_0 - J_{\nu_0}) \cdot J_0 \cdot J_{\nu_0} = \bar K \cdot C_0 - d_{\nu_0} \,,
\ee
so that altogether
\be
\bar K \cdot C_0 = 2 + d_{\nu_0} \,, \qquad \quad \quad N_{C_{\nu_0}/B_3} = {\cal O}_{C_{\nu_0}} \oplus  {\cal O}_{C_{\nu_0}}(d_{\nu_0}) \,.
\ee
Finally, the volume of $C_{\nu_0}$ is computed as 
\be
{\rm vol}(C_0) = b'_{\mu} d_{0 \nu_0 \mu} =  b_{\mu} t^{-a_{\nu_0} - \Delta_{\mu}} d_{0 \nu_0 \mu}     \quad \to 0 \qquad {\rm as } \quad t \to \infty \,.
\ee

It is interesting to note that $d_{\nu_0}=0$ is required if the only way to take the limit is for the exponent in $b'_{\nu_0} = b_{\nu_0} \, t^{a_{\nu_0}}$  to lie in the regime $a_{\nu_0} > \frac{1}{2}$: In this case
finiteness of ${\rm vol}(B_3)$ requires that $d_{0 \nu_0 \nu_0} = 0$. But  we cannot draw this conclusion in full generality,
and hence we cannot exclude situations where the only shrinking curves are rational, while having a non-trivial normal bundle.

\section{Example of a base $B_3$ admitting both Type A and Type B  weak coupling limits} \label{App_Illustration}

We will present in this Appendix a complete analysis of the global limits for F-theory models, where the base three-fold of the elliptic fibration has the form:
\beq
B_3=\IP^1 \times \IP^2 \,. 
\eeq
We are interested in this geometry because it supports global limits of both Type A and Type B, as defined in Section~\ref{subsec_geometry1}, 
 in two different regimes of its K\"ahler moduli space.

The K\"ahler cone of $B_3$ is generated by the hyperplane classes $j_0$ of $\IP^1$ and $j_1$ of $\IP^2$. Since the weak coupling  limit must involve at least one large parameter in the expansion of the K\"ahler form in terms of these two generators, there arise two qualitatively different scenarios by identifying the large-parameter direction, $J_0$, as $j_0$ and $j_1$, respectively. We will analyze these two cases in turn. 

\subsection{Type A: $J_0= j_1$}
The choice
\be
J_0 =j_1.,
\ee
 with $J_0 \cdot J_0 \neq 0$, realizes  the scenario of Type A in the classification of  Section~\ref{subsec_geometry1}. Our general discussion hence assures that the WGC is satisfied in the weak coupling limit thanks to the excitation modes of the emergent tensionless heterotic string. In what follows we illustrate the geometric underpinnings of this general result by verifying the relation (\ref{volSvolC0}) explicitly. The importance of this relation for the WGC is explained in Section \ref{subsec_WGCbounds}.

To this end, observe that  the remaining K\"ahler cone generator, $J'=j_0$, has the following  non-zero triple intersection number:
 $J_0 \cdot J_0 \cdot J' = 1$. According to~\eqref{Jnugenerators}, 
   it is of type $\cI_1$. 
The global limit must then take the form
\beq
J= t J_0 + \frac{a}{t^2} J' \,, 
\eeq
where $a$ is a finite order-one constant such that the base volume,
\beq
{\rm vol}(B_3) = \frac12 a \,,
\eeq
is fixed to be finite in the limit $t \to \infty$. 
The volume of a general gauge divisor 
\beq
{\bf S} = h_0 J_0 + h J' \,
\eeq
can be expressed as
\beq
{\rm vol}({\bf S}) = \frac12 t^2 h + \frac{a}{t} h_0 ~\to~ \frac12 t^2 h \quad \text{as~} t\to \infty \,.
\eeq
Let us now consider the curve 
\beq
C_0=J_0 \cdot J_0 \neq 0 \,, 
\eeq
which represents the first factor of the product $B_3 = \IP^1 \times \IP^2$. Its volume is 
\beq
{\rm vol}(C_0) = \frac{a}{t^2} \,,
\eeq
and the normal bundle of the curve is 
\beq
N = \cO_{C_0} \oplus \cO_{C_0}\,,
\eeq
as $J_0^3 = 0$. Therefore, the tensionless string obtained by wrapping a D3 brane on the shrinking curve $C_0$ must be the
heterotic string. Moreover, the $U(1)$ fugacity index of this string is expressed as
\beq
m=\frac12 C_0 \cdot {\bf S} = \frac{h}{2} \,,
\eeq
and this confirms the advertised relation (\ref{volSvolC0}) 
between the volumes in the weak coupling limit,
\beq
{\rm vol} ({\bf S}) \, {\rm vol}(C_0) = \frac{ah}{2} = 2m {\rm vol}(B_3) \,.
\eeq

\subsection{Type B: $J_0 = j_0$}
Alternatively, let
\be
J_0 = j_0 \,.
\ee
 With $J_0 \cdot J_0 = 0$, this case corresponds to a limit of Type B as defined in Section~\ref{subsec_geometry1}. As  discussed there, it is not guaranteed that a heterotic string emerges in this case, but we have shown generally that a tensionless effective string must nevertheless appear, which causes a breakdown of the effective theory description in the limit. To illustrate this, we now explicitly identify the asymptotically shrinking  holomorphic curve  which gives rise to such a tensionless non-critical string when wrapped by a D3 brane. 

The remaining K\"ahler cone generator, $J'=j_1$, satisfies $J_0 \cdot J' \cdot J' = 1$ and is thus of type $\cI_2$ according to~\eqref{Jmugenerators}.
In the global limit the K\"ahler generator must  take the form
\beq
J= t J_0 + \frac{b}{\sqrt{t}} J' \,,
\eeq
where $b$ is a finite order-one constant such that the base volume,
\beq
{\rm vol}(B_3) = \frac12 b^2 \,,
\eeq
is fixed to be finite in the limit. 
The volume of a gauge divisor 
\beq
{\bf S} = h_0 J_0 + h J' \,
\eeq
 is computed as
\beq
{\rm vol}({\bf S}) =  \sqrt{t} b h + \frac{b^2}{2t} h_0 ~\to~ \sqrt{t} bh \quad \text{as~} t\to \infty \,.
\eeq
Let us now consider the curve 
\be
C_0  = J_0 \cdot J' \neq 0 \,,
\ee
which represents the hyperplane $\IP^1$ in the $\IP^2$ of the product $B_3 = \IP^1 \times \IP^2$. Its volume is given as
\beq
{\rm vol}(C_0) = \frac{b}{\sqrt{t}} \,,
\eeq
and hence, the curve shrinks in the limit $t \to \infty$. 
The normal bundle on the other hand is not trivial but rather
\beq
N = \cO_{C_0} \oplus \cO_{C_0}(1) \,,
\eeq
because of $J_0^2 \cdot J' = 0$ and $J_0 \cdot J'^2 = 1$. Therefore, we expect that the resulting effective string is non-critical. 
Interestingly, upon defining the fugacity index $m$ as
\beq
m=\frac12 C_0 \cdot {\bf S} = \frac{h}{2} \,,
\eeq
we observe an analogous relationship amongst the volumes in the global limit
\beq \label{relationTypeII}
{\rm vol} ({\bf S}) {\rm vol}(C_0) = b^2 h  = 4m {\rm vol}(B_3) \,.
\eeq
Notably, this differs by a factor of $2$ as compared to the relation  (\ref{volSvolC0}), which has played a crucial role in our quantitative proof of the WGC for limits of Type A. 

No other curve class on $B_3$ exists which would contain a rational curve with trivial normal bundle, and which shrinks in the limit considered here. 
Unless the weak coupling limit is obstructed dynamically, the WGC would imply that the non-critical string has likewise massless excitations that satisfy the super-extremality bound.
The numerical discrepancy between (\ref{relationTypeII}) and (\ref{volSvolC0}) could e.g. be compensated by a modified mass-shell condition for the light excitations.
It would be very interesting to investigate this further.

\section{Example of an emergent critical string in a Type B weak coupling limit } \label{TypeIIExample-App}

While the tensionless string that emerges in weak coupling
limits of Type B as defined in Section~\ref{subsec_geometry1} is not necessarily the heterotic string,
this possibility is not excluded either.
In this Appendix we provide an explicit example where such a heterotic string becomes asymptotically tensionless  in a Type B weak coupling limit.
As we will see, in special regions in the moduli space, we can even encounter several of such `heterotic string' curves shrinking at the same time -- a phenomenon which deserves further investigation. 

Let us consider the base three-fold $B_3 = \IP^1_0 \times \IP^1_1 \times \IP^1_2$. The K\"ahler cone of $B_3$ is generated by the hyperplane classes $j_{i}$ of the $\IP^1_i$ factors ($i=0,1,2$), which satisfy
\be
j_i\cdot j_i = 0 \,,\qquad  j_0 \cdot j_1 \cdot j_2 = 1 \,. 
\ee
Since every generator self-intersects trivially, this geometry only admits a weak coupling of Type B.
Without loss of generality we identify
\bea
J_0 &=& j_0 \,,\\
J_\mu &=& j_\mu \,,\quad  \mu=1,2 \in \cI_2 \,.
\eea
In the global limit the K\"ahler form must then take the form
\beq
J=t J_0 +\sum_\mu b'_\mu J_\mu \,,
\eeq
where $b'_\mu$ should be chosen such that the base volume
\beq\label{Ex:volB}
{\rm vol}(B_3) = tb'_1b'_2
\eeq
remains finite in the limit $t \to \infty$. 
Denoting the gauge divisor $\bf S$ as
\beq
{\bf S} = h_0 J_0 + \sum_\mu h_\mu J_\mu \,,
\eeq
we can express its volume as
\beq
{\rm vol}({\bf S}) = b_1' b_2' h_0 + t(b_1' h_2 + b_2' h_1) ~\to~t(b_1' h_2 + b_2' h_1)  \quad \text{as~} t\to \infty \,,
\eeq
where the first term is dropped in the limit as the volume~\eqref{Ex:volB} of $B_3$ is finite. 

Let us now consider the two curves 
\be
C_0^{(\mu)}  = J_0 \cdot J_\mu \neq 0 \,,
\ee
i.e. the curves $\IP^1_2$ and $\IP^1_1$ for $\mu=1$ and $2$, respectively. Their volumes take the form
\beq
{\rm vol}(C_0^{(\mu)}) =
\left\{ \begin{array}{rl}
b_2' &\mbox{ for $\mu=1$} \\
b_1' &\mbox{ for $\mu=2$}
\end{array} \right.
\eeq
and they both have trivial normal bundle,
\beq
N^{(\mu)} = \cO_{C_0^{(\mu)}} \oplus \cO_{C_0^{(\mu)}}\,,
\eeq
because $J_0 \cdot J_\mu \cdot J_\mu = 0$. Therefore, two different types of heterotic strings arise by wrapping a D3 brane on $C_0^{(\mu)}$. Their respective $U(1)$ fugacity indices are given as
\beq
m^{(\mu)}=\frac12 C_0^{(\mu)} \cdot {\bf S} =
\left\{ \begin{array}{rl}
\frac{h_2}{2}&\mbox{ for $\mu=1$} \\
\frac{h_1}{2} &\mbox{ for $\mu=2$} \,.
\end{array} \right.
\eeq
To make the limiting behavior of the various volumes clearer, let us define the two ratios of the parameters
\beq
r_b \equiv \frac{b_2'}{b_1'} \,,\quad r_h \equiv \frac{h_2}{h_1} \,,
\eeq
and compute the product of the volumes of $\bf S$ and $C_0^{(\mu)}$ in the limit:
\beq\label{mismatched}
{\rm vol}({\bf S}) {\rm vol}(C_0^{(\mu)}) = 
\left\{ \begin{array}{ll}
t(b_1' h_2 + b_2' h_1) b_2' = 2 (1+ \frac{r_b}{r_h}) m^{(1)} {\rm vol}(B_3) &\mbox{ for $\mu=1$} \\
t(b_1' h_2 + b_2' h_1) b_1' = 2 (1+\frac{r_h}{r_b}) m^{(2)} {\rm vol}(B_3) &\mbox{ for $\mu=2$} \,.
\end{array} \right.
\eeq
Recall from Section \ref{sec_WGC} that a proof of the WGC relation for the excitation modes of these strings would require that the RHS takes the form $2m^{(\mu)} {\rm vol}(B_3)$. As we see, our geometric  findings so far differ from this by the 
 respective prefactors
\beq
\alpha^{(\mu)}=
\left\{ \begin{array}{ll}
1+r&\mbox{ for $\mu=1$} \\
1+\frac{1}{r}&\mbox{ for $\mu=2$} \,,
\end{array} \right.
\eeq
where $r \equiv \frac{r_b}{r_h}$ is the ratio of $r_b$ and $r_h$.

 The ratio $r_h$ of the two parameters $h_\mu$ in the divisor class ${\bf S}$ is some fixed positive number of order one, since $h_\nu$ are  positive integers 
 that are fixed in a given F-theory background.\footnote{Both $h_\mu$ have to be strictly positive for $\bf S$ to be the divisor of a height pairing, which corresponds to a gauge group ${\cal G} = U(1)$, whether or not there exist additional non-abelian gauge group factors. This is because the height pairing can be written as~\cite{Lee:2018ihr}
\beq
{\bf S} = \frac{1}{N^2} (2 \bar K + 2 \pi_* ({\rm div}(N s) \cdot S_0)) \,,
\eeq
where $s$ denotes the section to the elliptic fibration that generates the Mordell-Weil lattice. Morevoer $N$ is a fixed positive integer determined by the geometry of the F-theory background (note that $N=1$ in the absence of additional non-abelian group factors). Given that the cone of effective divisors is spanned by $J_0$, $J_1$ and $J_2$ and that $2\bar K = 4 J_0 + 4 J_1 + 4 J_2$, we immediately learn that $h_0$, $h_1$ and $h_2$ are all larger than or equal to $\frac{4}{N^2}$ and hence positive.  
}
On the other hand, the two parameters $b_1'$ and $b_2'$ are  positive numbers constrained only by the requirement that the product $t b_1' b_2' (= {\rm vol}(B_3))$ is finite in the limit. 
In a {\it generic} weak coupling limit, $b_1'$  and $b_2'$ do not scale with the same power of $t$, and the 
 ratio $r_b$ for a generic such limit therefore satisfies either $r_b \to 0$ or $r_b \to \infty$. 
 Hence in such a {\it generic} limit either $r$ or $\frac{1}{r}$  vanishes for $t \to \infty$, and precisely one of the two types of the curves $C_0^{(\mu)}$ asymptotically satisfies the relationship
\beq\label{desired}
{\rm vol}({\bf S}) {\rm vol}(C_0^{(\mu)}) =  2 m^{(\mu)} {\rm vol}(B_3) \,.
\eeq
The heterotic string associated with this curve is the one whose tension vanishes parametrically  faster and which is parametrically more weakly coupled. 
Furthermore, {there exists no other class} of a shrinking rational curve on $B_3$ with trivial normal bundle. 
Its quantisation thus gives rise to a required tower of massless particles, supporting the WGC at the quantitative level. 

In the light of F-/heterotic duality we interpret the respective curve $C_0^{(\mu)}$ which vanishes at the parametrically faster rate as the fiber of a (trivial) $\mathbb P^1$-fibration over a base $B_2$.
The D3-brane along this fiber dualises to the critical fundamental string in a dual heterotic  description. 
The other vanishing  curve then defines a $\mathbb P^1$ on the base $B_2$. The D3-brane along it dualises to an NS5-brane on the elliptic fibration over this curve on the heterotic side. Since this surface is a K3, we indeed recover a string whose worldsheet theory coincides with that of the heterotic string. However, the ratio of the tensions of the fundamental to this effective string vanishes, which determines the first to be the string defining the duality frame. 
In particular, the effective string is parametrically more strongly coupled than the fundamental string, and we cannot trust its spectrum. 
This explains why there is no doubling of e.g. massless modes in this regime in moduli space. The situation is  analogous to the co-existence of a fundamental Type I string and an effective SO(32) heterotic string in 10d Type I, where the role of the effective heterotic string is played by the D1-brane.

Interestingly, the moduli space does allow for a non-generic ``symmetric'' weak coupling limit where both $b_\mu'$ take the form,
\beq
b_\mu' = \frac{b_\mu}{\sqrt{t}}\,,
\eeq 
for finite positive numbers $b_\mu$. In this situation, $r$ is a finite positive number and hence, none of the two curves $C_0^{(\mu)}$ obey the desired relationship~\eqref{desired}. 
The dual heterotic model is at a self-dual point, where the tension and coupling of both the `fundamental' and the `effective' heterotic string are parametrically comparable and hence no distinction between the two makes sense.
It is clear that due to the appearance of such tensionless strings the effective theory description must break down, but a quantitative analysis of the WGC relation is not possible with our present methods. This is why the mismatch between (\ref{mismatched}) and the desired value does not yet indicate a violation of the WGC.
It would be very interesting to further study how such a non-generic regime in the moduli space has to be interpreted.

\bibliography{papers}
\bibliographystyle{JHEP}

\end{document}